\documentclass[useAMS,usenatbib]{mn2e}

\usepackage{graphicx}
\usepackage{natbib}

\newcommand{\km}{\,\mbox{km}\,\mbox{s}^{-1}}

\def\Ha{H$\alpha$}

\title[Ionized gas in dwarf galaxies]{Ionized gas velocity dispersion in nearby dwarf galaxies: looking
at supersonic turbulent motions.
\thanks{Based on observations obtained with the 6-m telescope of the
Special Astrophysical Observatory of the Russian Academy of
Sciences (SAO RAS). The observations were carried out with the financial
support of the Ministry of Education and Science of Russian Federation
(contracts no.~16.518.11.7073 and 16.552.11.7028).
}}
\author[Moiseev \& Lozinskaya]
{Alexei V. Moiseev,$^1$\thanks{moisav@gmail.com} and Tatiana A. Lozinskaya$^2$\\
$^1$Special Astrophysical Observatory,  Russian Academy of Sciences, Nizhnii Arkhyz, 369167  Russia\\
$^2$Sternberg Astronomical Institute of Lomonosov Moscow State  University, Moscow,
119992 Russia }

\begin{document}

\date{Accepted ....  Received ....}

\pagerange{\pageref{firstpage}--\pageref{lastpage}} \pubyear{2011}

\maketitle

\label{firstpage}

\begin{abstract}

We present the results of   ionized gas  turbulent motions study in several nearby
dwarf galaxies using a scanning Fabry-Perot interferometer with the 6-m telescope of
the SAO RAS. Combining the `intensity--velocity dispersion' diagrams ($I-\sigma$) with
two-dimensional maps of radial velocity dispersion we found a number of common patterns
pointing to the relation between the  value of chaotic ionized gas motions and
processes of current star formation. In five out of the seven  analysed galaxies we
identified expanding shells of ionized gas with diameters of 80--350 pc and kinematic
ages of 1--4 Myr. We also demonstrate that the $I-\sigma$ diagrams may be useful for
the search of supernova remnants, other small expanding shells or unique stars
in nearby galaxies. As an  example,  a  candidate  luminous blue variable (LBV) was
found in UGC 8508. We propose some additions to the interpretation, previously used by
Mu\~{n}oz-Tu\~{n}\'{o}n et al. to explain the $I-\sigma$ diagrams for  giant star
formation regions. In the case of dwarf galaxies, a major part of the regions with high
velocity dispersion belongs to the diffuse low  surface brightness  emission, surrounding the
star forming regions. We attribute this to the presence of perturbed  low density gas
with high values of turbulent velocities around the giant HII regions.
\end{abstract}


\maketitle

\section{Introduction}

The study of the gas component of dwarf galaxies is important and challenging for
several reasons. Firstly, due to the shallow potential well and a lack of spiral
density waves, such galaxies provide a good opportunity   for studying the  interaction
of young stellar groups with the interstellar medium. The  ionizing radiation of OB
stars, as well as the kinetic energy of stellar winds and supernova explosions are
heating the gas, forming  the cavities, bubbles, shells, ordered outflows and chaotic
turbulent motions in the gaseous disc. Secondly, it is important to be able to properly
account for the influence of these effects on the gaseous medium in order to estimate
the   circular rotation curve from the observed distribution of radial velocities.
The information about  an accurate rotation curve is critical to study the distribution of
dark matter and the mass function of dwarf galaxies within various cosmological tests.

Two-dimensional distributions of the parameters of neutral and ionized hydrogen yield
the most complete and detailed data on the structure and kinematics of the interstellar
medium. We managed to compare the kinematics of neutral  and ionized gas  for several
dwarf galaxies  of the Local Group using the `position--velocity' diagram, allowing to
identify  expanding  shells \citep{Lozinskaya2003, Lozinskaya2008}. For more distant
objects, the analysis of the velocity field and the shapes of \Ha\, and 21~cm line
profiles \citep{vanEymeren2009a, vanEymeren2009b} allows to identify gas outflows above
the galactic plane. The advantage of the radio interferometry method in the 21~cm line
is the capability to map the HI distribution far beyond the optical discs of galaxies,
see e.g. the results of the FIGGS \citep{Begum2008} and THINGS \citep{Oh2011} surveys.
However, the typical angular resolution of such observations (beam$=10-30$ arcsec)
often provides an excessively smooth picture and does not allow to study the motions of
gas on small spatial scales. In contrast to radio observations, with 3D spectroscopy in the
optical emission lines we can study the kinematics of ionized gas with a much higher
resolution $1-3$ arcsec. However, it is limited only to the regions, having enough UV-photons to
ionize the gas.

Until recently, most of the two-dimensional data on the kinematics of gas in dwarf
galaxies were obtained from the radio observations in the HI line. Lately though, the
situation is changing. The results of optical studies of the kinematics   not only for
individual objects, but also for small samples of galaxies are published
\citep*{Ostlin1999,Moiseev2010}. Velocity fields of ionized gas in two hundred galaxies
of different morphological types, 23 of which belong to the dIrr type,  were
constructed within the GHASP survey \citep*{Epinat2008} with the scanning
Fabry-Perot interferometer (FPI). The growing accuracy of measurements allowed to
build and analyze not only the radial velocity fields, but also the two-dimensional
maps of  ionized gas velocity dispersion ($\sigma$) in some dwarf galaxies with violent
star formation. The $\sigma$ maps of such galaxies were published, resulting from the
observations using the methods of integral field spectroscopy \citep[see,
e.g.,][]{Bordalo2009,Lagos2009,Monreal-Ibero2010} or interferometry with the scanning
FPI \citep{delgado,Lozinskaya2006,Moiseev2010}. Although such data as yet exist for
less than two dozen objects, the number of constructed $\sigma$ maps will be growing.
That way, we are preparing to publish the results of observations of another few dozen
of dwarf galaxies \citep*{Moiseev2012}.

The nature of the supersonic ($\sigma>10-20\km$) turbulent motions of ionized gas,
observed in dwarf galaxies is the subject of a longstanding debate \citep[see
discussion and references in][] {BordaloTelles2011}. Thus, a number of authors
\citep*{TerlevichMelnick1981,Melnick2000} believe that the intensity-weighted mean
velocity dispersion of ionized gas ($\sigma_m$) in dwarf galaxies and in giant HII
regions is determined by virial motions, i.e. by the total mass of the system. On the
other hand, in their recent paper \citet{Green2010} show that in a wide range of galaxy
luminosities $\sigma_m$ is determined only by the current star formation rate
(estimated from the flux in the \Ha\  emission line) and does not correlate with the mass of
the galaxy. In this case, $\sigma_m$ is a characteristic of mechanical energy,
inferred to the gas by the stellar winds and supernova explosions in the process of
current star formation. For  neutral gas velocity dispersion, a similar conclusion was
earlier made by \citet*{Dib2006}.

To address this and other problems, related to the interaction of
stellar populations and interstellar medium, we must be able to
correctly interpret the structures, observed in the ionized gas
velocity dispersion maps. Among the few studies on this subject,
let us mention the papers by \citet{Yang1996} and
\citet{MunozTunon1996}, who thoroughly examined two giant star
formation regions NGC604 and NGC 588 in the disc of the M33
galaxy. The authors proposed a method for interpreting the
observed distributions of points on the `line intensity--velocity
dispersion' diagram ($I-\sigma$) in terms of evolution of ionized
shells in stellar groups. Later, with the aid of these and other
similar diagrams \citet{delgado} examined the behaviour of ionized
gas in three more distant BCD galaxies.

How universal is this method? This paper is a subsequent attempt to thoroughly consider
the nature of the structures, observed in the ionized gas velocity dispersion maps. The
observational data presented here were obtained at the 6-m telescope of the SAO RAS
within the project devoted to the internal kinematics of ionized gas in dwarf galaxies
of the Local Volume ($D<10$ Mpc). To correctly interpret the results of this project,
 it seems to be necessary to consider the most typical features, observed in the
$\sigma$ maps. In order to make a detailed analysis, we have selected seven most nearby
galaxies ($D=1.8-4.3$ Mpc) from the total sample, where we managed to fairly well study
both the bright HII regions, and  faint diffuse \Ha\, emission between them. It is
important that there are  HI distribution maps published for all the selected galaxies,
which allows to compare the distributions of ionized and neutral fractions of the
interstellar medium. In order to better understand the features of the $\sigma$
distribution, we employ the data on two more nearby dIrr galaxies, IC 10 and IC 1613,
we previously studied in detail.

\section{Observations and data reduction}

\label{obs}

\subsection{Observations with the scanning FPI}

The observations were made at the prime focus of the 6-m telescope
of SAO RAS using a scanning FPI,
installed inside the SCORPIO focal reducer
\citep{AfanasievMoiseev2005}. The operating spectral range around
the  H$\alpha$ line was cut by a narrow-band filter with a
bandwidth of  $FWHM=15-21$\AA. Most of the observations were made
with the FPI501 interferometer, providing in the H$\alpha$ line  a
free spectral range between the neighboring interference orders
$\Delta\lambda=13$\,\AA\, and spectral resolution ($FWHM$ of the
instrumental profile) of about $0.8$\AA\, ($35\km$), with the
scale of 0.36\AA\, per channel. In November 2009, a new
interferometer FPI751  was used, having $\Delta\lambda=8.7$\,\AA\,
and a spectral resolution of  $0.4$\AA\, ($18\km$) at the scale of
0.21\AA\, per channel.

Between 2005--2009  we used the EEV~42-40 and E2V~42-90 CCD
detectors, providing the image scale of $0.71\,\mbox{arcsec}\,
\mbox{pixel}^{-1}$ in   $4\times4$ on-chip binned mode. In
2002 a TK1024 CCD detector was used operating in on-chip
binned $2\times2$ pixel mode, yielding the image scale of $0.56\,\mbox{arcsec}\, \mbox{pixel}^{-1}$.

During the scanning process, we have consistently obtained 36 interferograms of a given object (40 for FPI751) at different
distances between the FPI plates. The seeing at different nights ranged from 1 to 3 arcsec. The data reduction was performed using the software package running in the IDL environment. Following the primary reduction, airglow lines subtraction, photometric and seeing corrections using the reference stars and wavelength calibration, the observational data were combined into the data cubes, where each pixel in the field of view contains a 36- (or a 40-) channel spectrum.  The detailed description of data  reduction algorithms and software is presented  in   \citet{Moiseev02ifp,MoiseevEgorov2008}. All galaxies except VII~Zw~403 were observed  in two scanned cycles in order to remove parasitic ghost refllection, as described by \citet{MoiseevEgorov2008}.

The log of observations is given in Table~\ref{tab_obs}, listing the following data: the name of the galaxy; the distance, scale
($\mbox{pc}\,\mbox{arcsec}^{-1}$) and absolute stellar magnitude $M_B$ according to \citet{Karachentsev2004}; the epoch of observations; the type of interferometer; total exposure time; resulting angular resolution ($\omega$) after the  smoothing by a two-dimensional  Gaussian (typical $FWHM=1.0-2.0$~pix)  in order to increase the signal-to-noise ratio in the regions of low surface brightness.  This value was estimated from foreground stars images in final data cube.

\begin{table*}
\caption{Log of the observations}\label{tab_obs}
\begin{tabular}{lcccccccc}
\hline
Name           & $D$     & scale                     & $M_B$          & Date            & FPI    & Exp. time          &\multicolumn{2}{c}{$\omega$}  \\
                    &  (Mpc)  & (pc\,arcsec$^{-1})$&                     &                    &          &  time (s)            &  (arcsec)  & (pc)\\
\hline
DDO~53      & 3.56     & 17                          &-13.21  & 26.02.2009 & FPI501& $200\times36$&  3.3        & 57 \\
DDO~99      & 2.64     &  13                         & -13.41 & 26.02.2009 & FPI501& $180\times36$&  3.8        &49\\
DDO~125    & 2.54     &  12                         & -14.07 & 18.05.2005 & FPI501& $180\times36$&  3.0        &37\\
DDO~190    & 2.79     &  14                         & -14.13 & 04.03.2009 & FPI501& $100\times36$&  3.3        &45\\
UGC~8508   & 2.56     & 12                         & -12.92 & 16.05.2005 & FPI501& $200\times36$&  3.0        &37\\
UGCA~92     & 1.80     & 9                          &  -14.48 & 10.11.2009 & FPI751& $180\times40$&  2.5        &22\\
VII~Zw~403 &  4.34    &  21                        & -13.87 & 29.11.2002 & FPI501& $300\times36$&  2.2        &46\\
  \hline
\end{tabular}
\end{table*}

\subsection{Velocity dispersion measurements}

Referring to the velocity dispersion of ionized gas ($\sigma$) in the present paper we
mean the standard deviation of the Gaussian, describing the profile of the H$\alpha$
emission line after accounting for the FPI instrumental profile and subtracting the
contribution of thermal broadening in the HII regions. We used the method described in
detail in \citet{MoiseevEgorov2008}.

The width of the instrumental profile of the FPI was measured from the calibration lamp
line spectrum. The observed H$\alpha$ line profiles were fitted by the Voigt function,
which gives a good description of the observed profile. From the results
of profile fitting, we constructed two-dimensional radial velocity fields of ionized
gas, radial velocity dispersion maps free from the effects of the instrumental profile
($\sigma_{real}$), as well as the galaxy images in the    H$\alpha$ emission line and in the
continuum. The flux maps were calibrated  to the absolute scale of surface brightness
($\mbox{ergs}\,\mbox{s}^{-1}\,\mbox{cm}^{-2}\,\mbox{arcsec}^{-2}$)
 by normalizing to the total flux of the galaxy in the H$\alpha$ line, known from the literature
\citep[see][]{Moiseev2012}.

The accuracy of velocity dispersion was estimated from the measurements of the
signal-to-noise ratio, using the relations, given in Fig.~5, \citet{MoiseevEgorov2008}.
On the $\sigma$ maps, we masked the regions with a weak signal, where the formal error
of velocity dispersion measurements exceeded $6-9\km$ (which corresponds to the
$S/N\leq 7$). Meanwhile  the line flux maps were constructed confidently
enough even for weaker lines, down to $S/N\approx2-3$.

The transition from the measured $\sigma_{real}$ to the needed
$\sigma$ was done according to the relation from
\citet{Rozas2000}:
\begin{equation}
\sigma^2=\sigma_{real}^2-\sigma_{N}^2-\sigma_{tr}^2,
\label{eq1}
\end{equation}
where $\sigma_{N}\approx3\km$ and  $\sigma_{tr}\approx9.1\km$
correspond to the natural width of the emission line and its
thermal broadening at $10^4$ K.

In the case of  UGC~8508, using the  formula (\ref{eq1}) we get  to  $\sigma^2<0$ in the centre
of several HII regions. This is most likely caused by a smaller value of thermal broadening  there due to
$T_e<10^4$ K. For this galaxy  we adopted $\sigma_{tr}=7.5\km$ that corresponds to the $T_e\approx6800$ K.

In most of the objects, the observed line profile is very well described by a single-component Voigt profile.

Of all the galaxies considered, only UGC 8508 and UGCA~92 have regions in which the  \Ha\ profile reveals two components with ther velocity separation larger than,  or comparable to the spectral resolution limit ($35$ and $18\km$ respectively).

\section{$I-\sigma$ diagrams}

Figure~\ref{fig1_1}  shows the monochromatic images in the \Ha~ line, the
$\sigma$ maps, as well as the $I-\sigma$ diagrams for all the galaxies. Our diagrams
differ from those  built earlier \citep{Yang1996,MunozTunon1996,delgado}:  instead
of the peak line intensity ($I_{peak}$), we use the total flux in the line ($I$),
having a clearer physical meaning (the luminosity per area unit). Since the $\sigma$
maps were masked by the fixed $S/N$ level, and the noise level in the outer regions is
mainly determined by the sky background, the left-hand side boundary of the point cloud in
the diagrams is an inclined line $I\sim \sigma I_{peak}$, where $I_{peak}=const$. The main advantages  of the $I$  against $I_{peak}$ is that the total flux obviously does not depend on spectral resolution. In order to  understand the difference of result using $I$ instead of $I_{peak}$ we have created both types diagrams for  the sample galaxies and we conclude that all the shell-like features are preserved on the diagrams. We also abandoned the linear scale of intensities and use the $\log I$, since the observed  surface brightness difference is 2-3 orders of magnitude, which is significantly larger than the brightness range inside the giant HII regions in \citet{MunozTunon1996}.

The characteristic features, commented on below, are illustrated with different colours
in the diagrams. The horizontal  lane in dark blue has a relatively low velocity dispersion
$\sigma$ and high surface brightness. The boundaries  of brightness are chosen so that
the highlighted regions contain 50\% of the total luminosity of the galaxy in the \Ha\,
emission line.  By red, yellow and green colours  we show the regions with
increased $\sigma$, trying to wherever possible identify the shell structures.
The red line in the diagrams marks the level of mean velocity dispersion across the galaxy, intensity-weighted in the $i$-th pixel:
$\sigma_m= \sum \sigma_i I_i/\sum I_i $. When we further refer to  `high' or `low'
velocity dispersion, we have in mind $\sigma>\sigma_m$ and $\sigma\leq \sigma_m$,
respectively. For instance, other regions with $\sigma>\sigma_m$ are shown by orange, whereas
low-luminosity regions with  low $\sigma$ are marked in gray.

Figure~\ref{fig1_1} also contains  maps indicating the locations of regions, marked on the corresponded $I-\sigma$ diagrams.

\begin{figure*}
\centerline{
\includegraphics[width=0.5\textwidth]{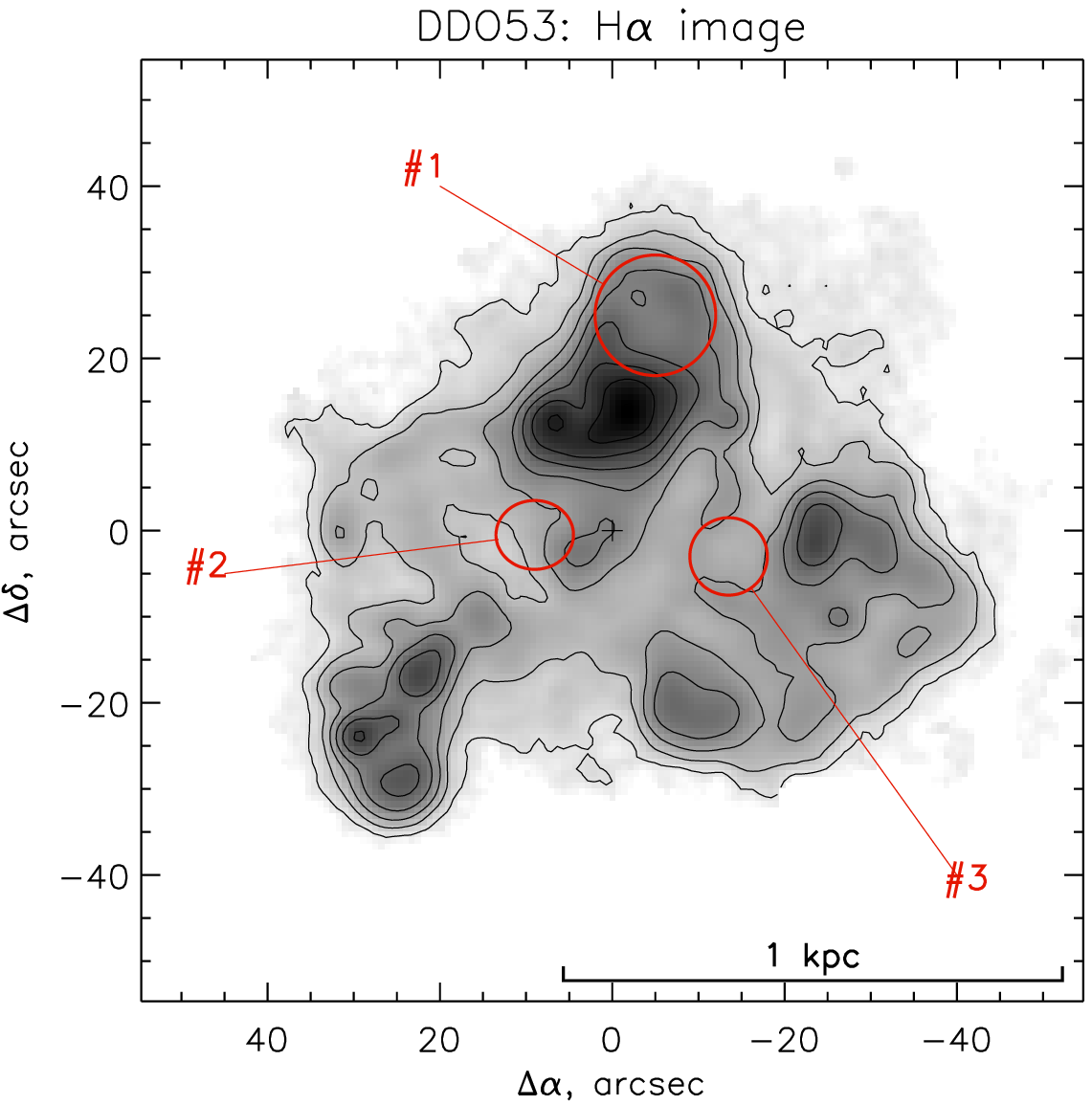}
\includegraphics[width=0.5\textwidth]{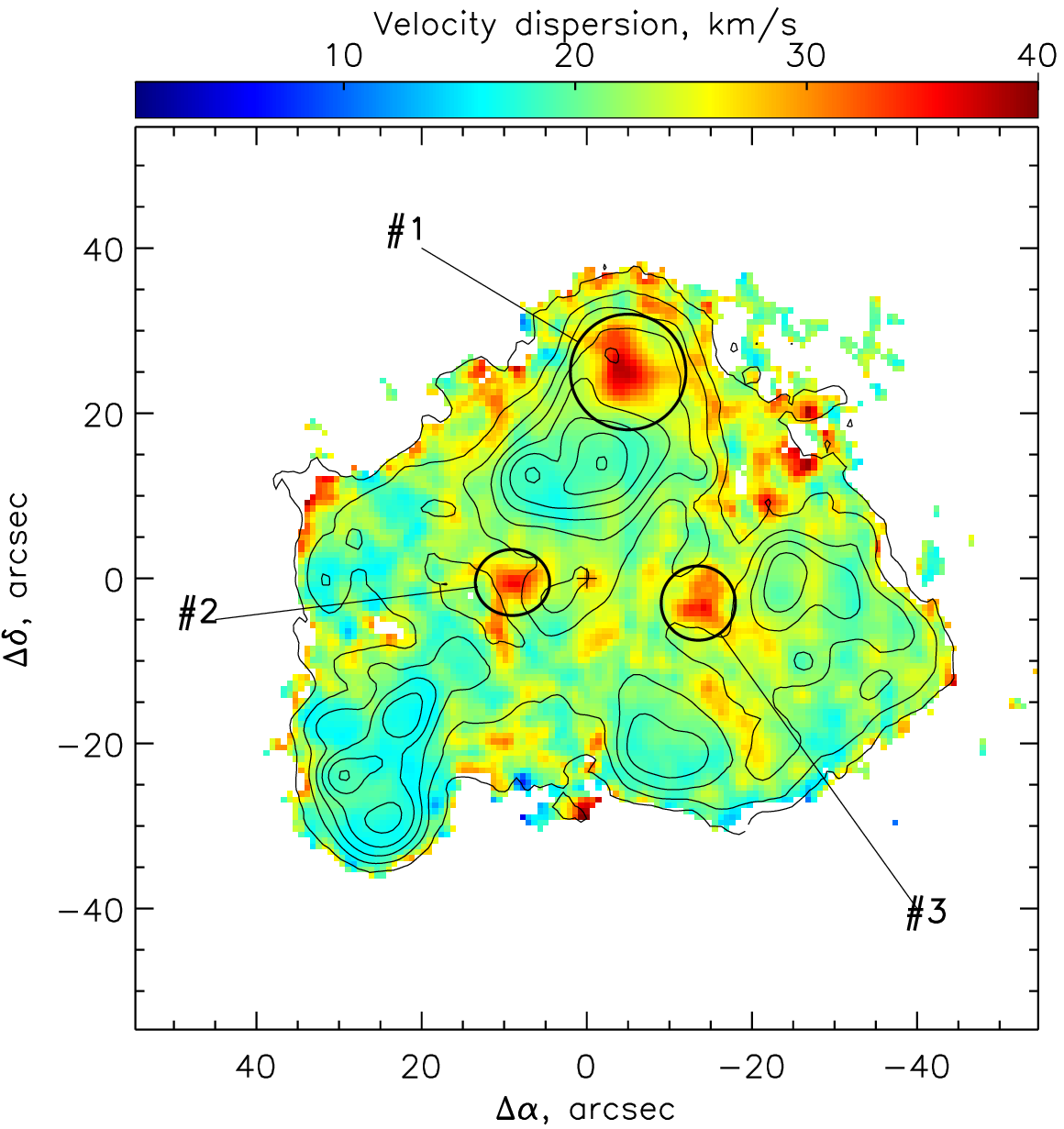}
}
\centerline{
\includegraphics[width=0.5\textwidth]{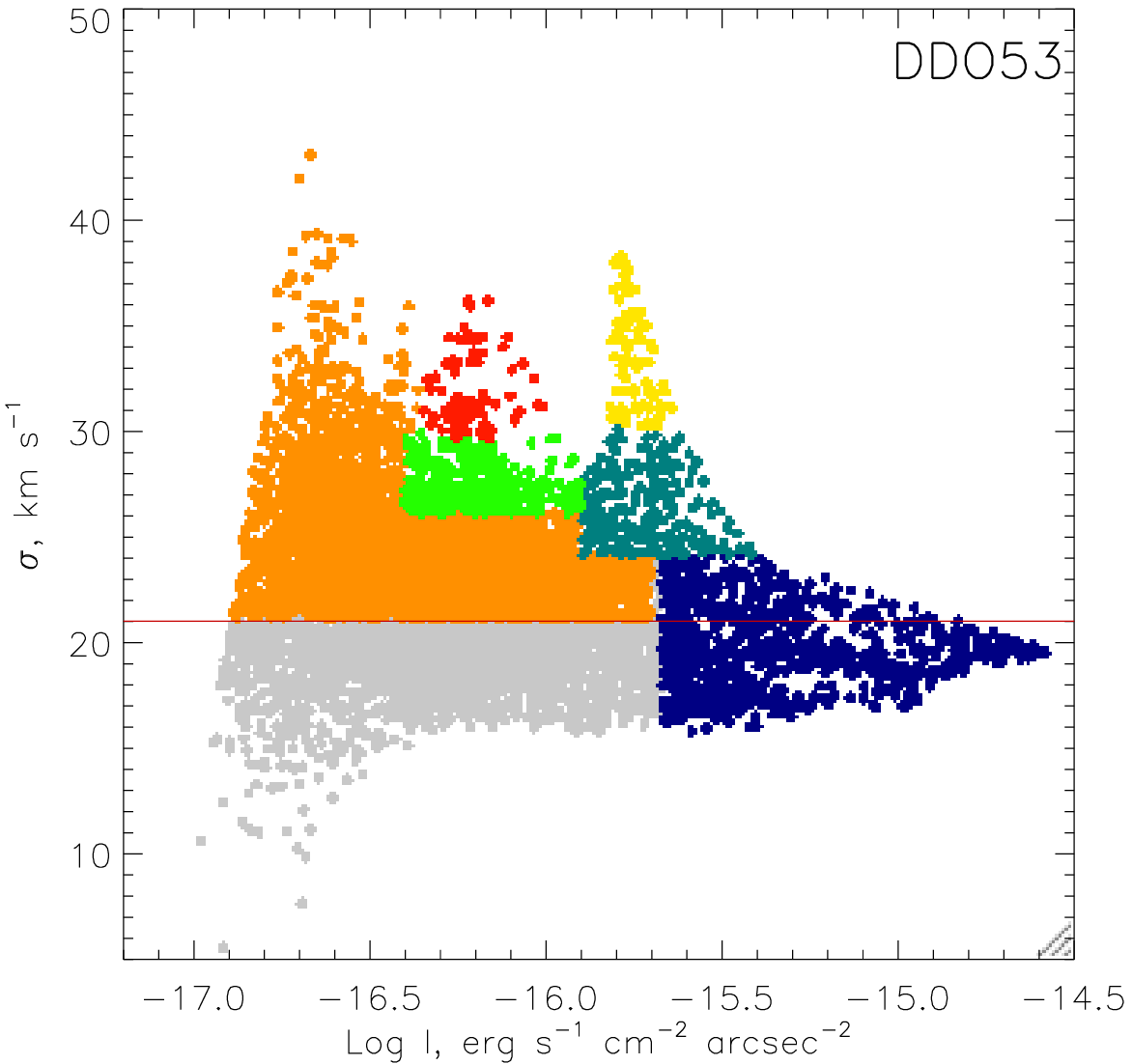}
\includegraphics[width=0.5\textwidth]{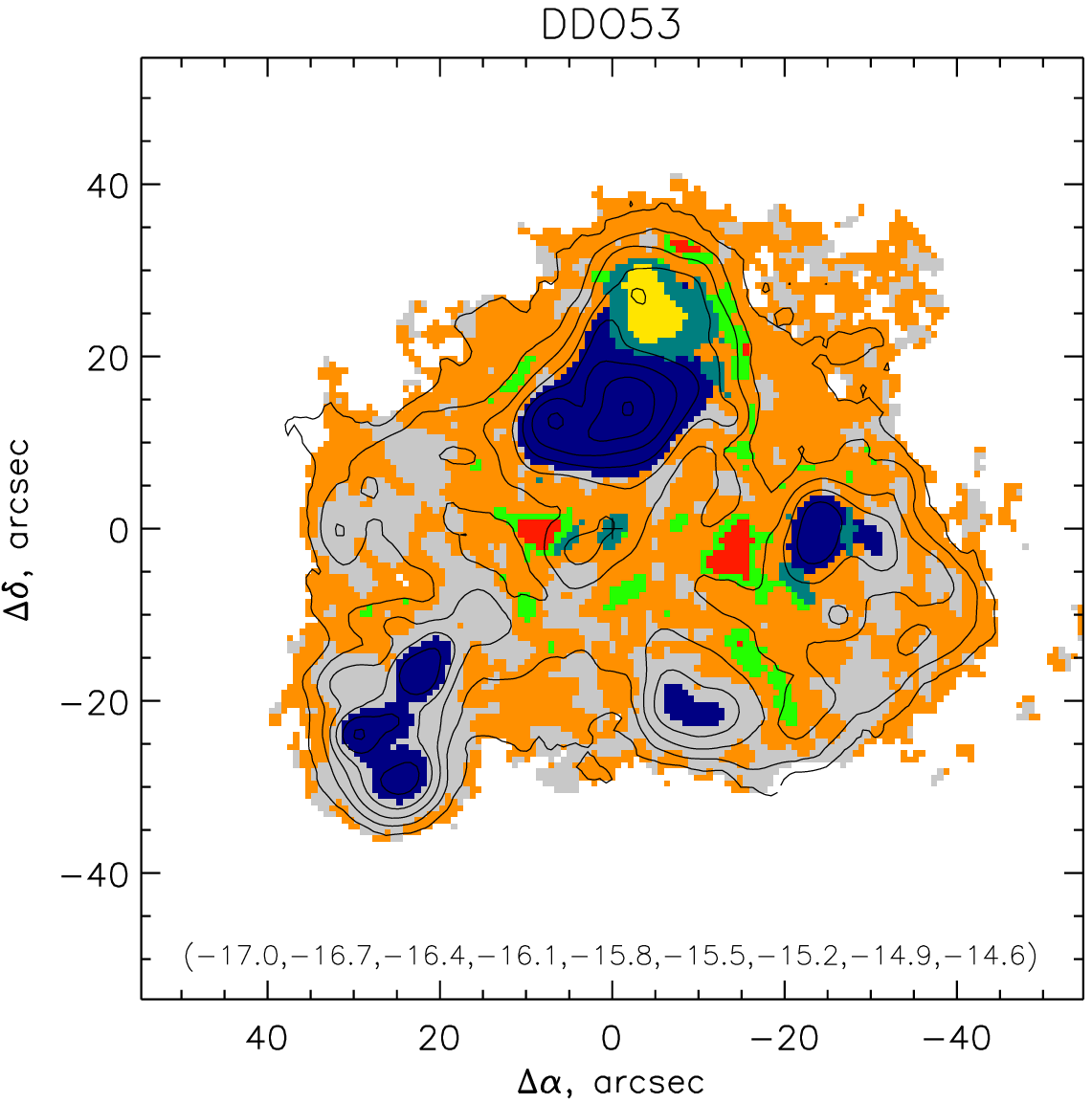}
} \caption{Results of observations with the FPI for each galaxy. The top row: the image
in the \Ha\, line  in logarithmic values of intensity  (left) and the velocity
dispersion map with superimposed contours of the \Ha\, image (right). The ellipses with
numbers mark the  shell-like structures, the parameters of which are given in
Table~\ref{tab_shells}.  For DDO~125 and UGC~8508   the ellipses without numbers show large-scale cavern, discussed in the text.
The cross marks the centre of the image in the continuum. The
bottom row: the intensity--velocity dispersion diagram (left). The red horizontal line
marks the intensity-weighted mean velocity dispersion $\sigma_m$. The right plot shows
the location of regions, marked by different colours on the $I-\sigma$ diagram. The
contours correspond to the isophotes in the \Ha\, line, their values (in $\log$ of $\mbox{ergs}\,\mbox{s}^{-1}\,\mbox{cm}^{-2}\,\mbox{arcsec}^{-2}$) are printed in brackets at the right bottom figures.} \label{fig1_1}
\end{figure*}

\begin{figure*}
\centerline{
\includegraphics[width=0.5\textwidth]{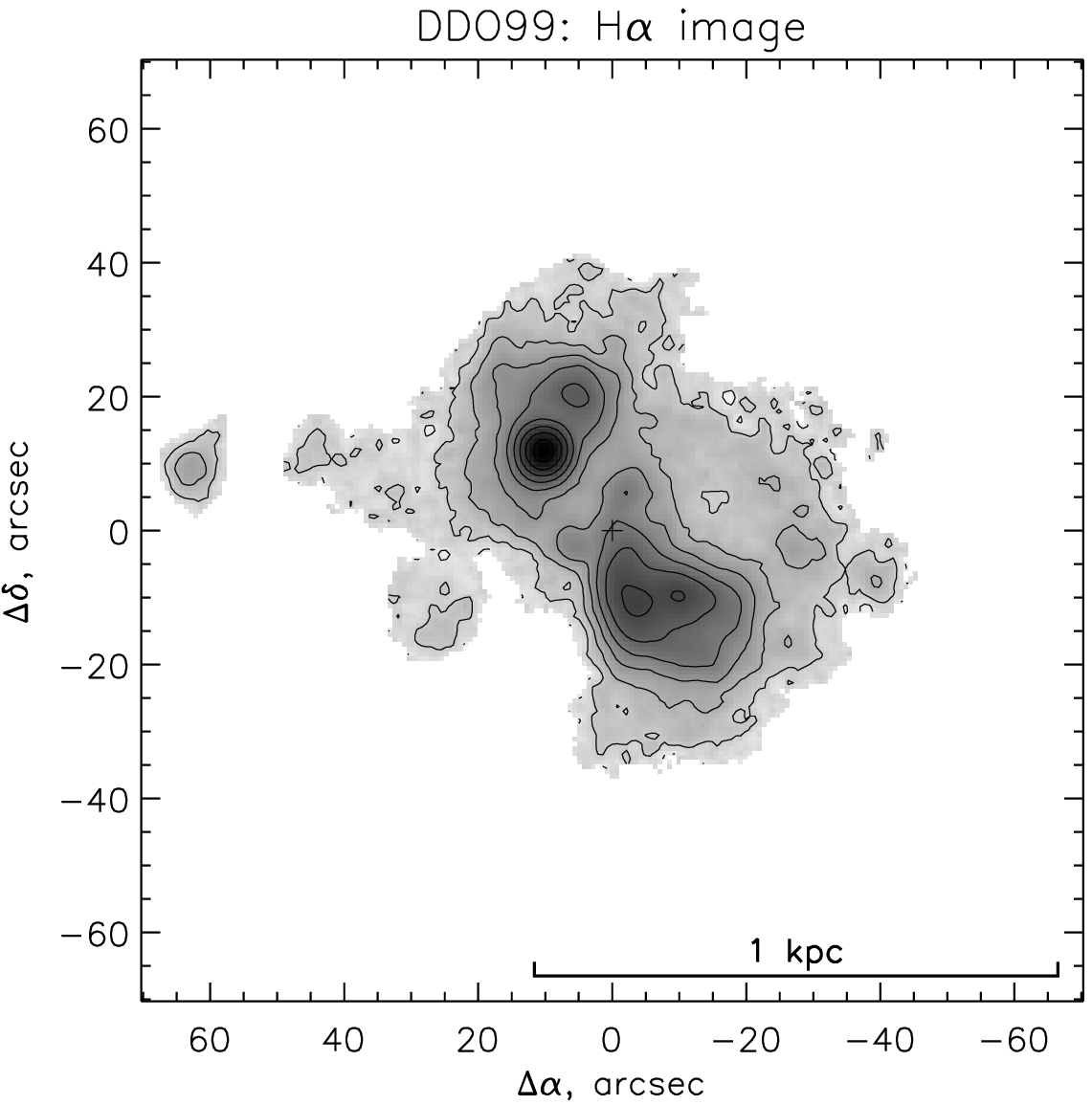}
\includegraphics[width=0.5\textwidth]{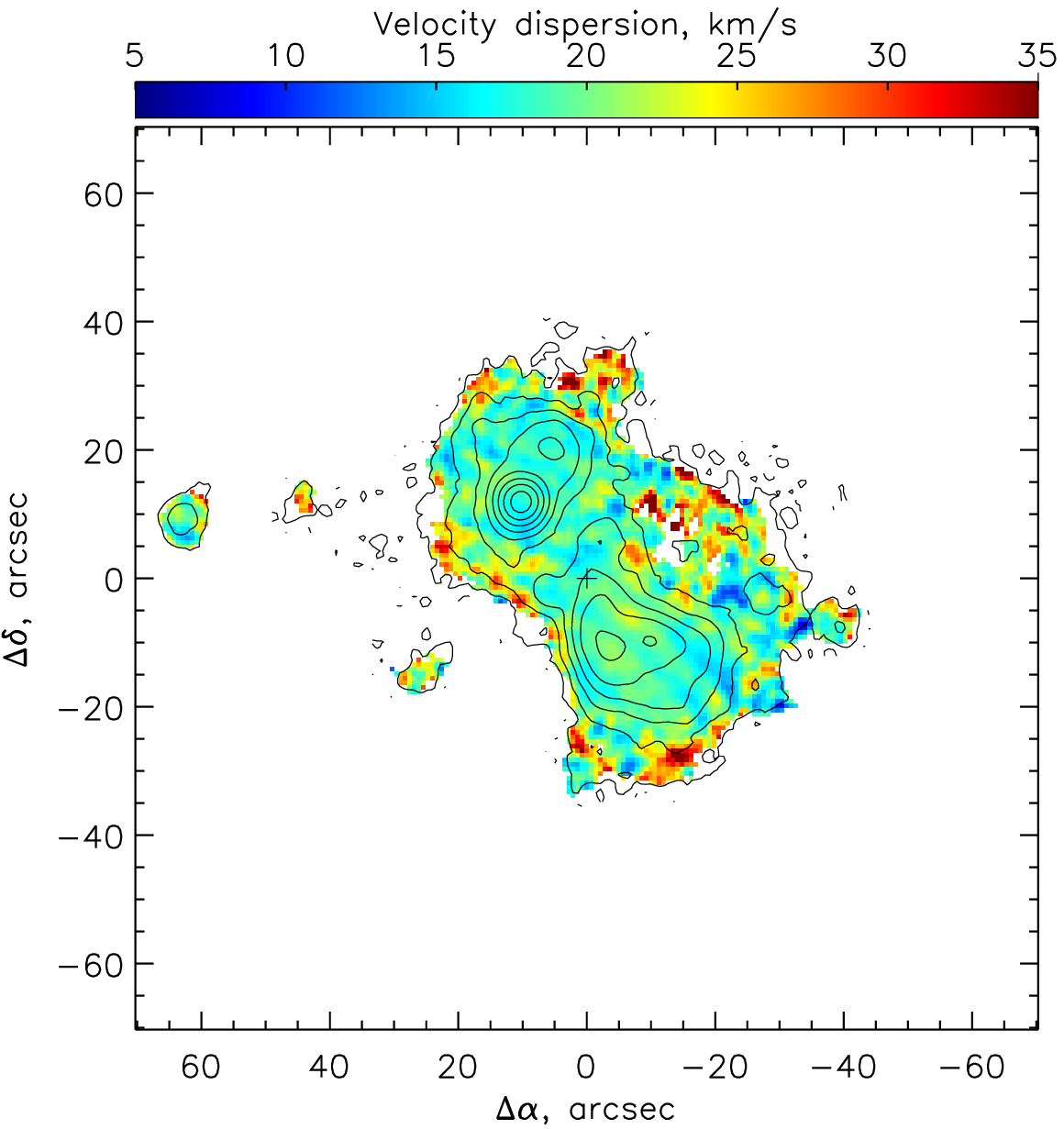}
}
\centerline{
\includegraphics[width=0.5\textwidth]{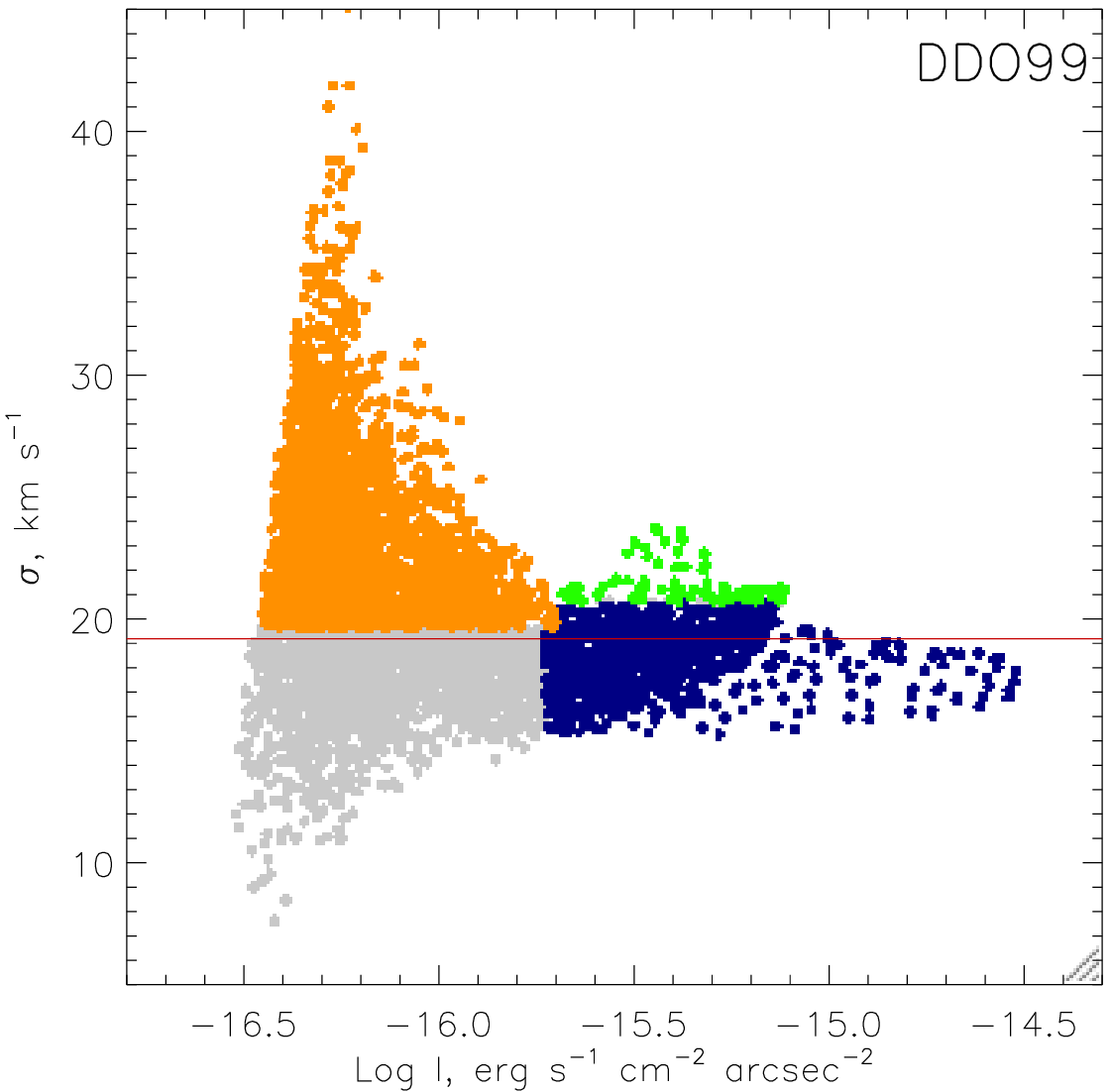}
\includegraphics[width=0.5\textwidth]{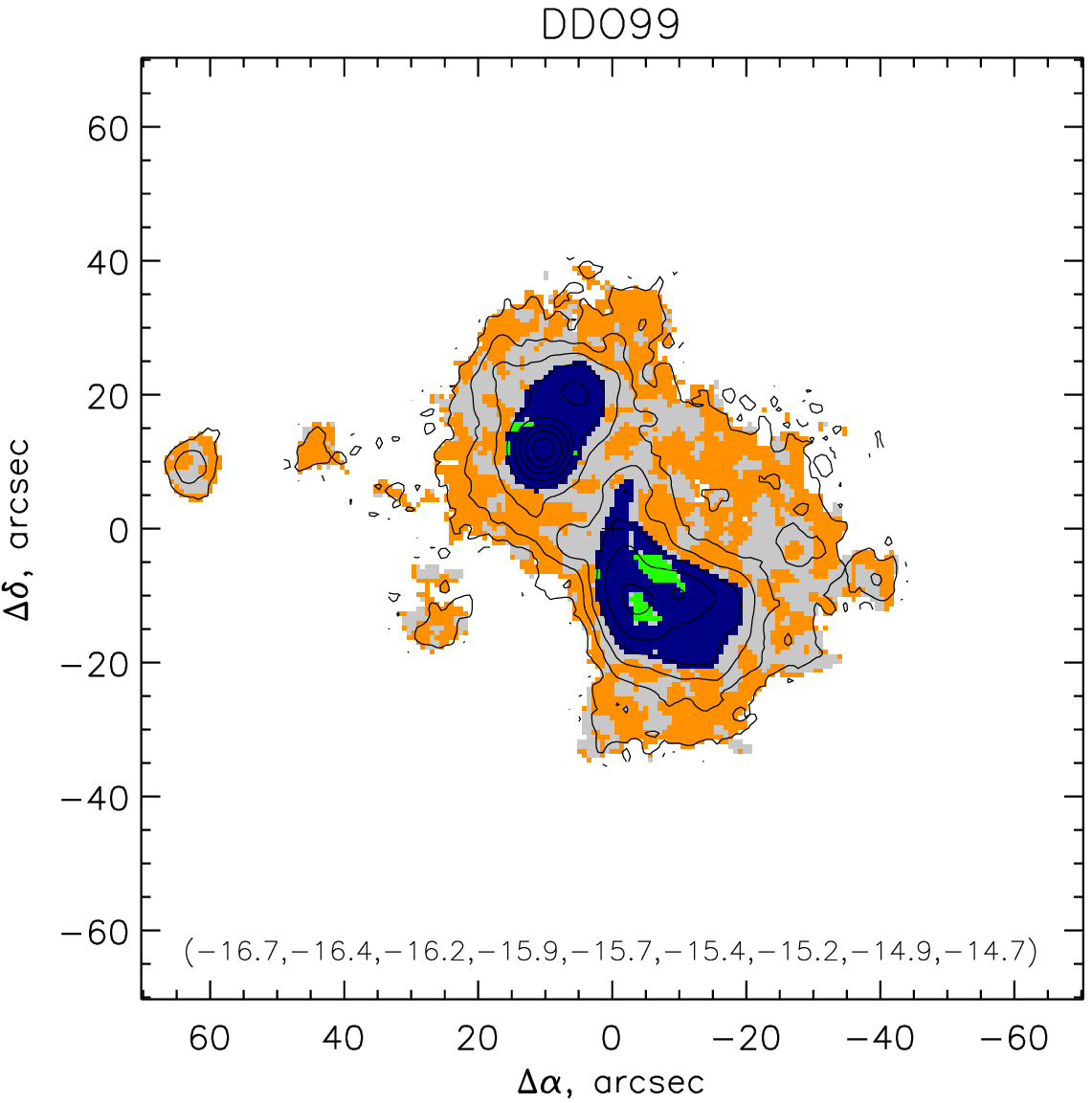}
}
\contcaption{}
 \end{figure*}

\begin{figure*}
\centerline{
\includegraphics[width=0.5\textwidth]{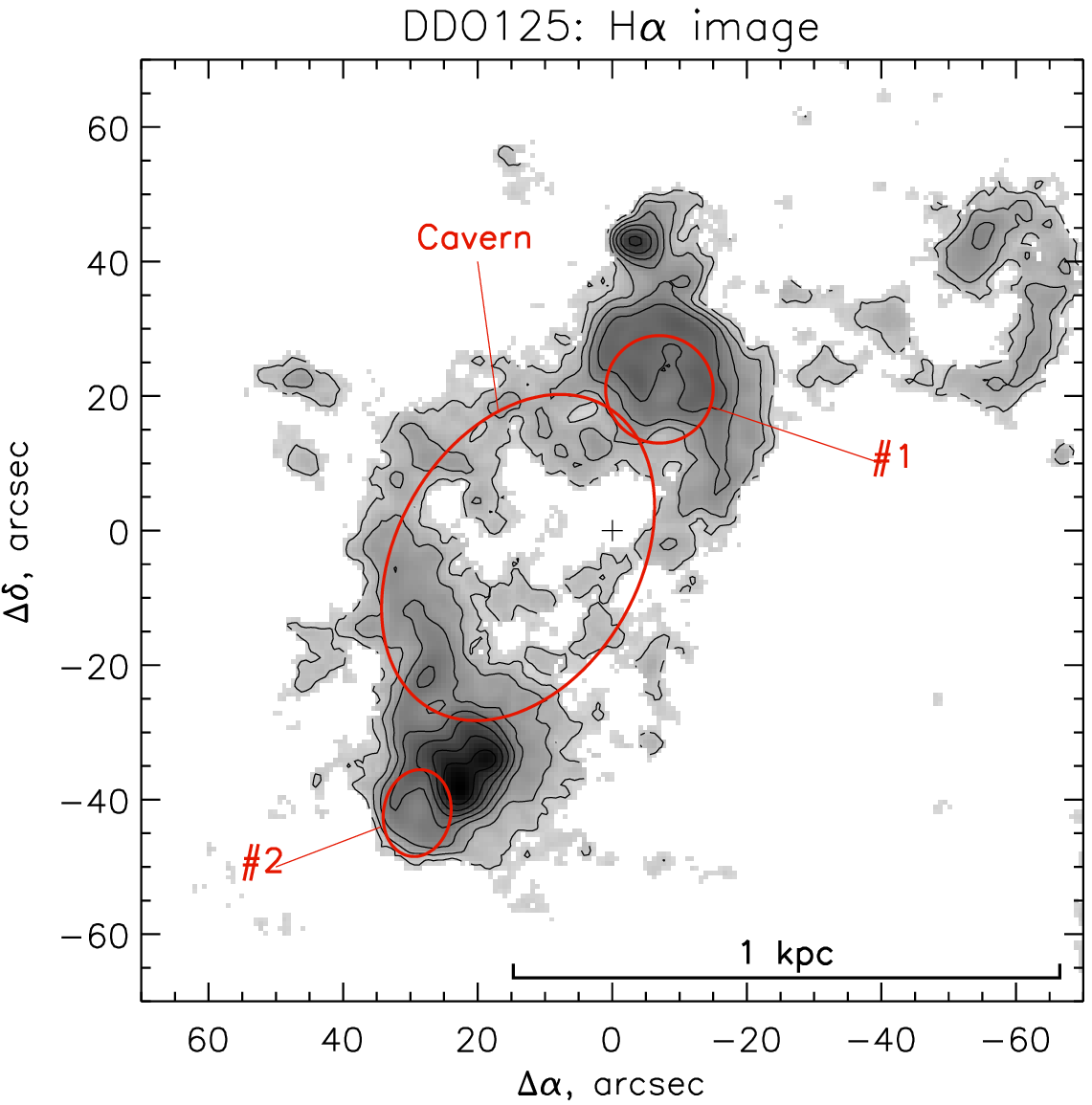}
\includegraphics[width=0.5\textwidth]{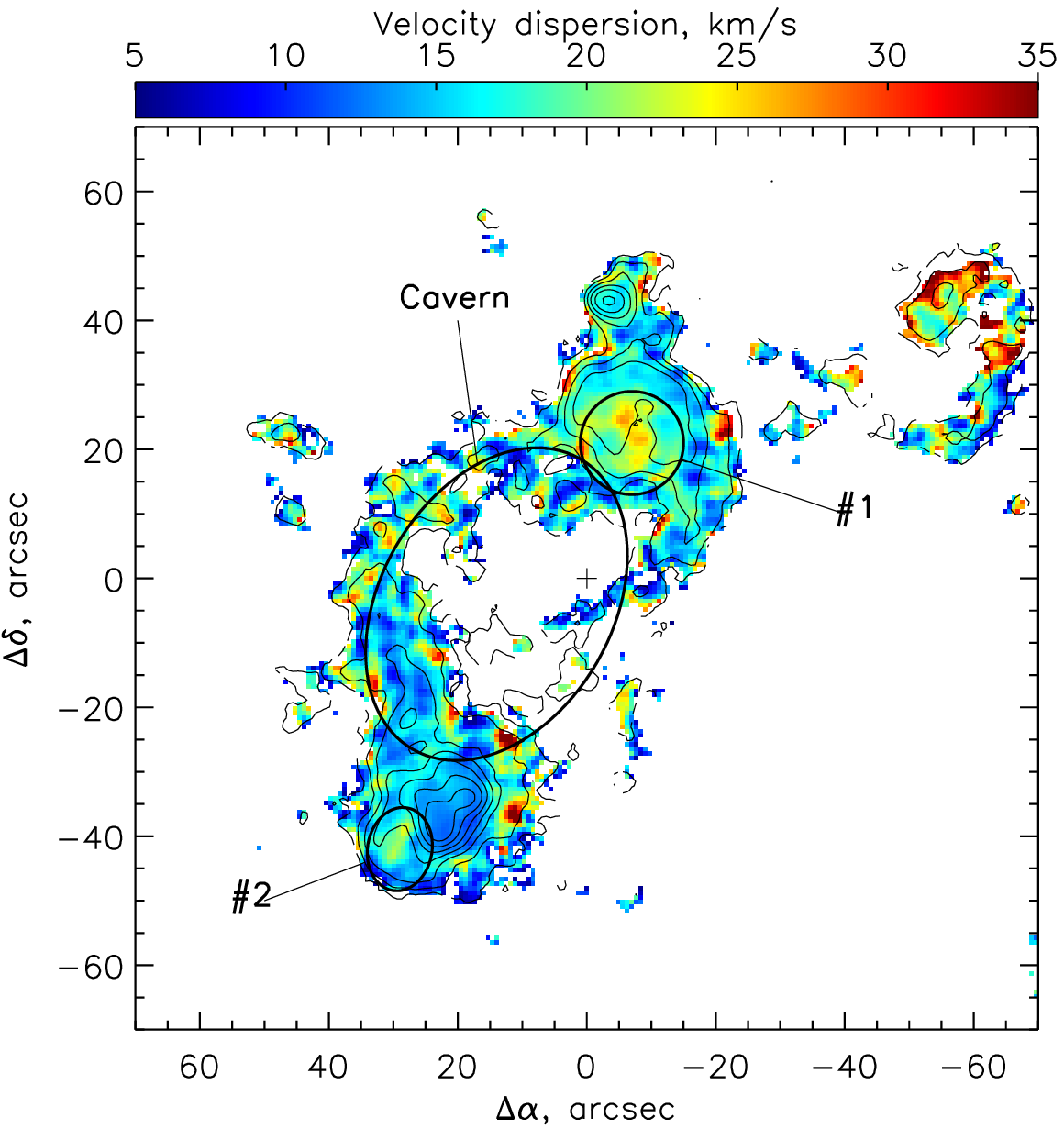}
}
\centerline{
\includegraphics[width=0.5\textwidth]{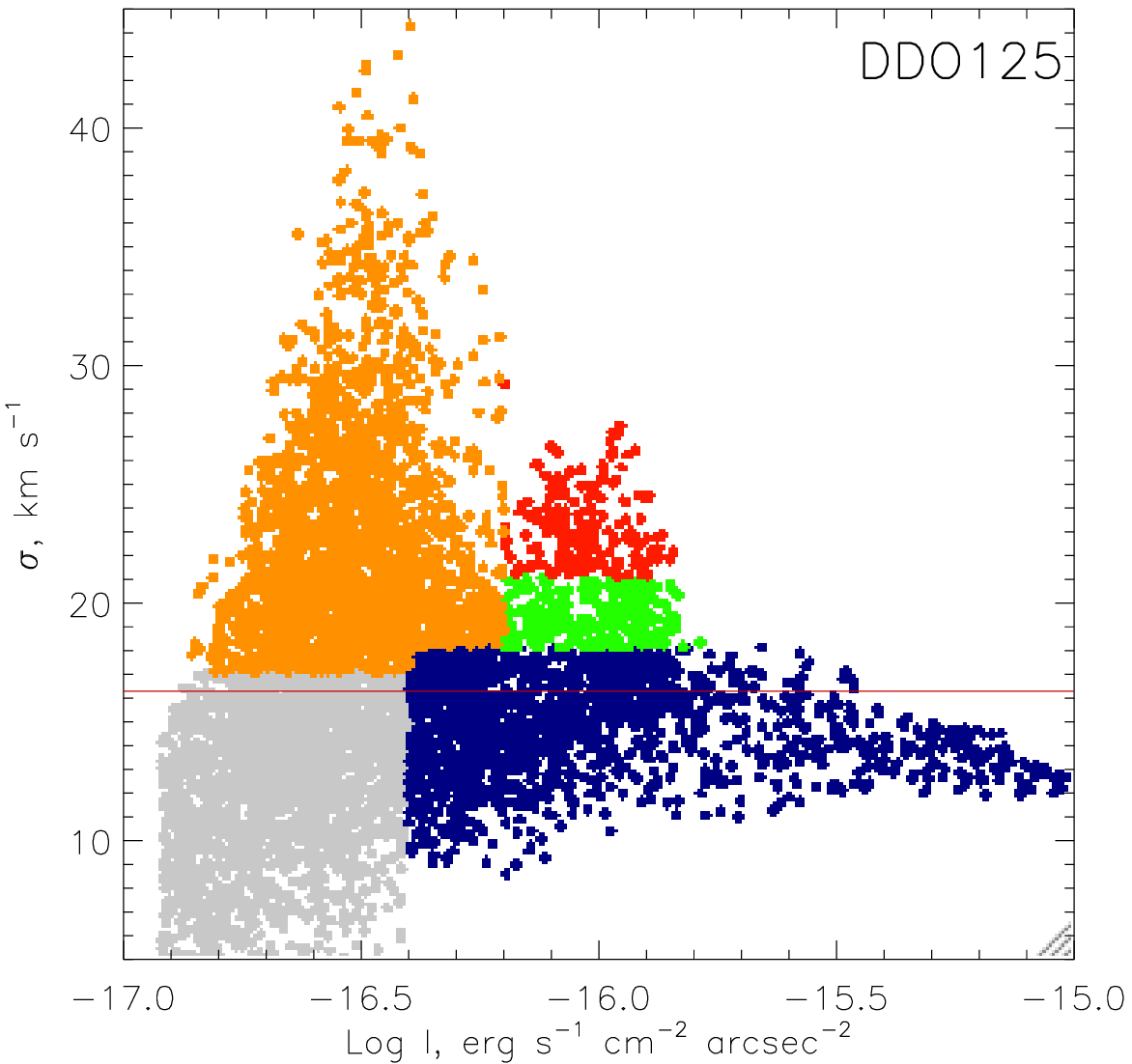}
\includegraphics[width=0.5\textwidth]{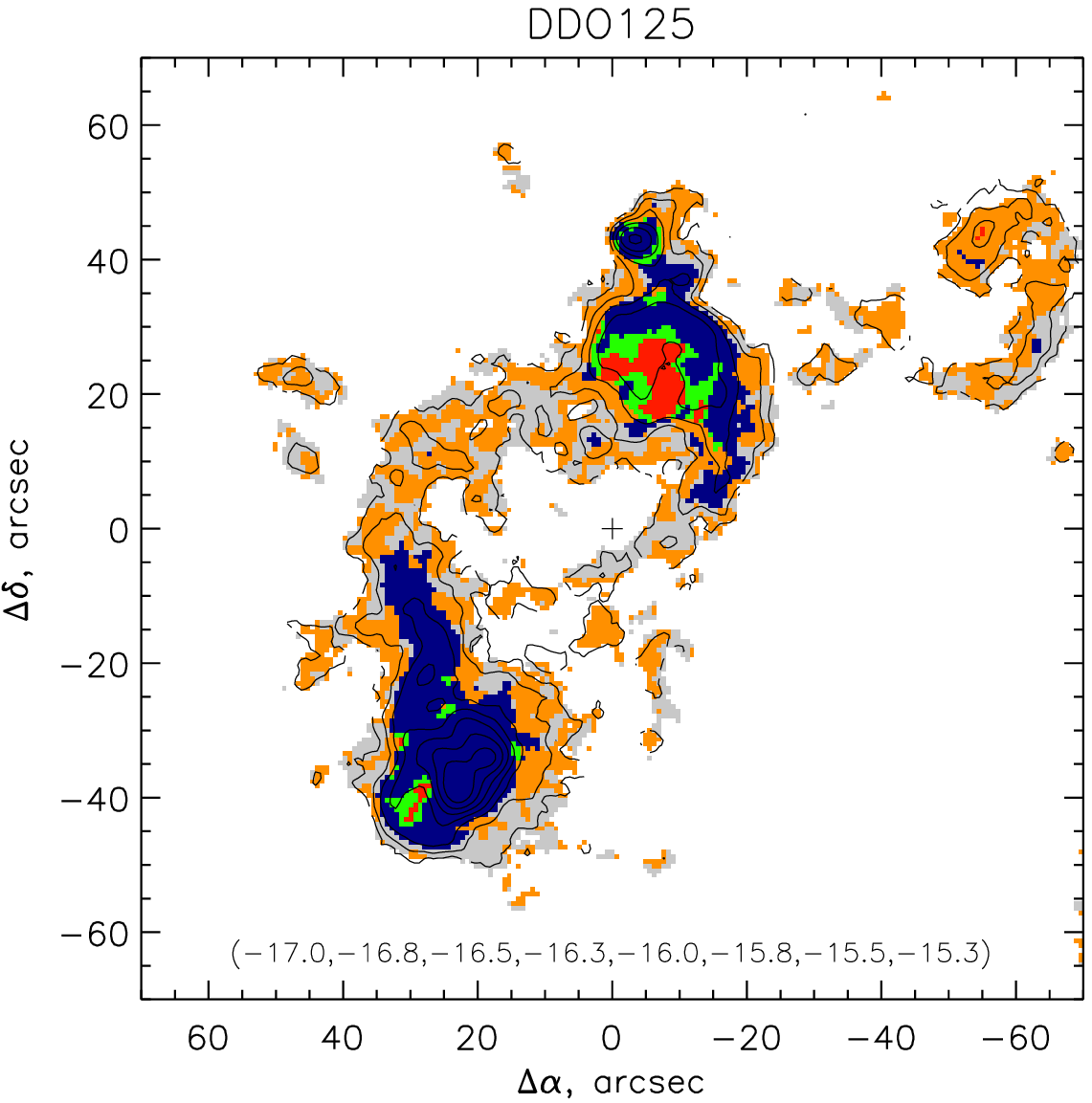}
}
\contcaption{}
 \end{figure*}

\begin{figure*}
\centerline{
\includegraphics[width=0.5\textwidth]{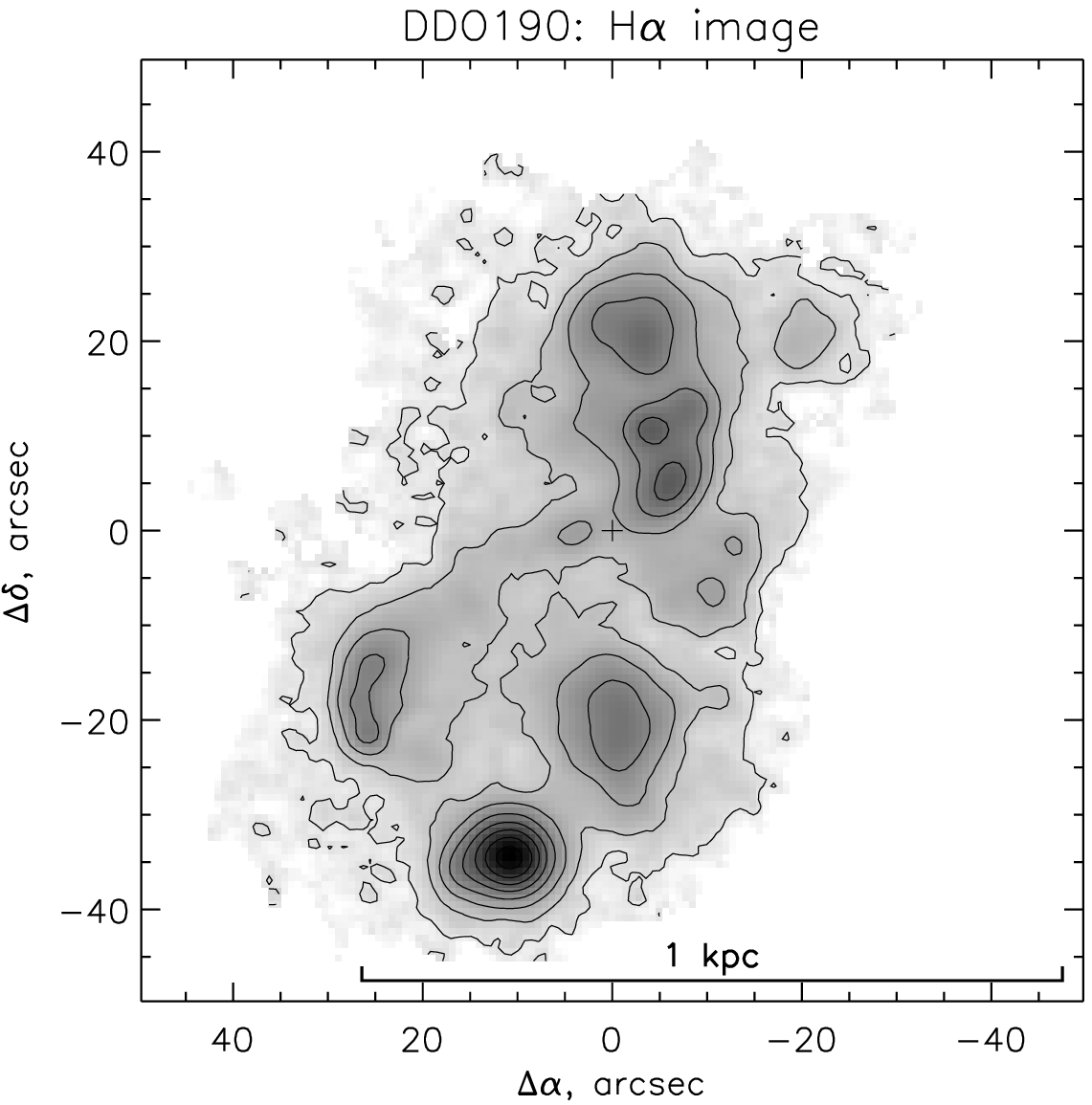}
\includegraphics[width=0.5\textwidth]{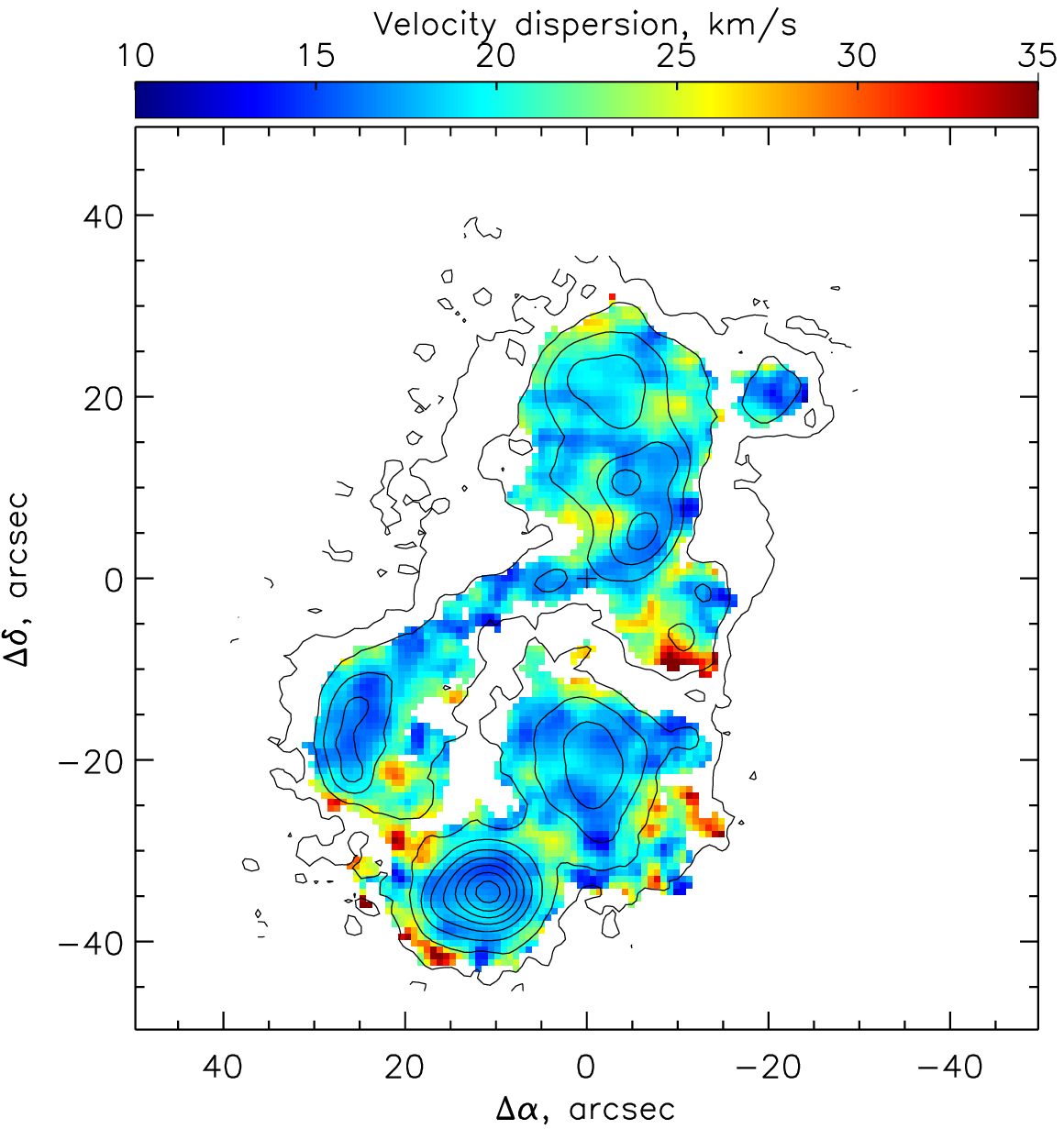}
}
\centerline{
\includegraphics[width=0.5\textwidth]{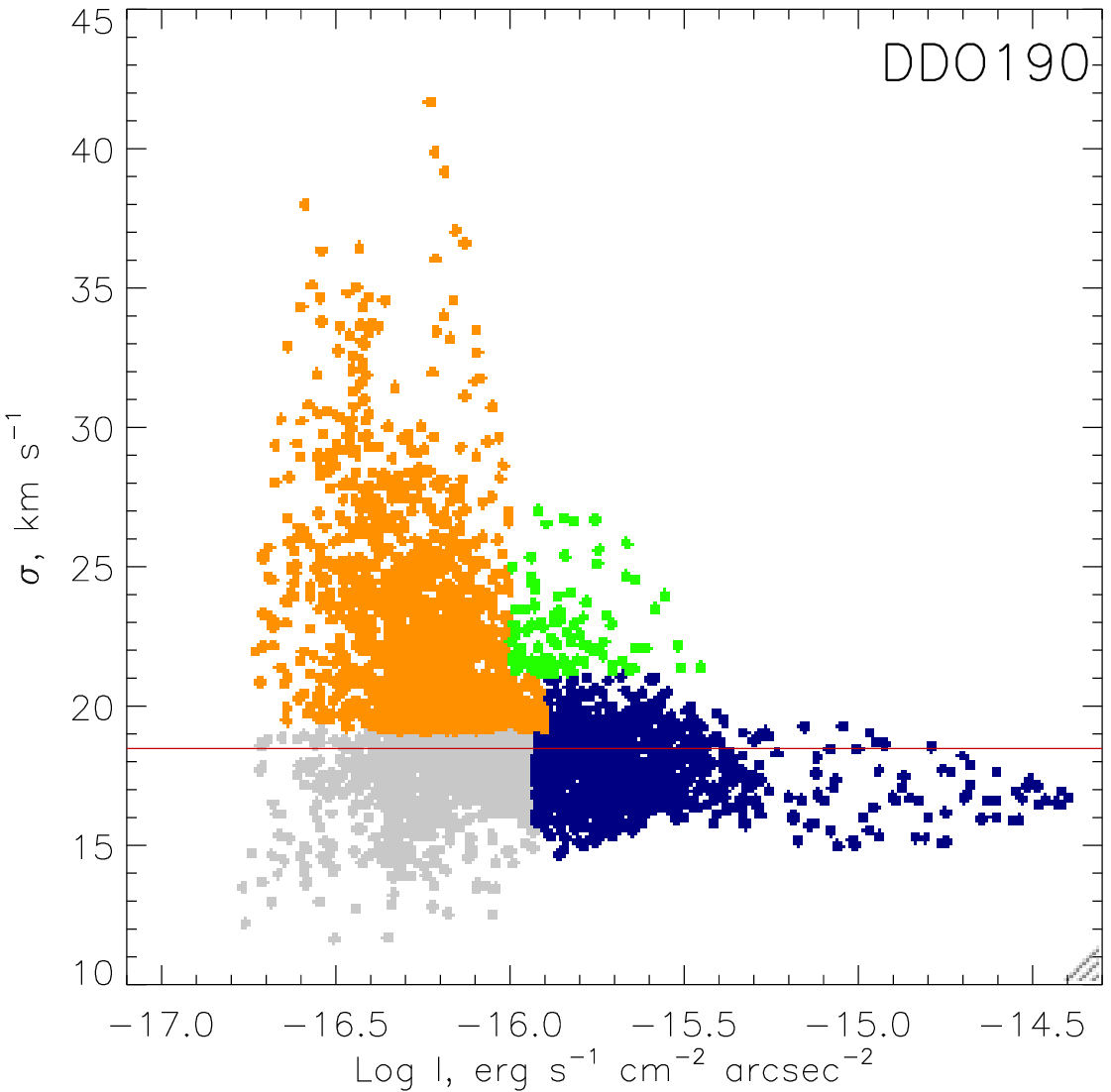}
\includegraphics[width=0.5\textwidth]{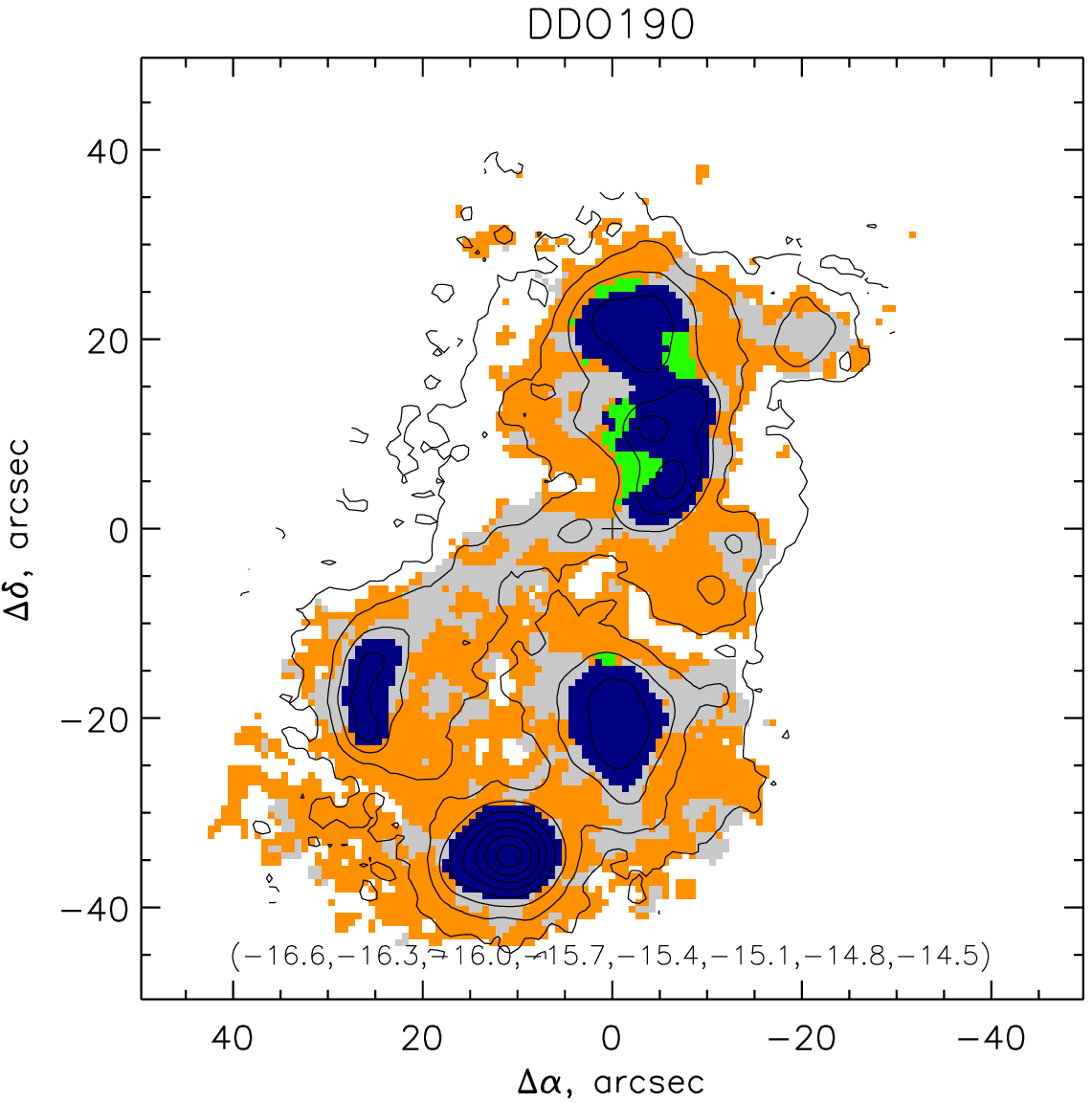}
}
\contcaption{}
 \end{figure*}

\begin{figure*}
\centerline{
\includegraphics[width=0.5\textwidth]{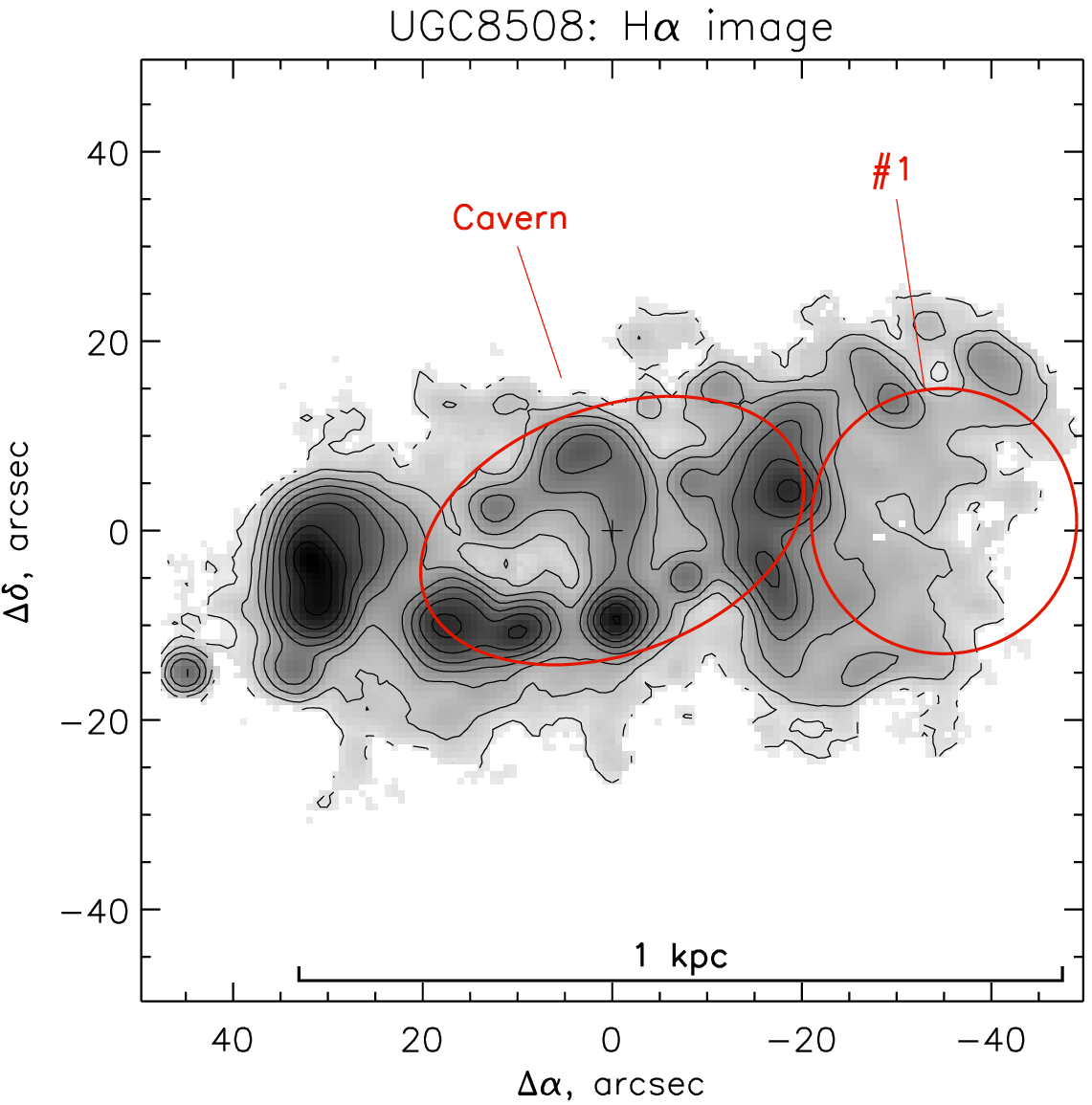}
\includegraphics[width=0.5\textwidth]{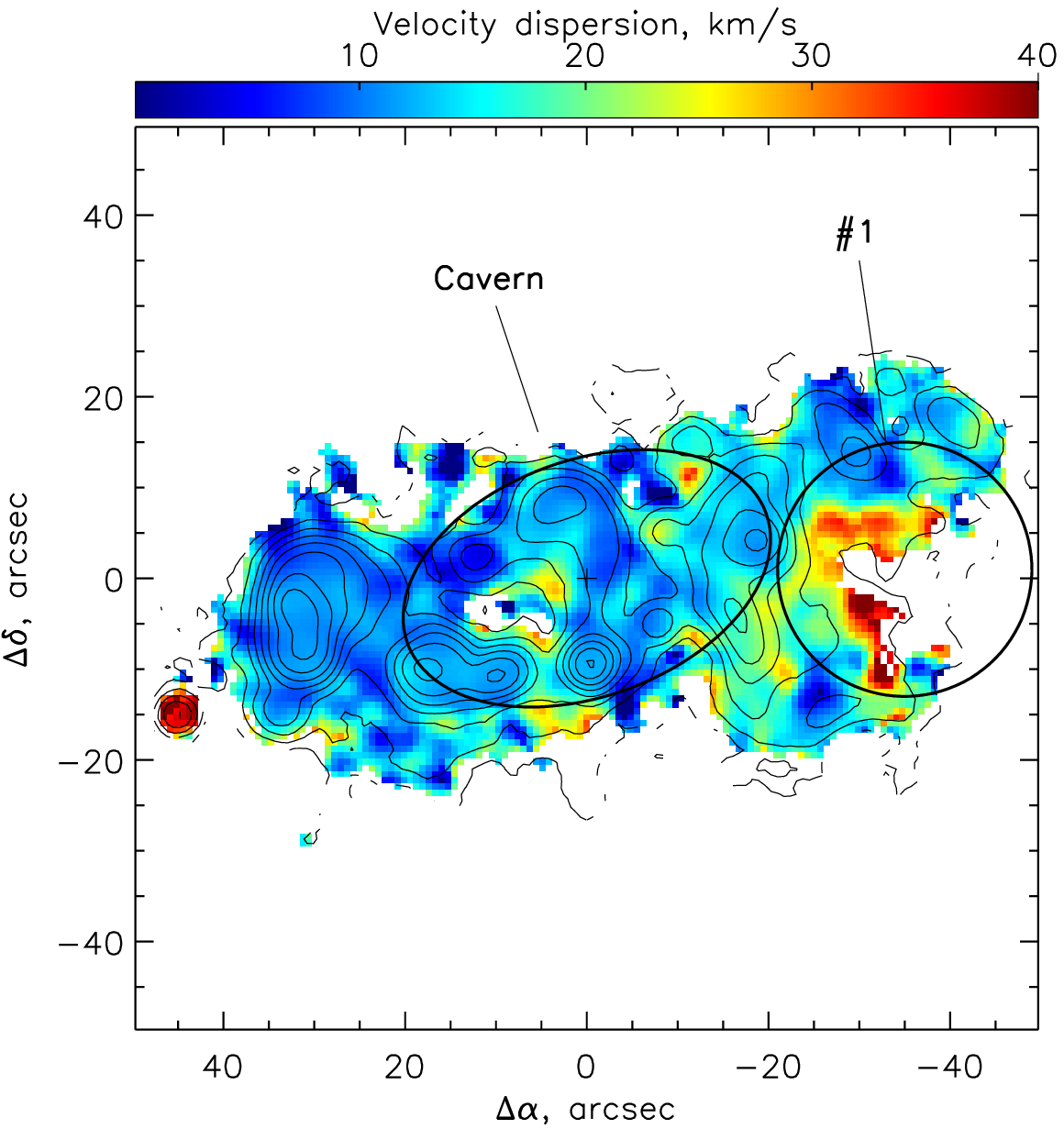}
}
\centerline{
\includegraphics[width=0.5\textwidth]{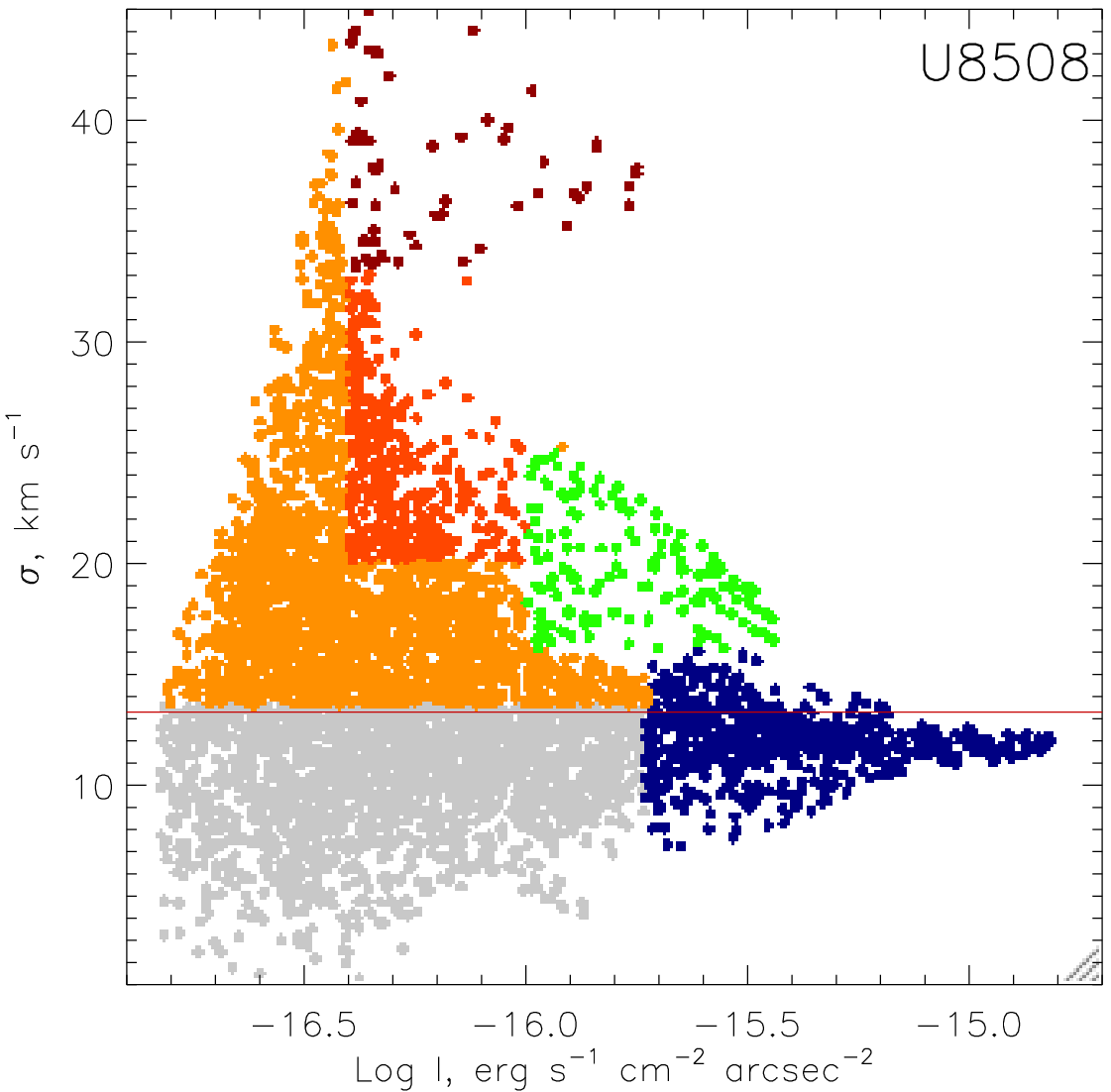}
\includegraphics[width=0.5\textwidth]{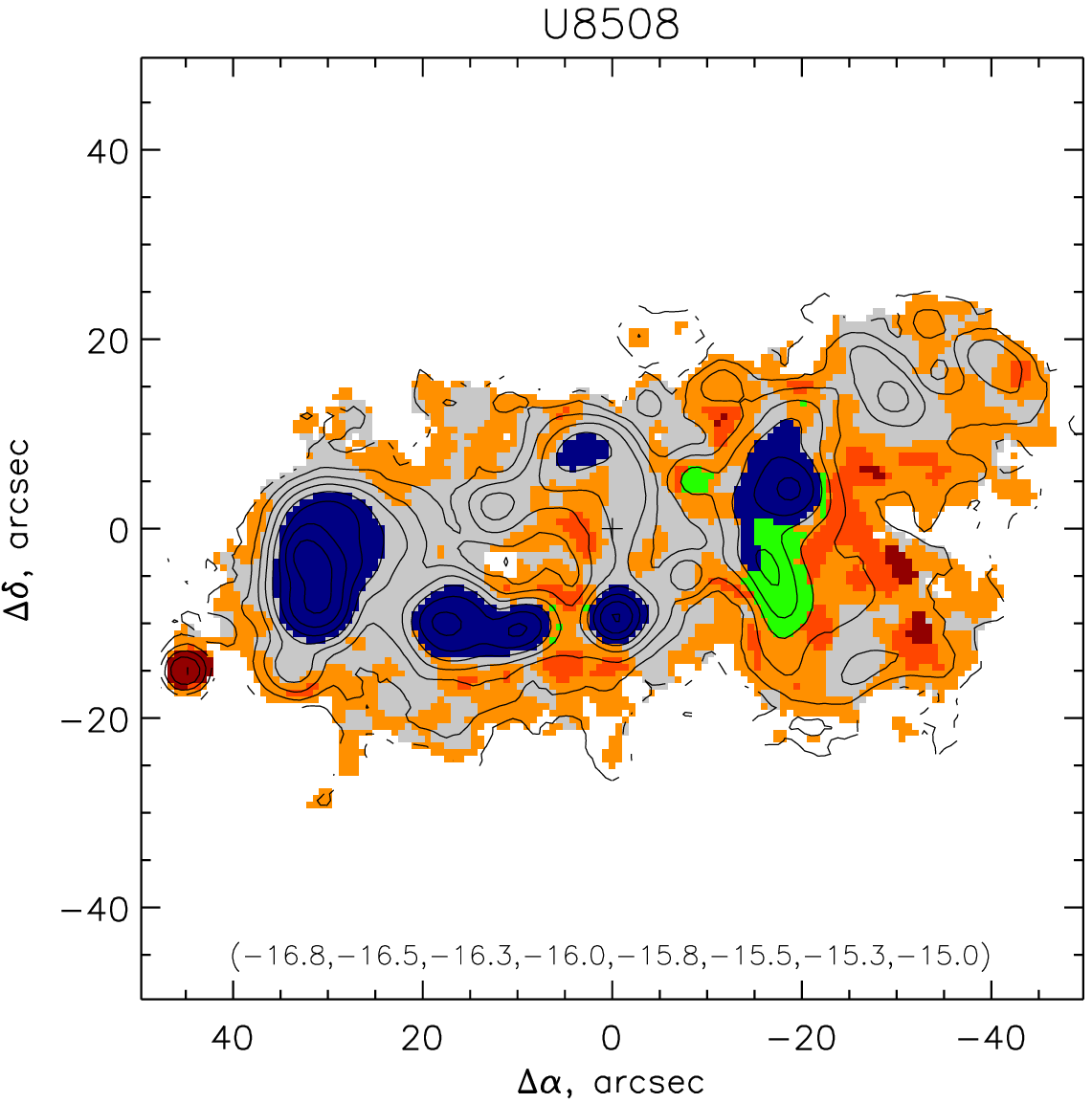}
}
\contcaption{}
 \end{figure*}

\begin{figure*}
\centerline{
\includegraphics[width=0.5\textwidth]{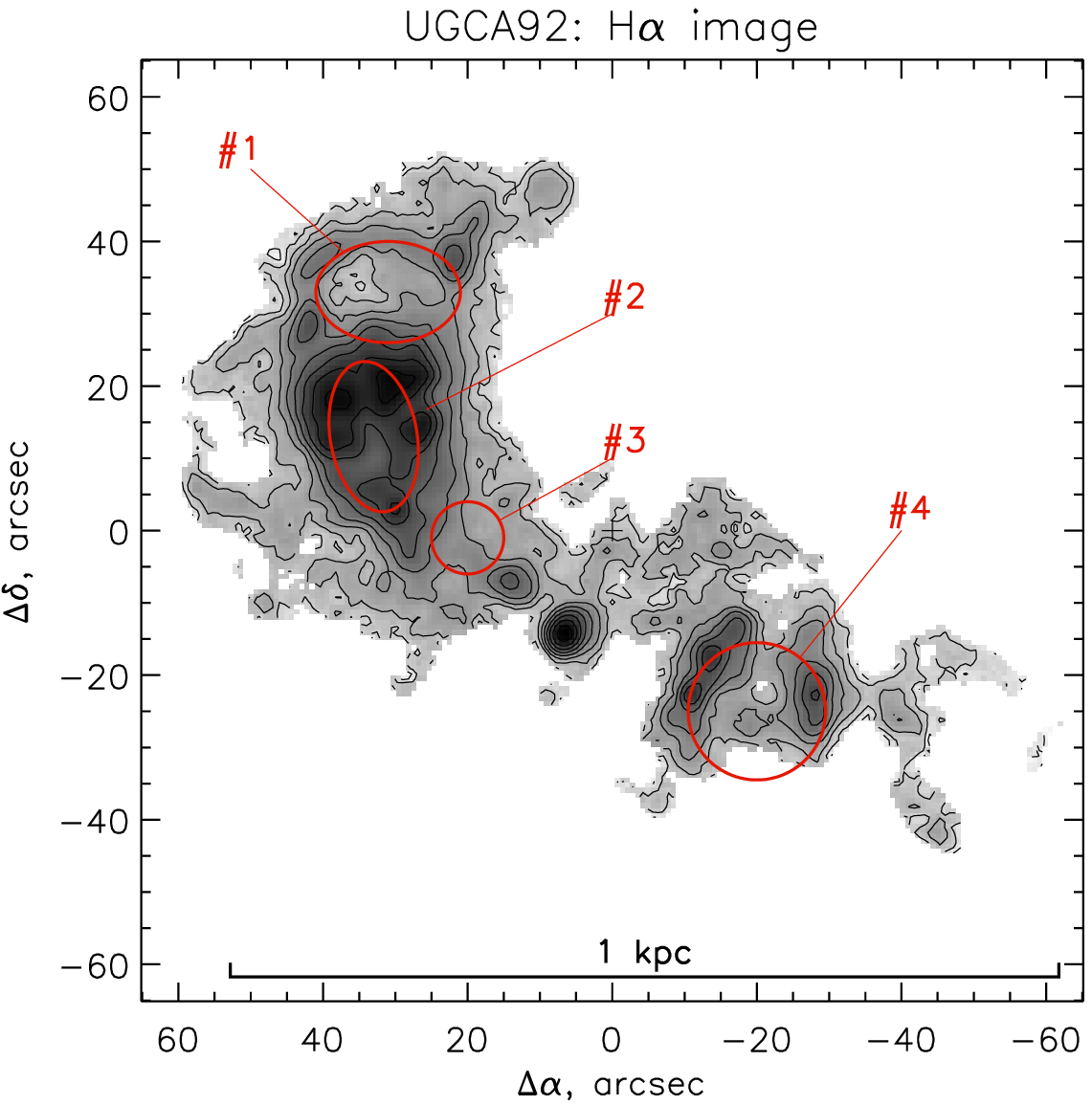}
\includegraphics[width=0.5\textwidth]{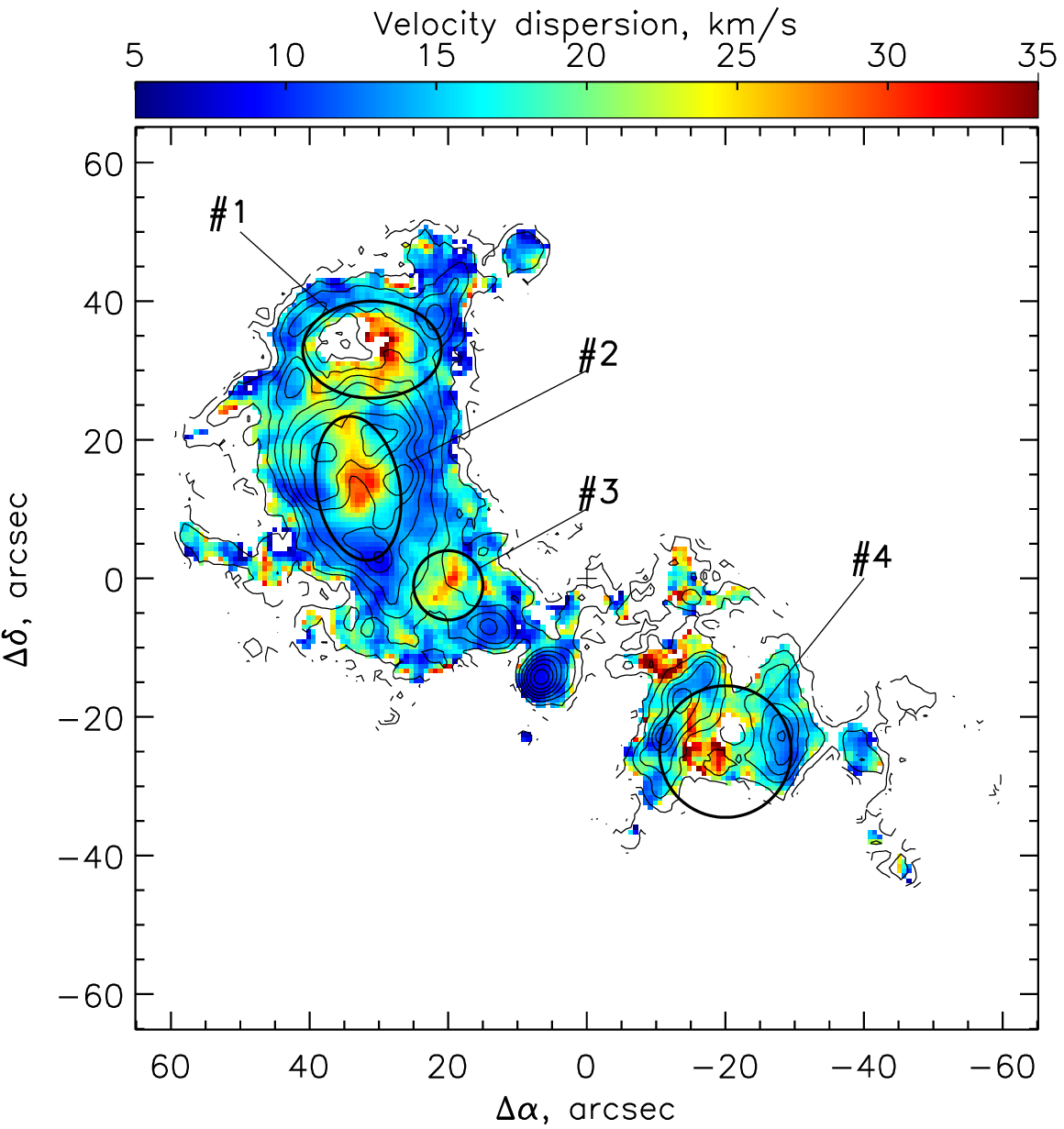}
}
\centerline{
\includegraphics[width=0.5\textwidth]{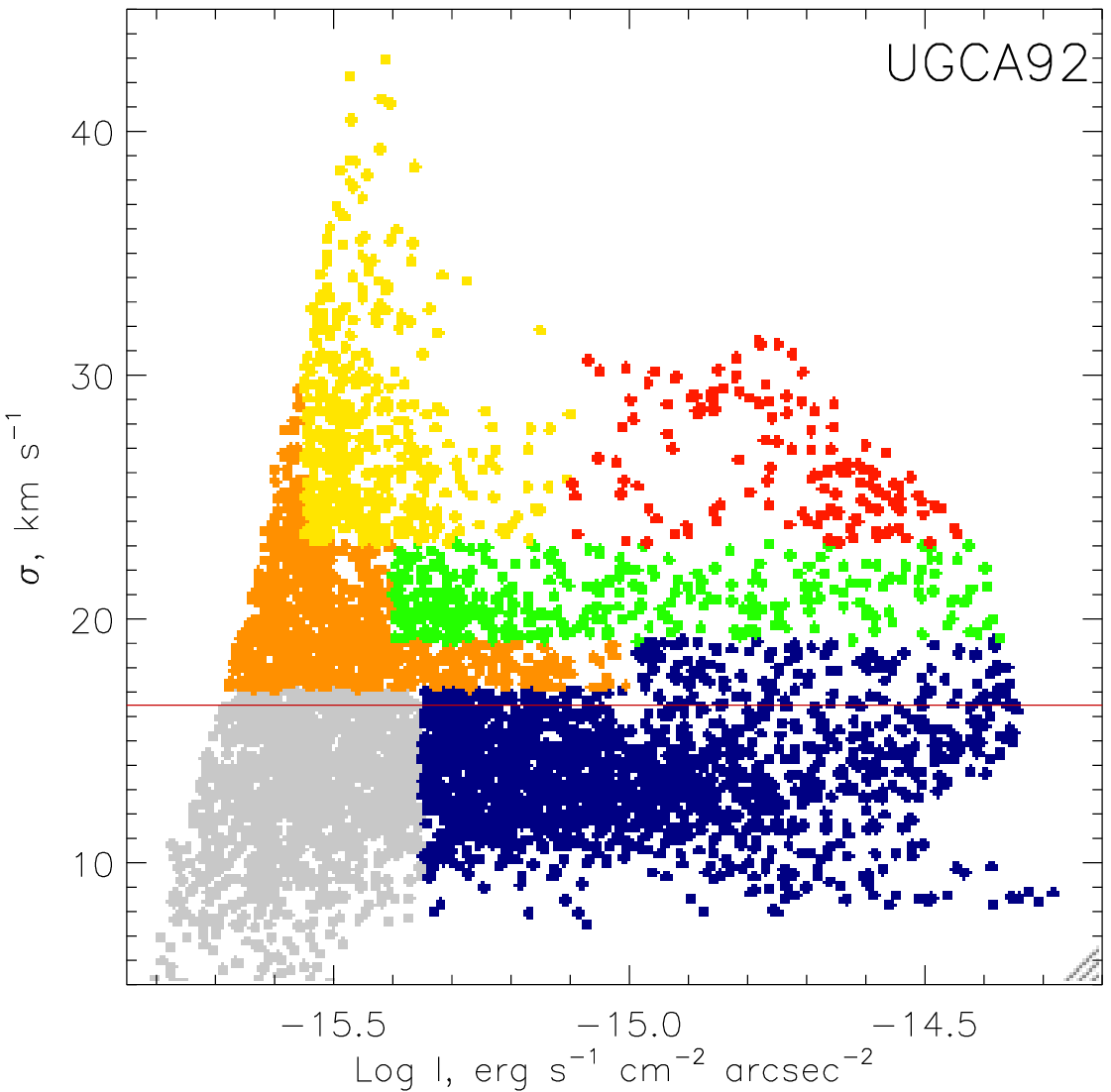}
\includegraphics[width=0.5\textwidth]{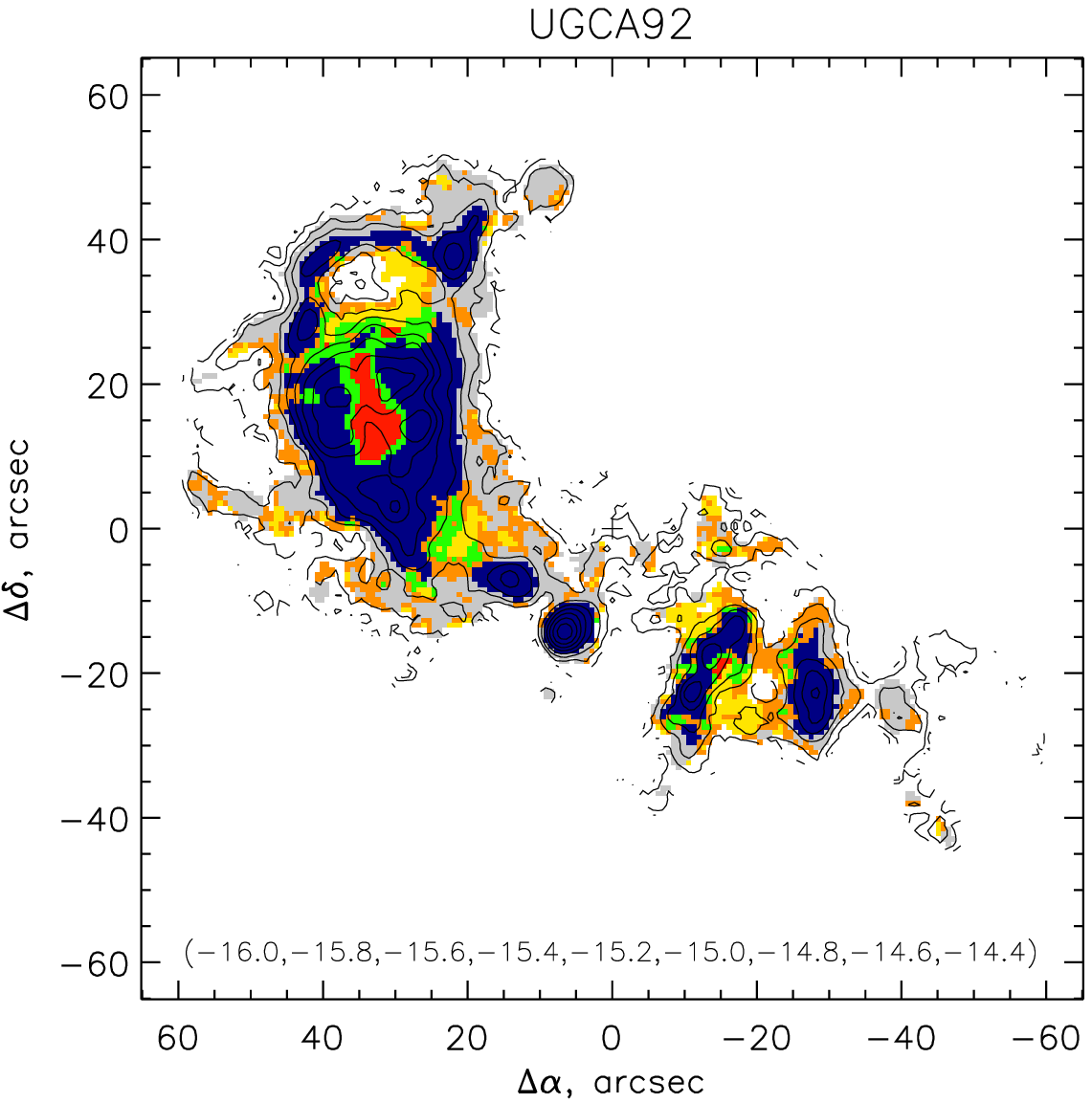}
}
\contcaption{}
 \end{figure*}

\begin{figure*}
\centerline{
\includegraphics[width=0.5\textwidth]{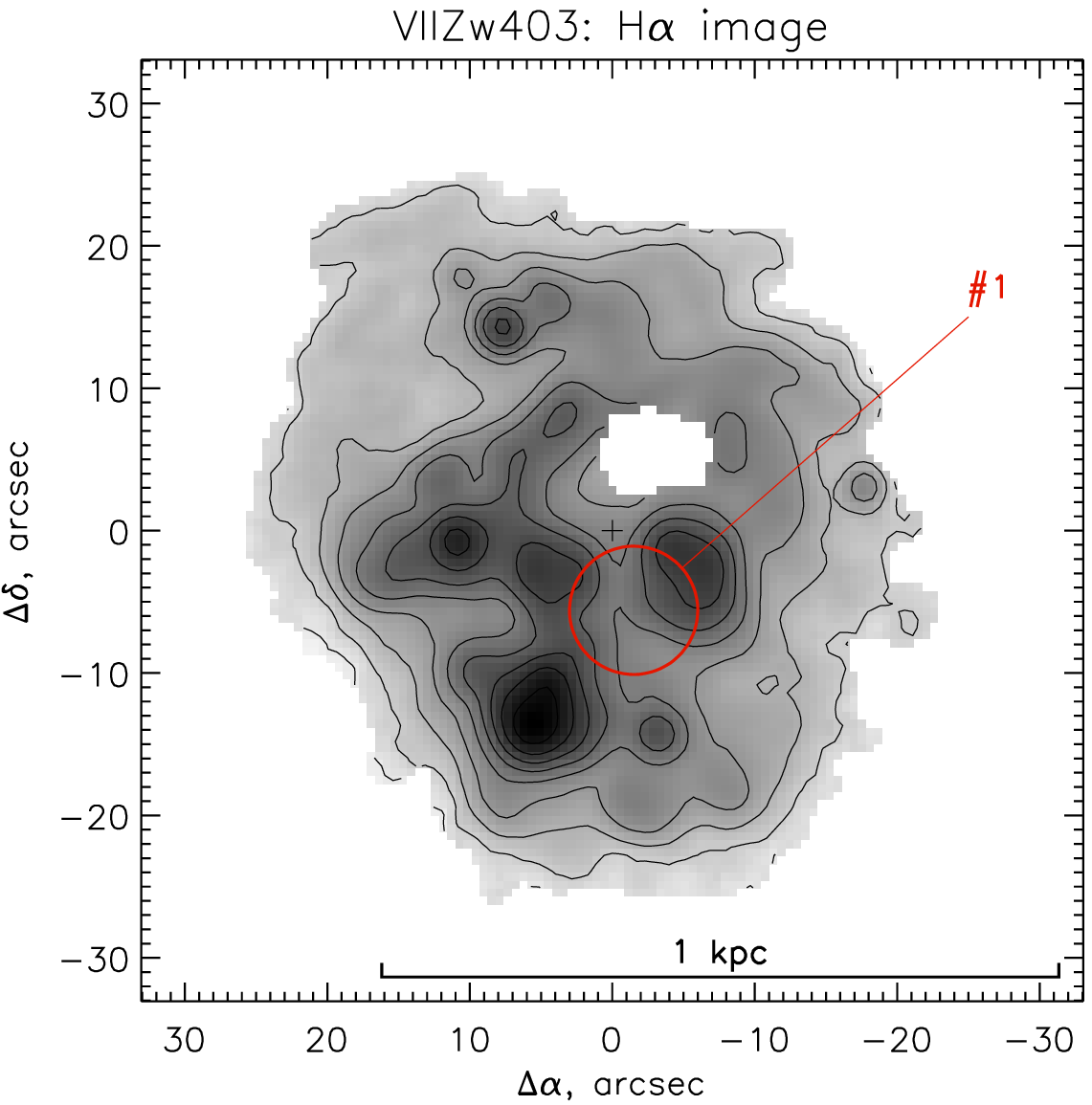}
\includegraphics[width=0.5\textwidth]{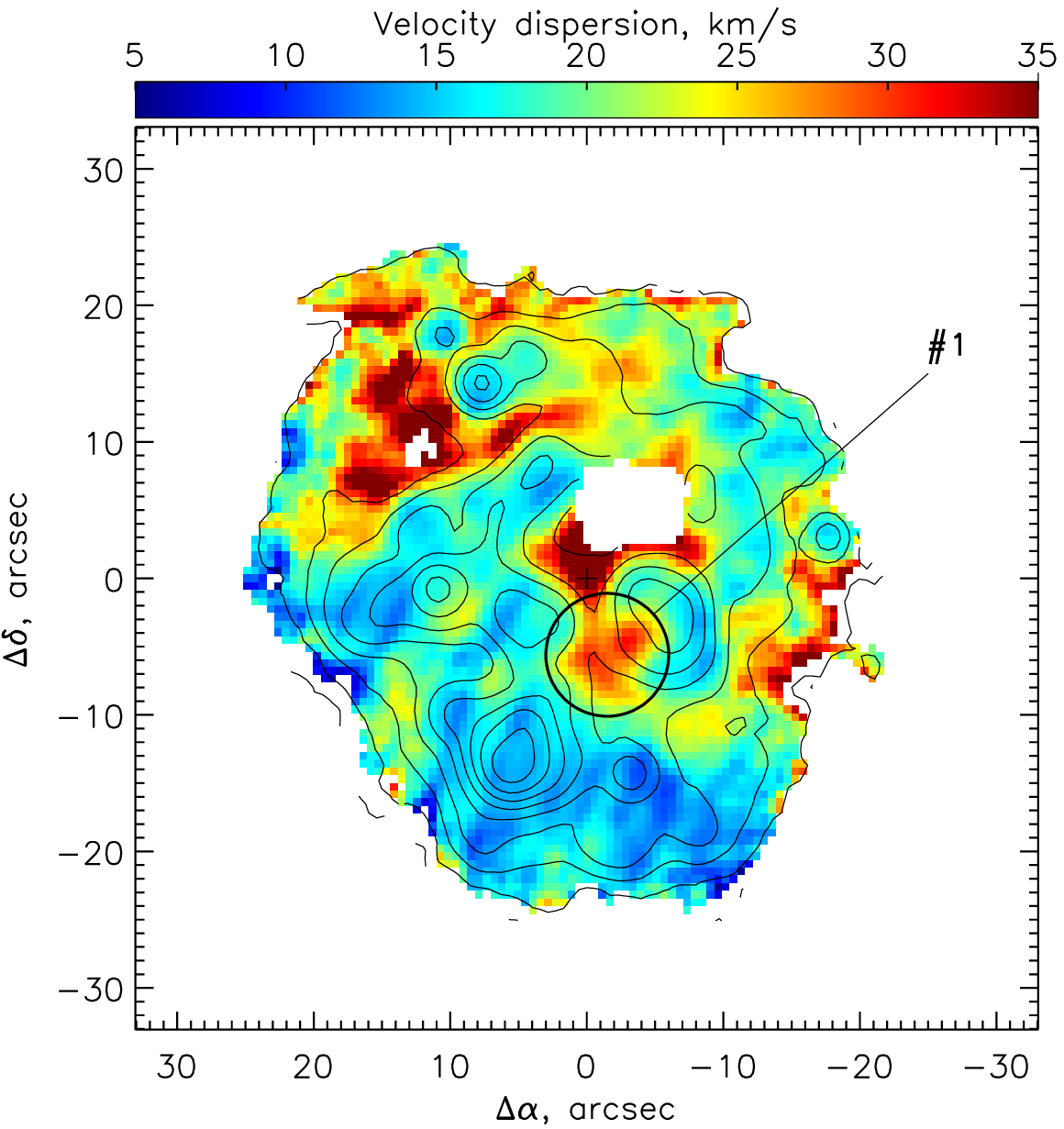}
}
\centerline{
\includegraphics[width=0.5\textwidth]{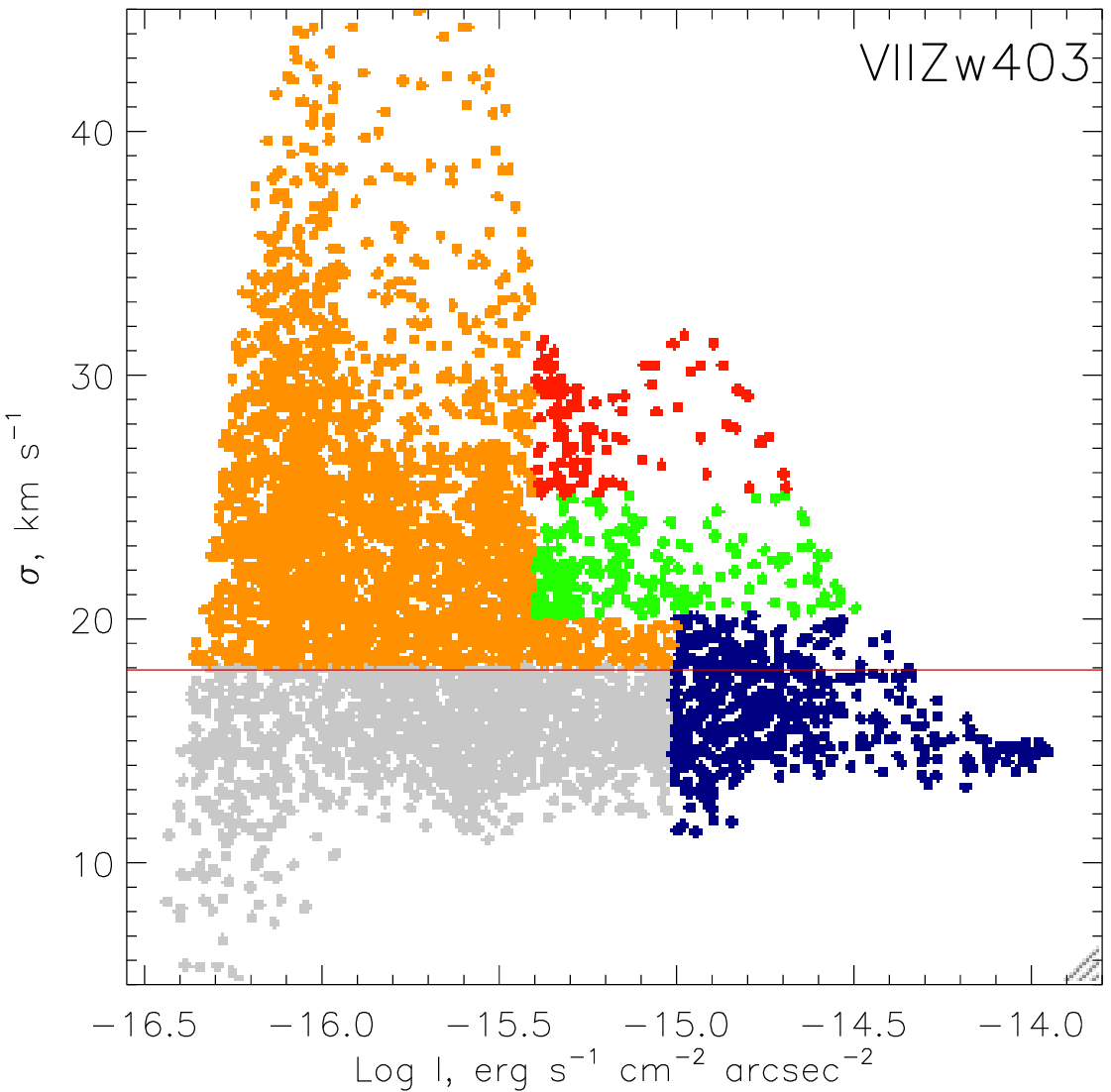}
\includegraphics[width=0.5\textwidth]{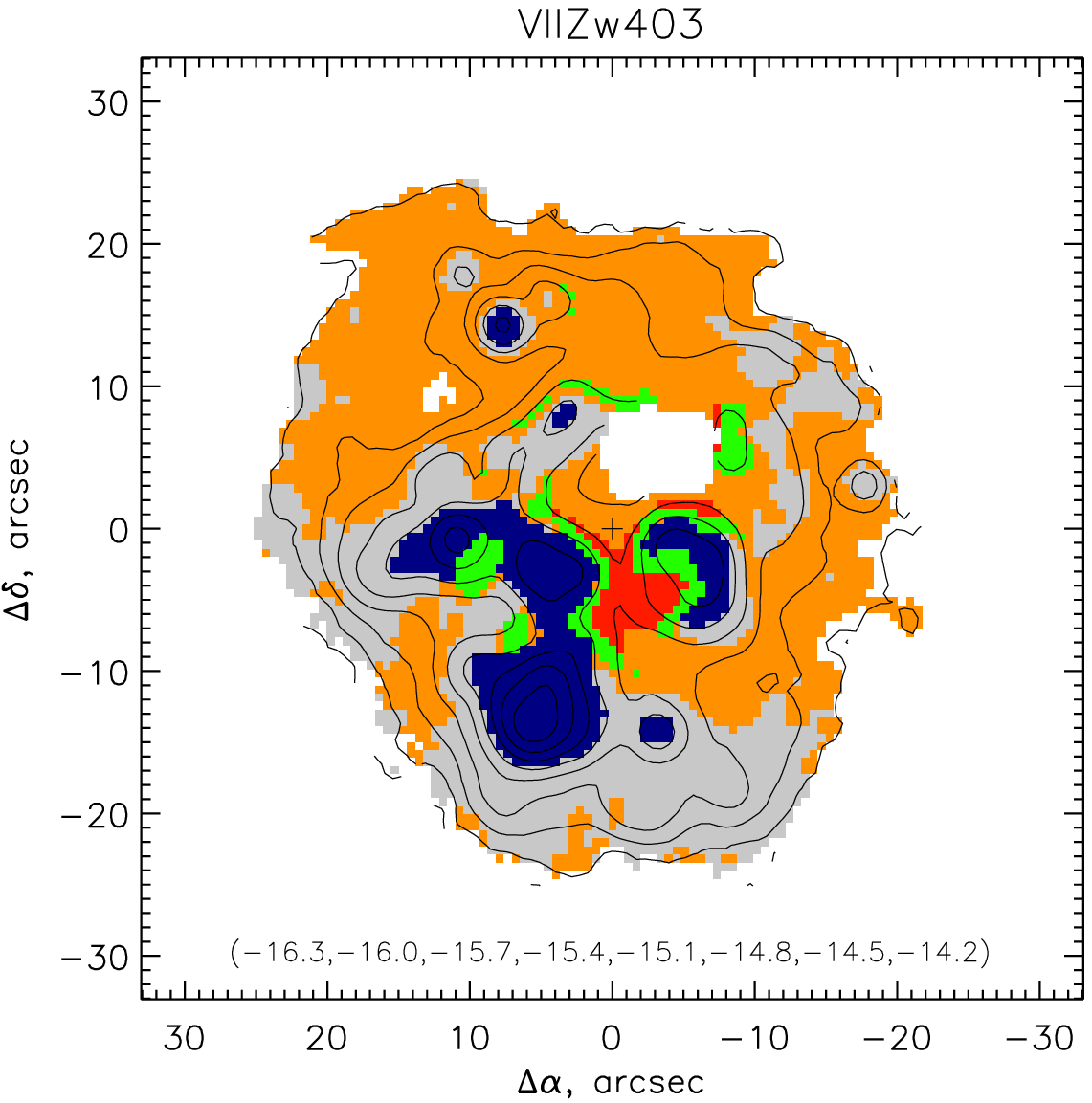}
}
\contcaption{}
 \end{figure*}

\begin{table}
\caption{Shell parameters}\label{tab_shells}
\begin{tabular}{lccc}
\hline
name           & size            & $\sigma_max$    & $t_{kin}$          \\
               &  pc             & $\km$        &  (Myr)             \\
\hline
DDO~53\,\#1    & $240\times240$  & 39           & 3.0 \\
DDO~53\,\#2    & $150\times140$  & 37           & 2.0 \\
DDO~53\,\#3    & $150\times150$  & 37           & 2.0 \\
DDO~125\,\#1   & $200\times200$ & 27           & 3.6 \\
DDO~125\,\#2   & $160\times120$ & 25           & 3.1 \\
UGC~8508\,\#1  & $350\times350$ & 40           & 4.2 \\
UGCA~92\,\#1   & $180\times130$ & 39           & 2.2 \\
UGCA~92\,\#2   & $190\times110$ & 31            & 3.1 \\
UGCA~92\,\#3   & $90\times90$  & 31            & 1.4 \\
UGCA~92\,\#4   & $170\times170$ & 40           & 2.1 \\
VII~Zw~403\,\#1   & $190\times190$ & 33           & 2.8 \\
  \hline
\end{tabular}
\end{table}

\subsection{DDO~53}

The kinematics of ionized gas in this galaxy was previously investigated with the
scanning FPI by \citet{Dicaire2008}, but the authors were only able to measure the
distribution of radial velocities in the bright HII regions, whereas our data include
the regions of faint diffuse emission. The $\sigma$ map clearly reveals a pattern,
typical of all galaxies  we observed: the minimum value of velocity dispersion is
seen in  the centers of bright HII regions, while in the space
between the star formation regions and in the outer parts of the galaxy $\sigma$
reaches its maximal values  up to $40-50\km$, which clearly points to the supersonic
nature of motions.

The shape of the $I-\sigma$ diagram resembles the scheme, given in the paper by
\citet{MunozTunon1996} -- a horizontal   lane (marked in dark blue in the
Fig.~\ref{fig1_1}) with $\sigma \approx \sigma_m$ and several inclined lanes with much
larger $\sigma$ values. According to the interpretation of \citet{MunozTunon1996}, such
lanes must correspond to individual thin expanding shells: the maximal velocity
dispersion (or even a double profile) and the minimal  surface brightness  at the centre of the
shell;  increasing $I$ and decreasing  $\sigma$  with the growing distance from the
centre until the inner edge of the shell is reached. The best example of such a shell
is given by a structure, adjacent to the north of the brightest HII region, in the
diagram it is coloured yellow (centre) and dark green (periphery). This shell-like structure is also
evident in the \Ha{}  surface  brightness distribution. In the figure we marked it as \#1, its diameter is
approximately $d\approx14$ arcsec (240 pc). For all the shells we identified
Table~\ref{tab_shells} gives the estimates of the sizes along the major and minor axes
($d_1,d_2$), the maximal velocity dispersion $\sigma_{max}$ and the kinematic ages
$t_{kin}=0.5d_{1}/\sigma_{max}$.

The points from the second, less contrast peak in the diagram (marked in red and green)
have to correspond to a younger shell according to the model by \citep{MunozTunon1996}.
On our maps the points of this peak mainly lie in two shells, designated nos. \#2 and
\#3. According to Table~\ref{tab_shells}, their kinematic age is in fact  about 1.5
times shorter than that of the shell \#1. On the other hand, the points from the most
apparent peak on the diagram (orange colour, high $\sigma$ and relatively low  surface  
brightness) do not belong to any individual knots or shells, but form a common diffuse envelope
which surrounds bright  HII regions.

\subsection{DDO~99}

Unlike the previous case, the $I-\sigma$ diagram  for DDO~99 has a simpler shape --
there are no pronounced features, except for the horizontal lane and a broad triangular
region with a high $\sigma$. Some kind of a hint for a possible shell is  a tiny cloud
of points in the diagram with  $\log I>-15.7 ,\,\sigma>21\km$ (marked in
green). On the map of the galaxy, these points are grouped mainly on the edge of the
southern star formation region, the size of this possible shell is comparable to the
spatial resolution of $\sim 50$ pc.  The remaining regions of increased $\sigma$
(plotted in orange in the figure) are distributed on the periphery of HII
regions.

\subsection{DDO~125}

The distribution of ionized gas here reveals several shells and
loops, associated with two bright star formation regions  to the
north and south of the galactic centre. These two HII regions are
located diametrically on the edge of a giant cavity, sized around
600\,pc  (denoted in Fig.\ref{fig1_1} as `Cavern'), inside of which
there are almost no sources of gas ionization. In the centre of
the cavity the HI density is also reduced (see
Section~\ref{sec_discuss} below), indicating that the gas was
depleted or swept-out during the previous burst of star formation.
The size of the cavity and the features of mutual distribution of
HI and HII are very similar to the complex of giant shells and
arcs in the galaxy IC 1613 \citep{Lozinskaya2003, Silich2006},
with a diameter from 300 pc to 1 kpc. We were unable to find any
kinematic evidence for the expansion of shell DDO125\,\#1, as it
is not identified on the $I-\sigma$ diagram. At once the
$\sigma$ distribution  easily reveals a more compact shells
DDO125\,\#1, and  ,\#2 which coincides with the northern and southern HII regions
(coloured red and green in the figure). Our estimate of its age
is $t_{kin}=3.1-3.6 Myr$. Other regions with high $\sigma$,
marked in the map with orange, are embordering the bright
star forming regions.

\subsection{DDO~190}

The distribution of the ionized gas velocity dispersion and the $I-\sigma$ diagram here
resemble the picture, observed in DDO~99. Namely, the points with high $\sigma$
 surround bright HII regions; velocity dispersion in HII regions is comparable to
$\sigma_m$. Only a few low-contrast peaks of the velocity dispersion around the
elongated HII region in the northern part of the galaxy can be regarded here as a vague
hint to the shell \citep[in terms of the model by][]{MunozTunon1996}. These possible
shells with sizes comparable to the spatial resolution of our observations are coloured
green in the diagram.

\subsection{UGC~8508}

\label{U8508}

The observed distribution of $\sigma$ is very similar to  previous objects: a minimal
velocity dispersion inside the HII regions   surrounded by the diffuse gas  with increased $\sigma$.
In addition, the $I-\sigma$ diagram  reveals an inclined lane, which should
correspond to a shell according to \citet{MunozTunon1996}.

Indeed, most of the points of this lane (marked in Fig.~\ref{fig1_1} with
red and green)  are concentrated within a huge arch of HII
regions on the edge of the western part of the disc of ionized
gas. Morphologically and kinematically this resembles a half of an
unclosed  expanding shell with a diameter of
approximately $350$ pc. The kinematic age of this shell, we
designated as UGC~8508\,\#1, is about 4 Myr
(Table~\ref{tab_shells}).

In a recent paper \citet{Warren2011} showed that the eastern half
of the galaxy has a cavity in the HI distribution with a diameter
of about 550 pc, so that the bright HII regions are located along
its borders.  However, this region  (denoted as `Cavern' in our figure) is not distinguishable   by the kinematics
of ionized gas -- velocity dispersion is small, except for several
spots with increased $\sigma$ in the heart of this cavity, where
almost no \Ha{} emission is present. They are difficult to be
interpreted as a separate shell, most likely we are talking about
an increase of turbulent velocities on the border of HII regions.

A group of points, forming a `horizontal sequence' in the upper right part of the
$I-\sigma$ diagram caught our eye, since the surface brightness here is considerably
greater than for the other points with a large velocity dispersion ($\sigma=35-40\km$).
All the points, marked brick-red in our scheme, are grouped within the single \Ha\,
knot on the eastern edge of the disc. Its $FWHM$ in the image is equal to the size of
stellar images, i.e. it can not be spatially resolved.  The \Ha\, line profile has  a
distinct two-peak structure here with the distance between peaks of about $80\km$. The
examples of emission line profiles are shown in Fig.\ref{fig_prf}. The mean
systemic velocity of the knot is about $100\km$ in agreement with line-of-sight
velocities of the nearest side of the galaxy disc ($\sim80\km$), i.e. it seems to
belong the UGC 8508. Our first  explanation, based on the \Ha\, double component
profiles was that we observe a single remnant of the supernova explosion  (similar
to the SNR  S8 in IC1613, see below). However, our new  spectroscopic observations
revealed that it is a good candidate to unique  emission star -- luminous blue variable
(see Sect.~\ref{sec_app}).

\subsection{UGCA~92}

\begin{figure*}
\centerline{
\includegraphics[width=0.5\textwidth]{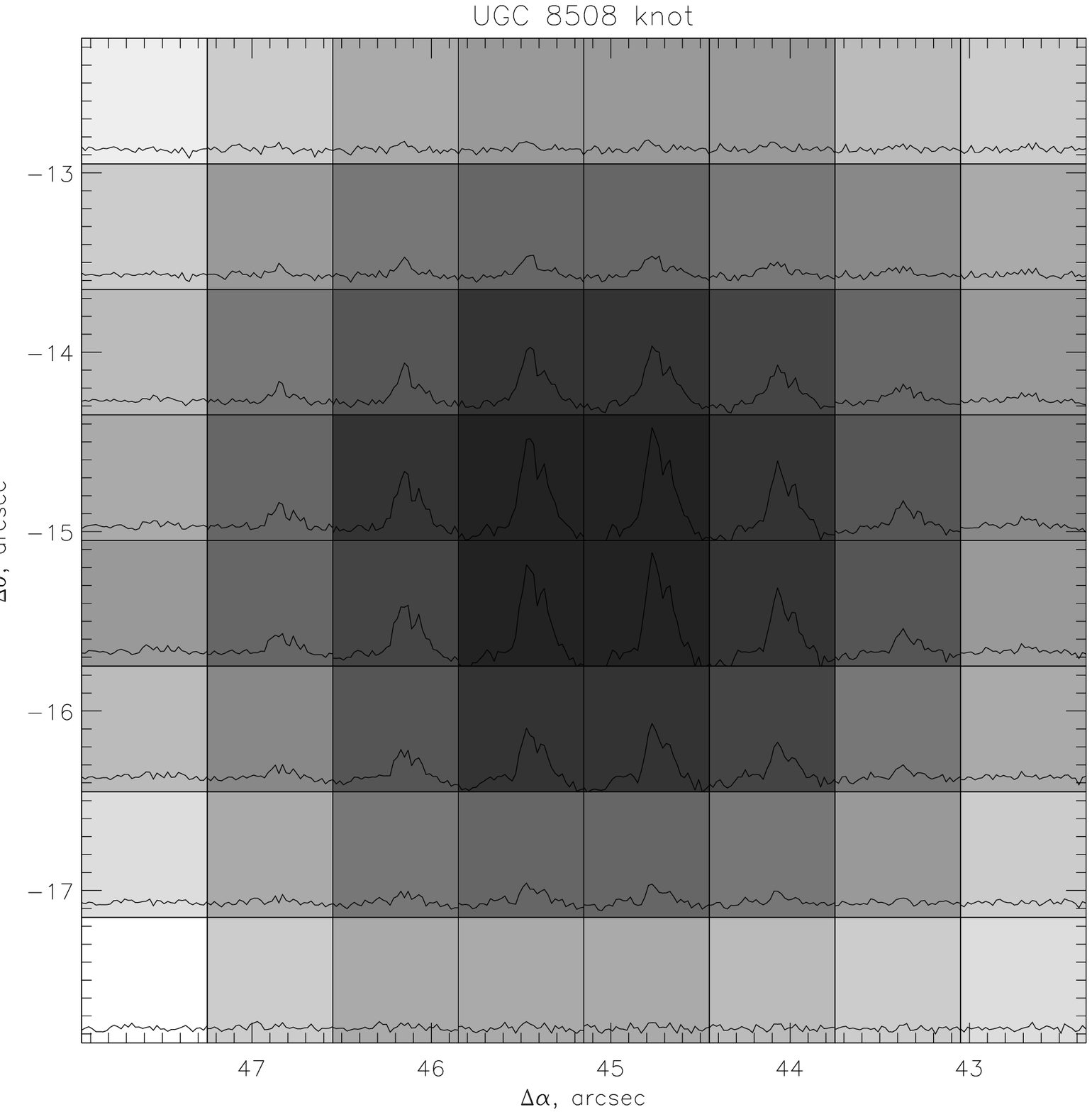}
\includegraphics[width=0.5\textwidth]{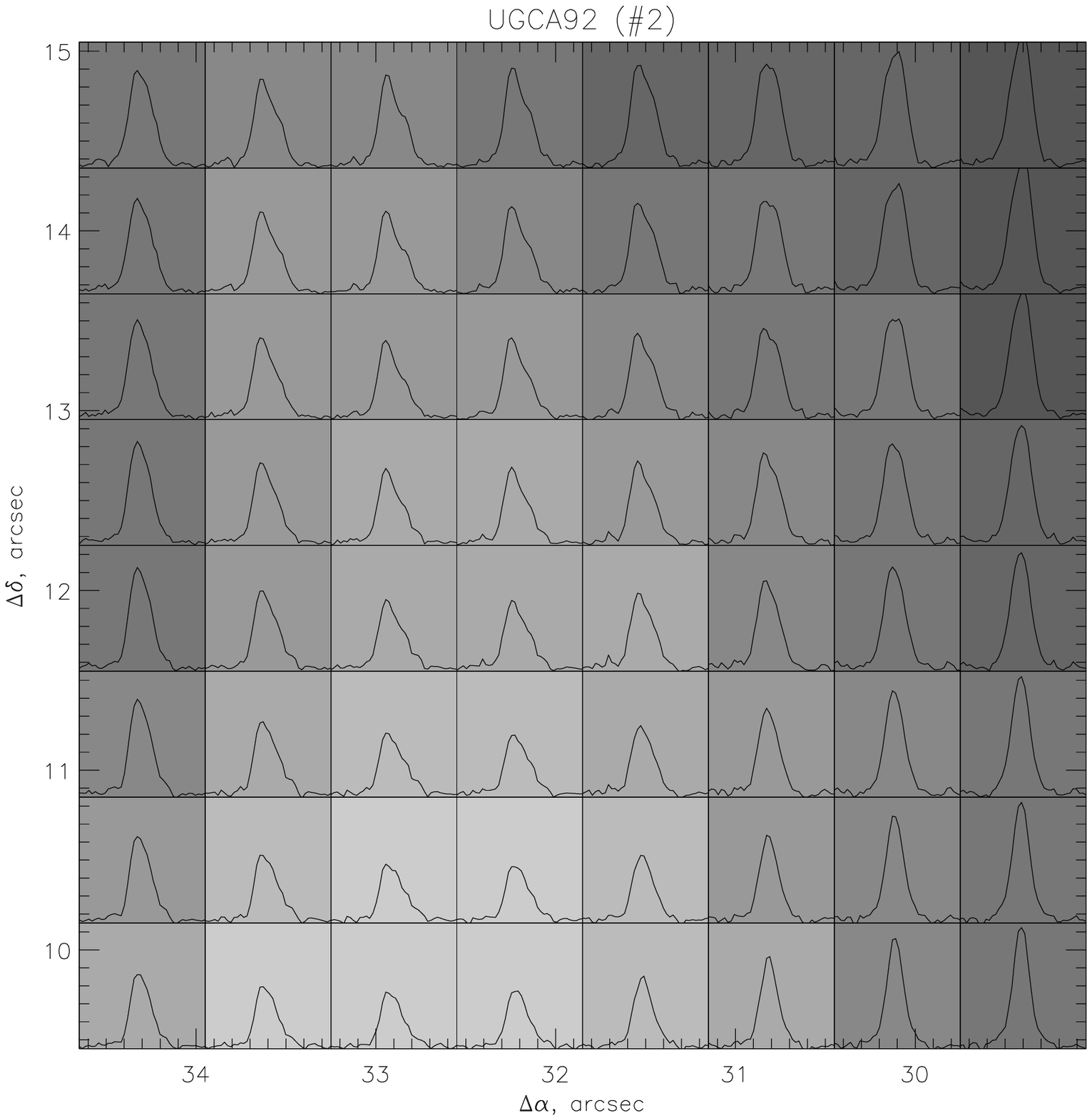}
} \caption{The regions possessing emission line profiles with a
two-component structure, overlapped on  the
 \Ha\, images: UGC~8508 (left)  and UGCA~92
(right).} \label{fig_prf}
\end{figure*}

We observe here the most significant asymmetry in the distribution of ionized gas
relative to the centre among all the galaxies of our sample. In the \Ha\, line image,
several shells and  arcs are seen, most of which are related to the bright star
formation region in the eastern part of the galaxy. There, five HII regions form a
closed elongated loop in the centre of which a maximum in the $\sigma$ distribution is
observed. One can discern a broad inclined lane on the $I-\sigma$ diagram (coloured
green and orange), the corresponded  points lying inside the above-mentioned loop of HII regions.
Meanwhile, in the centre of the loop, where the velocity dispersion is maximal and
intensity is minimal, the \Ha\, line profiles have a noticeable right wing, so that
they can be decomposed into two components -- a more intense blue one and a 1.5-2
times fainter red one (see Fig.\ref{fig_prf}). The separation between the components
is $\Delta V=45-57\,\km$, decreasing from the centre to edges.
This multicomponent profile structure could not be a result of spatial smoothing procedures (`beam smearing'), because the  peak velocity in the  HII  regions, that formed this loop is constant $\pm10\km$.
The observed features
indicate that we are dealing with a shell (designated as UGCA~92\,\#2), expanding at
a rate of $v_{exp}=\frac{1}{2}\Delta V\approx30\km$, this value  coincides with
the maximal velocity dispersion in this region (see Tab.~\ref{tab_shells}).
The velocity dispersion distribution easily reveals three more expanding shells of similar size and age. It is possible that
the external arch structures in the western and eastern parts of the galaxy have the
same nature, however, we failed to measure their   expansion velocity.

\subsection{VII~Zw~403}

We present here the re-analyzed data first described in \citet{Lozinskaya2006}. An
empty spot near the centre is caused by masking the trace of a bright parasitic reflex (see \citet{MoiseevEgorov2008} for details on the  ghost reflex). Galaxy VII~Zw~403 is one of the most well-described in the
literature objects of our sample. According to \citet{Lynds1998} the age of the latest
burst of star formation here is 4--10 Myr, which closely coincides with the estimates
of the kinematic age of shells, associated with bright regions of star formation
\citep[at least 3--4 Myr old, according to][]{Lozinskaya2006}. On the other hand, the
velocity dispersion of ionized gas in the HII regions themselves is small:
$\sigma<(\sigma_m=18\km$). The $I-\sigma$ diagram  confidently reveals an inclined lane
(marked with green and orange) which, according to the model by \citet{MunozTunon1996},
must point to the presence of an expanding shell.

The distribution of points on the map of the galaxy shows that
this shell (VII~Zw~403 \#1) is located between the bright HII
regions in the centre of the galaxy and is adjacent to region 4
according to the numbering of \citet{Lynds1998}. The kinematic age of the shell coincides
with the estimates by \citet{Lozinskaya2006} given
above\footnote{Note that the use of the approximation by the Voigt
profile in our work allows making more accurate estimates of  the
expansion rate than the Gaussian approximation in
\citet{Lozinskaya2006}.}. Other regions of increased velocity
dispersion are located at the periphery of the HII regions and in
the outer parts of the disc of ionized gas. The local gas  kinematics
 is difficult to explain in terms of individual shells. Thus,
several spots with high velocity dispersion in the northern part of the galaxy are more
likely associated with the emission arcs of relatively low surface brightness, detected
by \citet{Silich2002} and \citet{Lozinskaya2006}.  According to the cited authors,
these arcs were caused by the feedback of the stellar population
aged $t\approx10$ Myr, formed at the early stages of the last burst of star formation.

\section{The effect of spatial resolution}

\citet{Yang1996} and \citet{MunozTunon1996} have  discussed and interpreted the
$I-\sigma$ diagrams, constructed for the giant complexes of star formation in the
nearby galaxy M33. The spatial resolution of their  observations was $\omega\approx3.5$
pc. In this work we study more distant galaxies with a much rougher resolution (see
the last column in Table~\ref{tab_obs}). It is hence clear that we are loosing the data
on the small-scale kinematics of gas. How seriously may this  affect the shape of the
$I-\sigma$ diagrams and the distribution of velocity dispersion? In order to analyze
the effects of spatial resolution on the observed kinematics, we examined two nearby
and well-studied dwarf irregular galaxies of the Local Group: IC 10 and IC 1613. We
have earlier observed both galaxies at the 6-m telescope with the same equipment,
described in Section~\ref{obs}. The spatial size of regions, encompassed by current
star formation is about 1 kpc, just like in the  galaxies described above (hereinafter
-- the `main sample'). A detailed description of the distributions of ionized and
neutral gas, the identification of the expanding HI and HII bubbles, and a discussion
of the relation between these bubbles and young stellar groups are given in
\citet{Lozinskaya2008} and \citet*{Egorov2010} for IC 10, and in \citet{Lozinskaya2003}
for IC 1613. The spatial resolution of  our observations of ionized hydrogen in the
\Ha\ emission  line  was 5 pc and 8 pc  correspondingly.

\begin{figure*}
\centerline{
\includegraphics[width=0.5\textwidth]{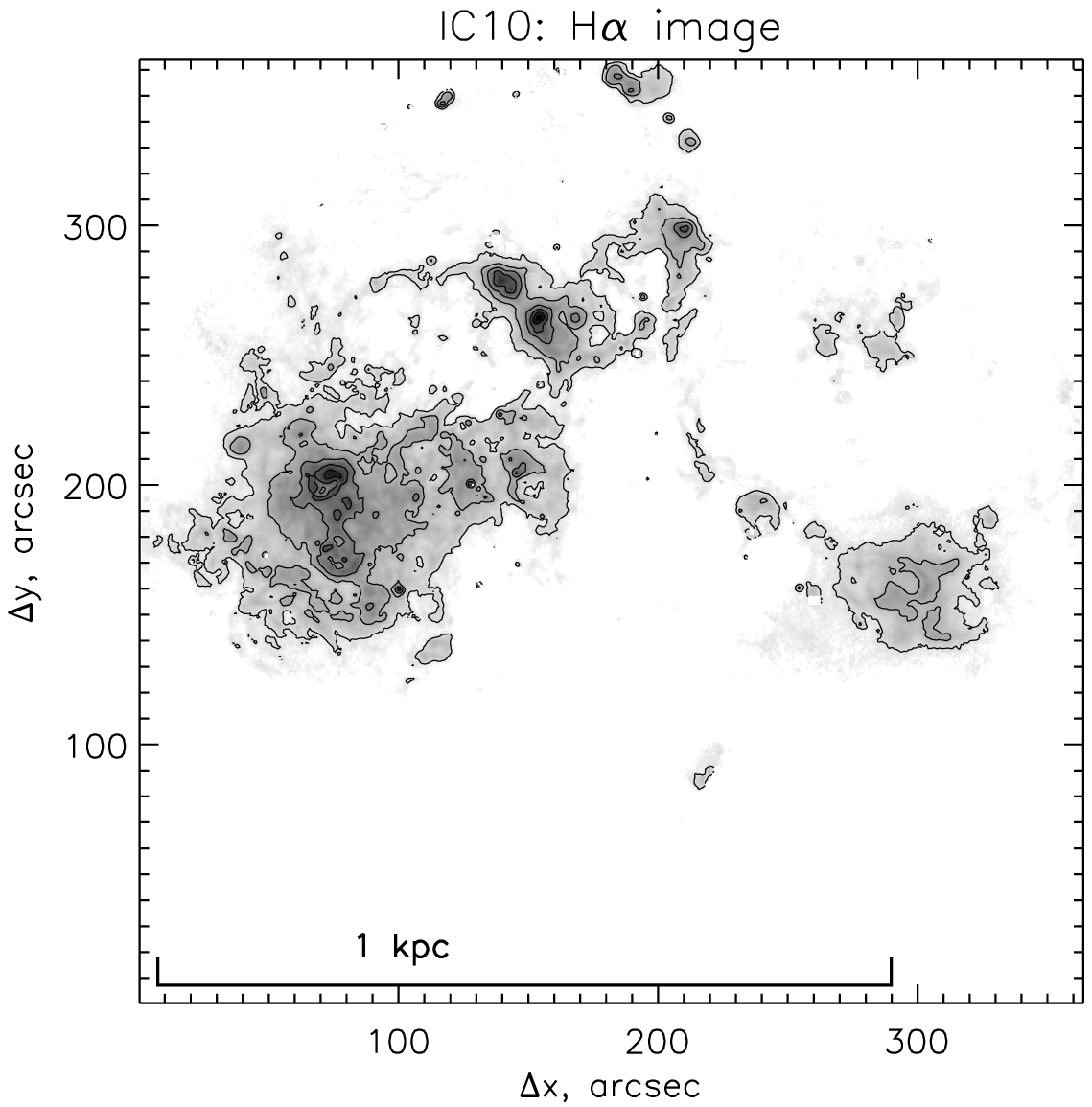}
\includegraphics[width=0.5\textwidth]{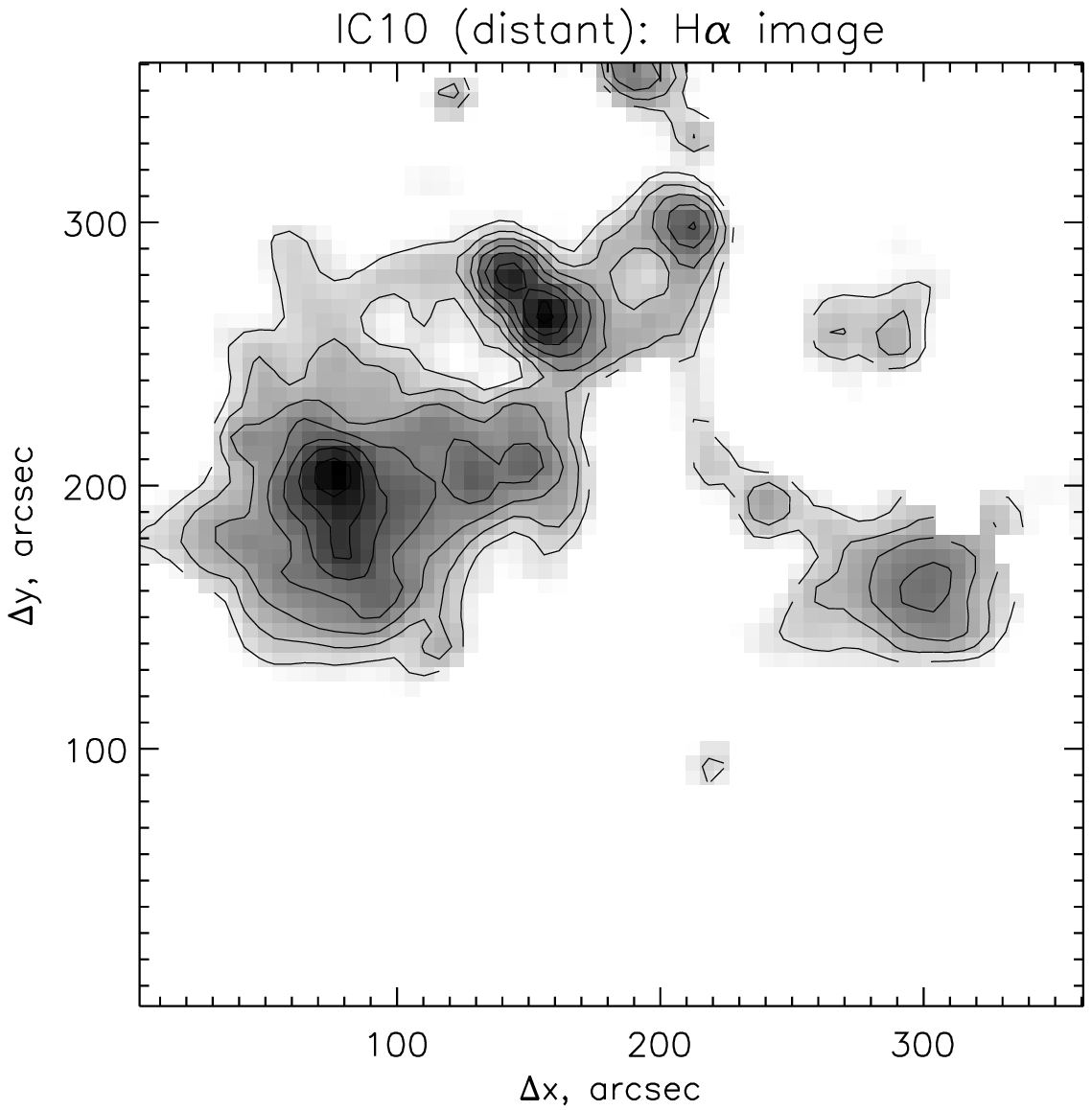}
}
\centerline{
\includegraphics[width=0.5\textwidth]{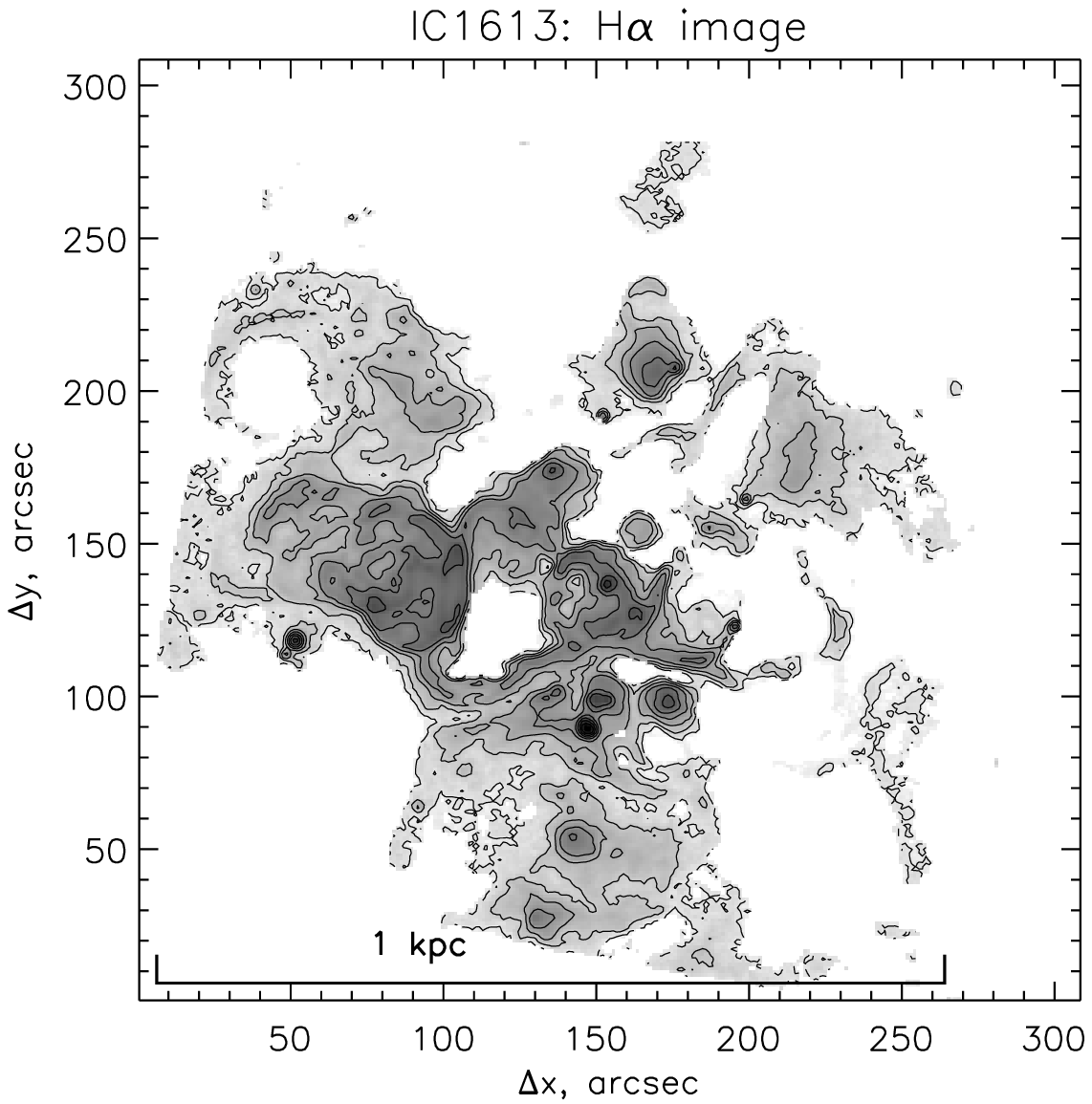}
\includegraphics[width=0.5\textwidth]{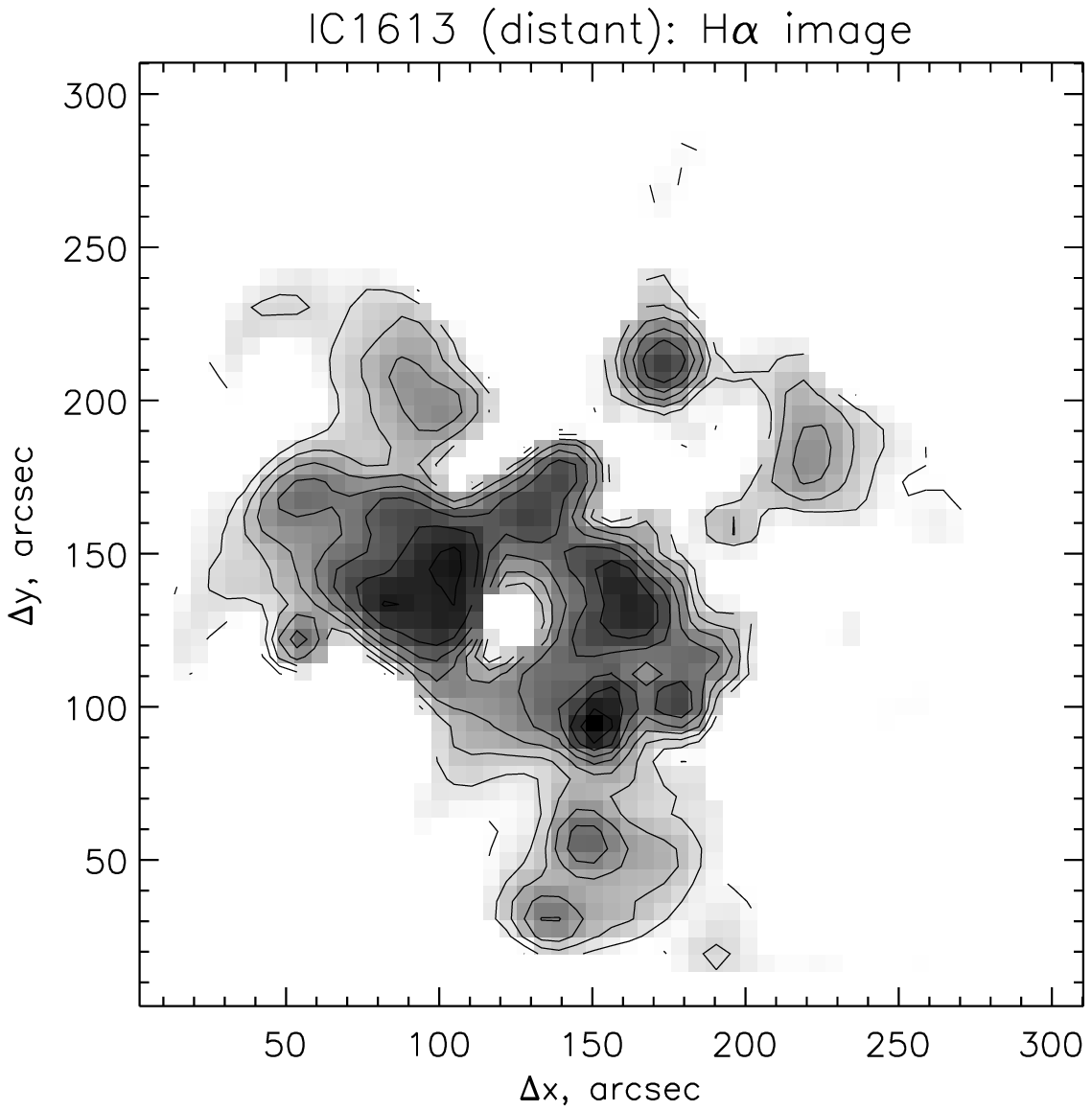}
} \caption{The \Ha\,   images of the galaxies IC 10 (top) and
IC1613 (bottom). Left column -- the original results of observations
with the scanning FPI, right column -- the same data smoothed to the
spatial resolution of $\sim40$ pc} \label{fig_IC}
\end{figure*}

\begin{figure*}
\centerline{
\includegraphics[width=0.5\textwidth]{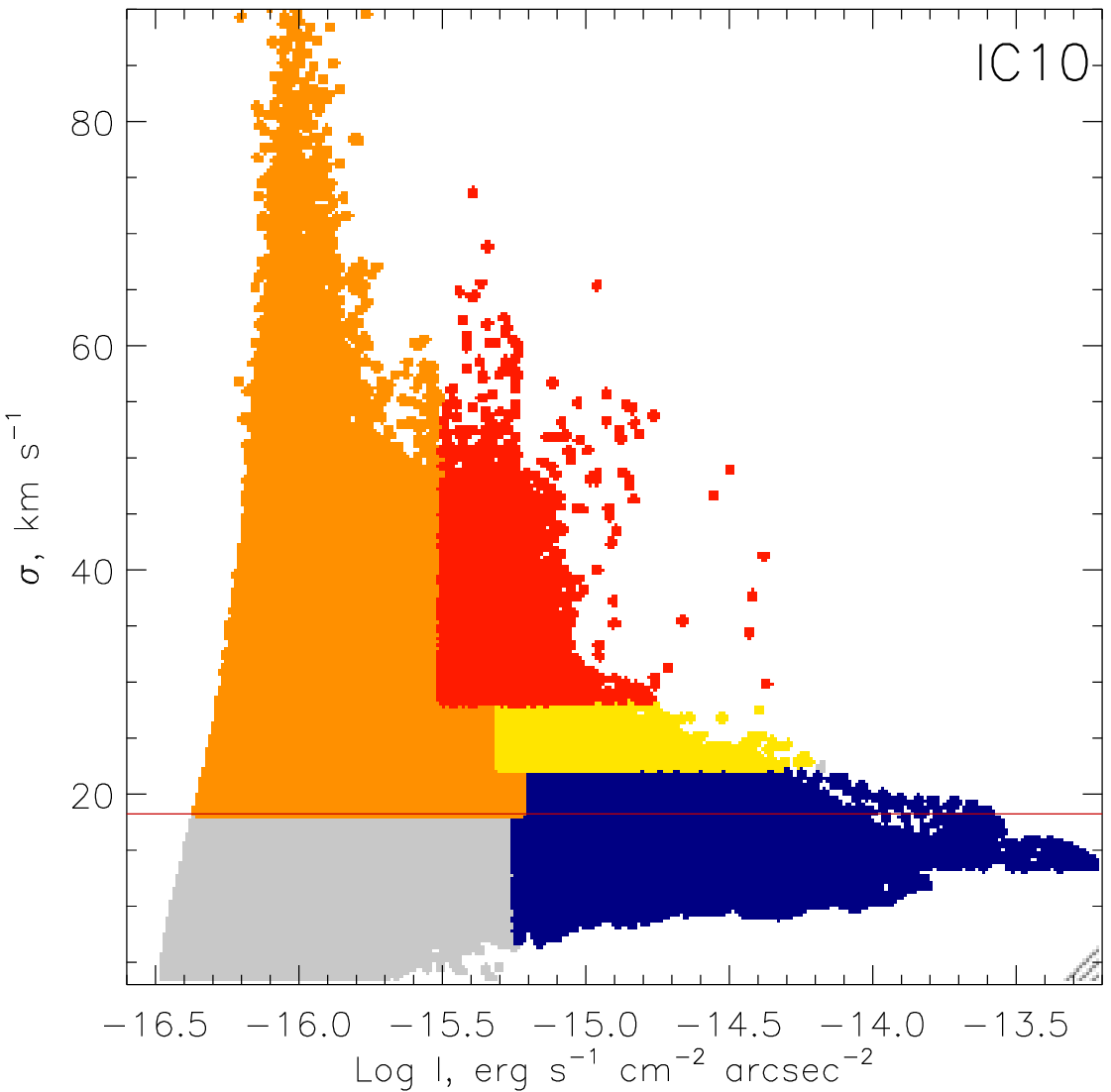}
\includegraphics[width=0.5\textwidth]{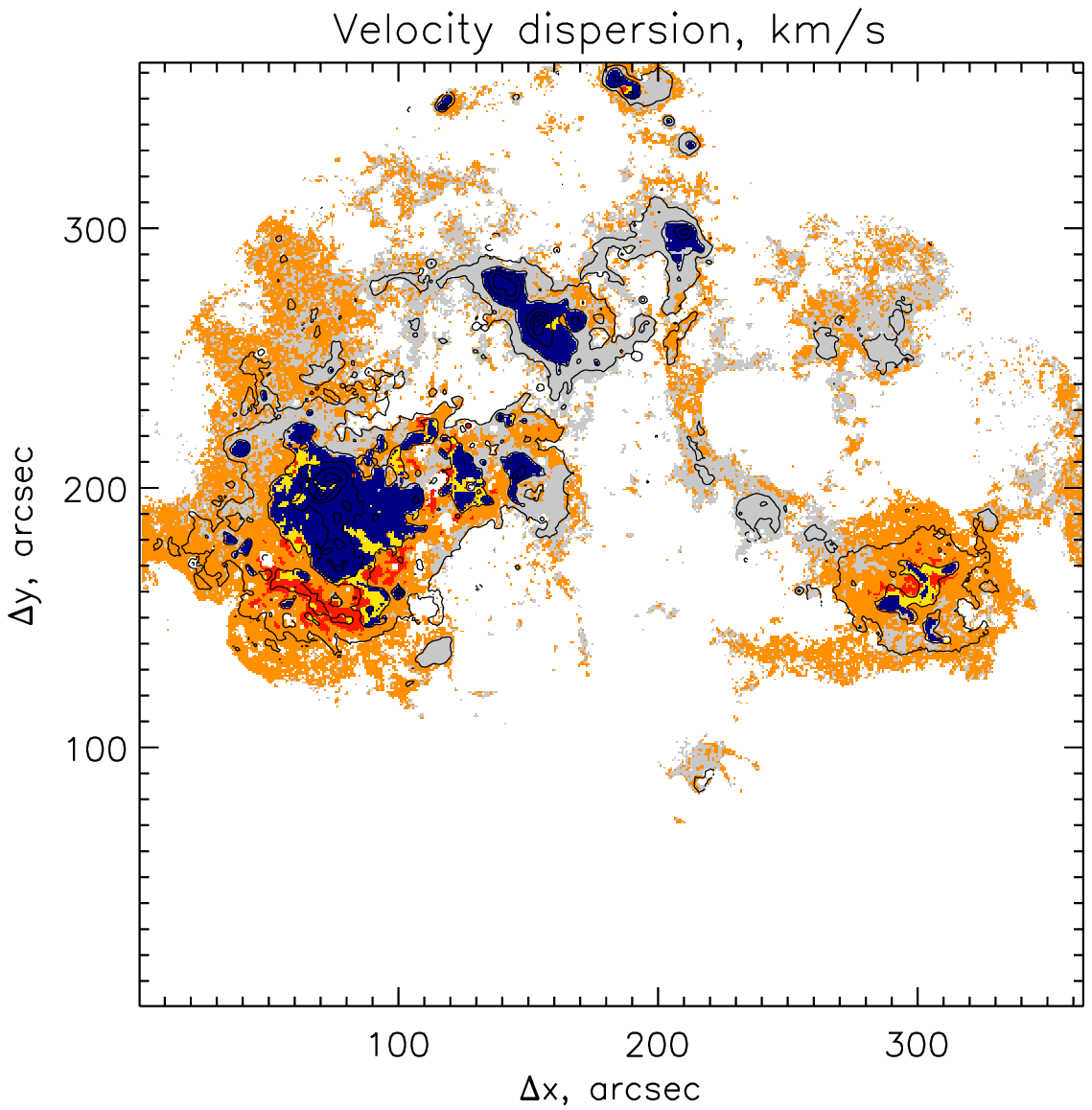}
}
\centerline{
\includegraphics[width=0.5\textwidth]{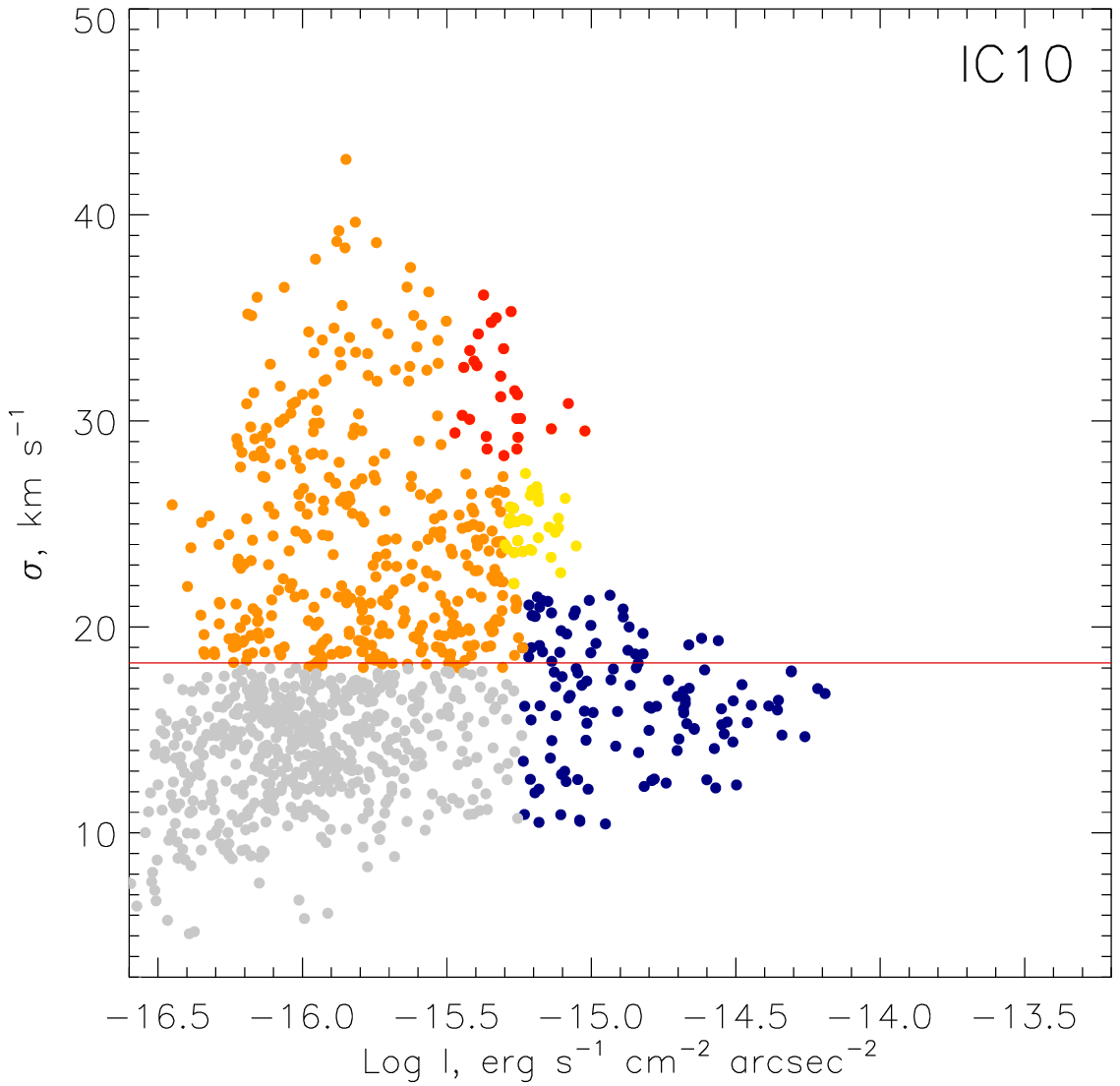}
\includegraphics[width=0.5\textwidth]{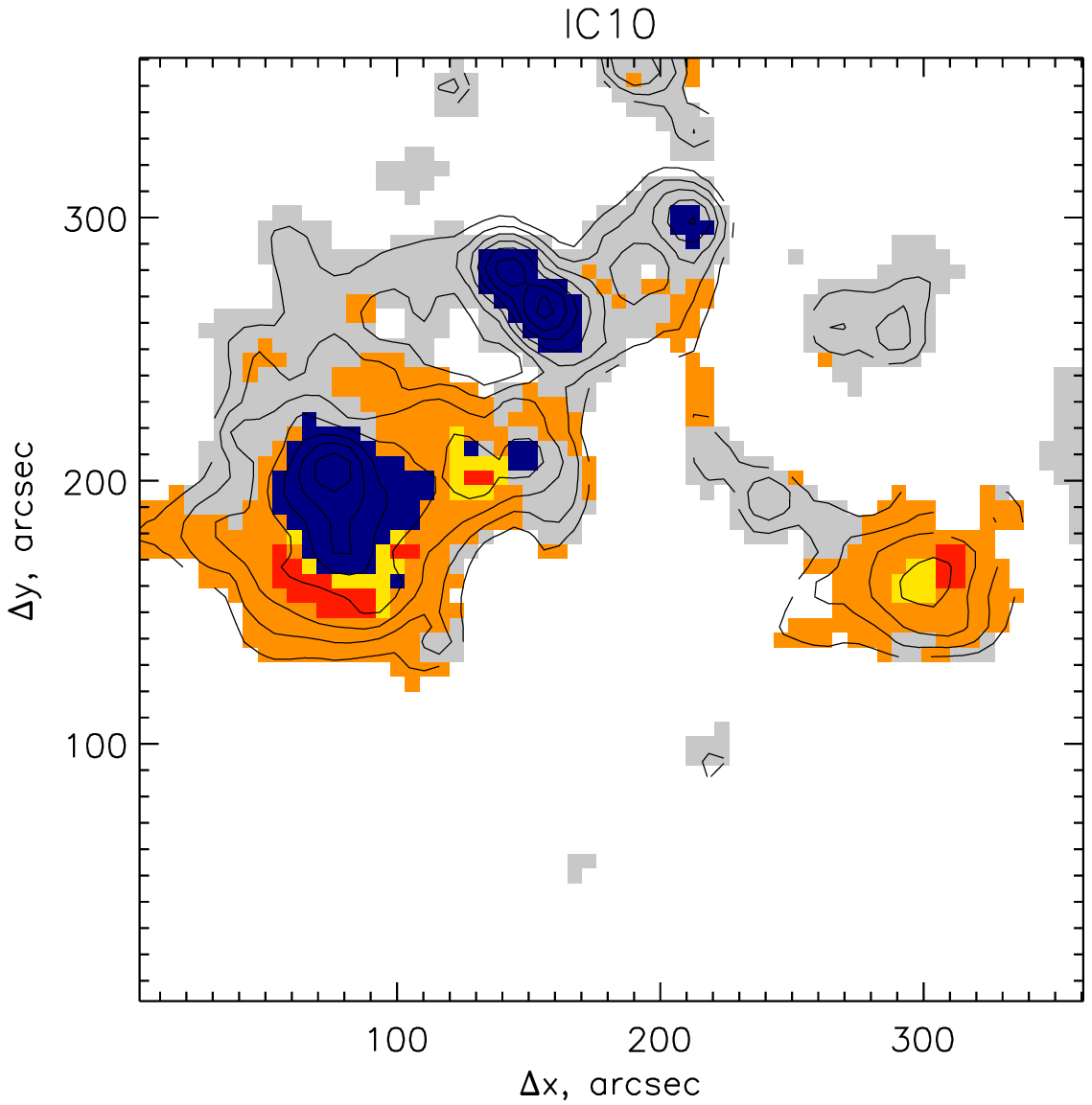}
} \caption{IC10: the original (top) and smoothed (bottom) data. The left plot shows the
$I-\sigma$ diagrams. The red horizontal line marks the value of the intensity-weighted
mean  velocity dispersion  $\sigma_m$. The right plot shows the location of regions
identified by different colours on the $I-\sigma$ diagram. The contours correspond to
the isophotes in the  \Ha{} line.} \label{fig_IC10}
\end{figure*}

\begin{figure*}
\centerline{
\includegraphics[width=0.5\textwidth]{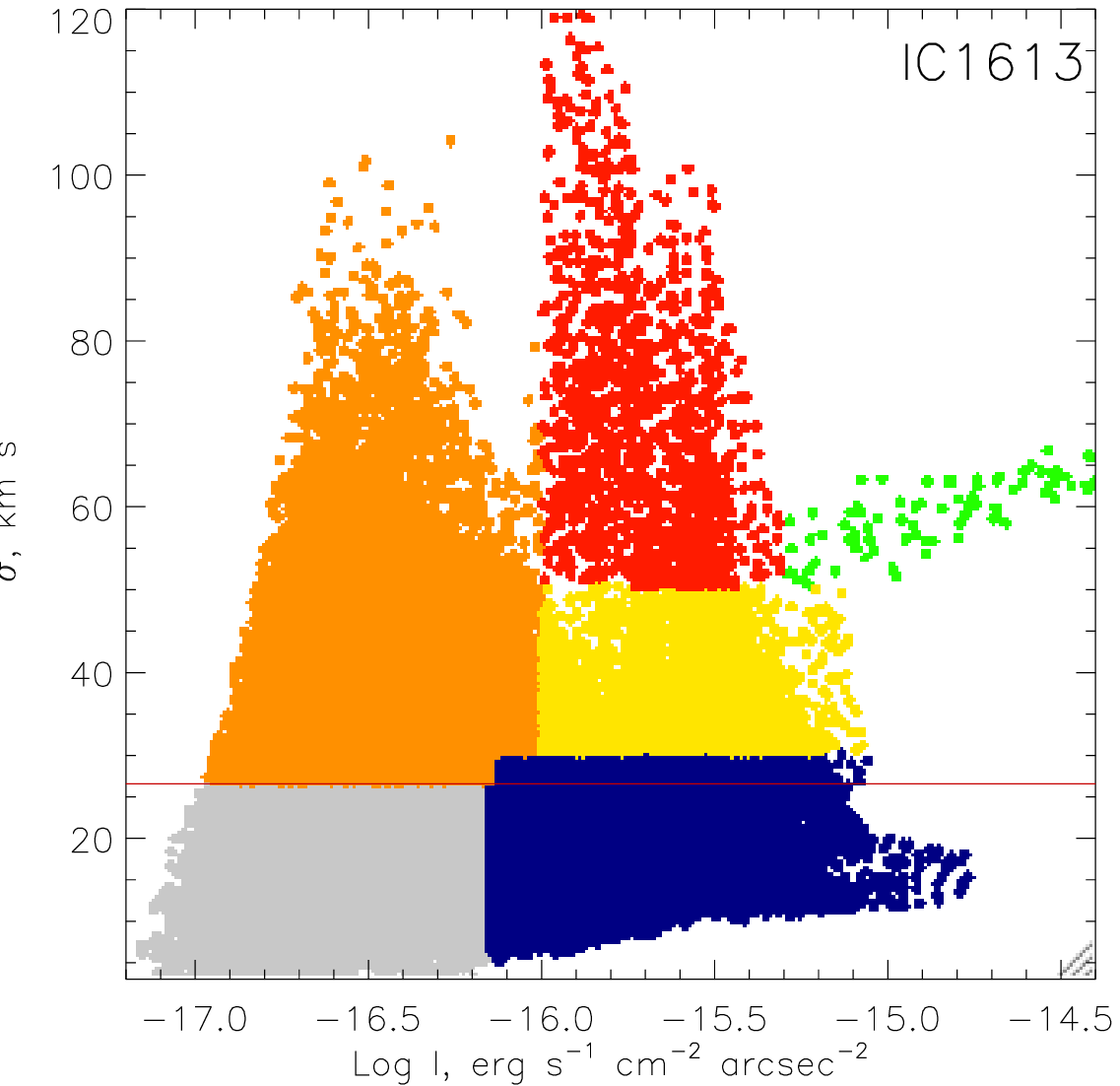}
\includegraphics[width=0.5\textwidth]{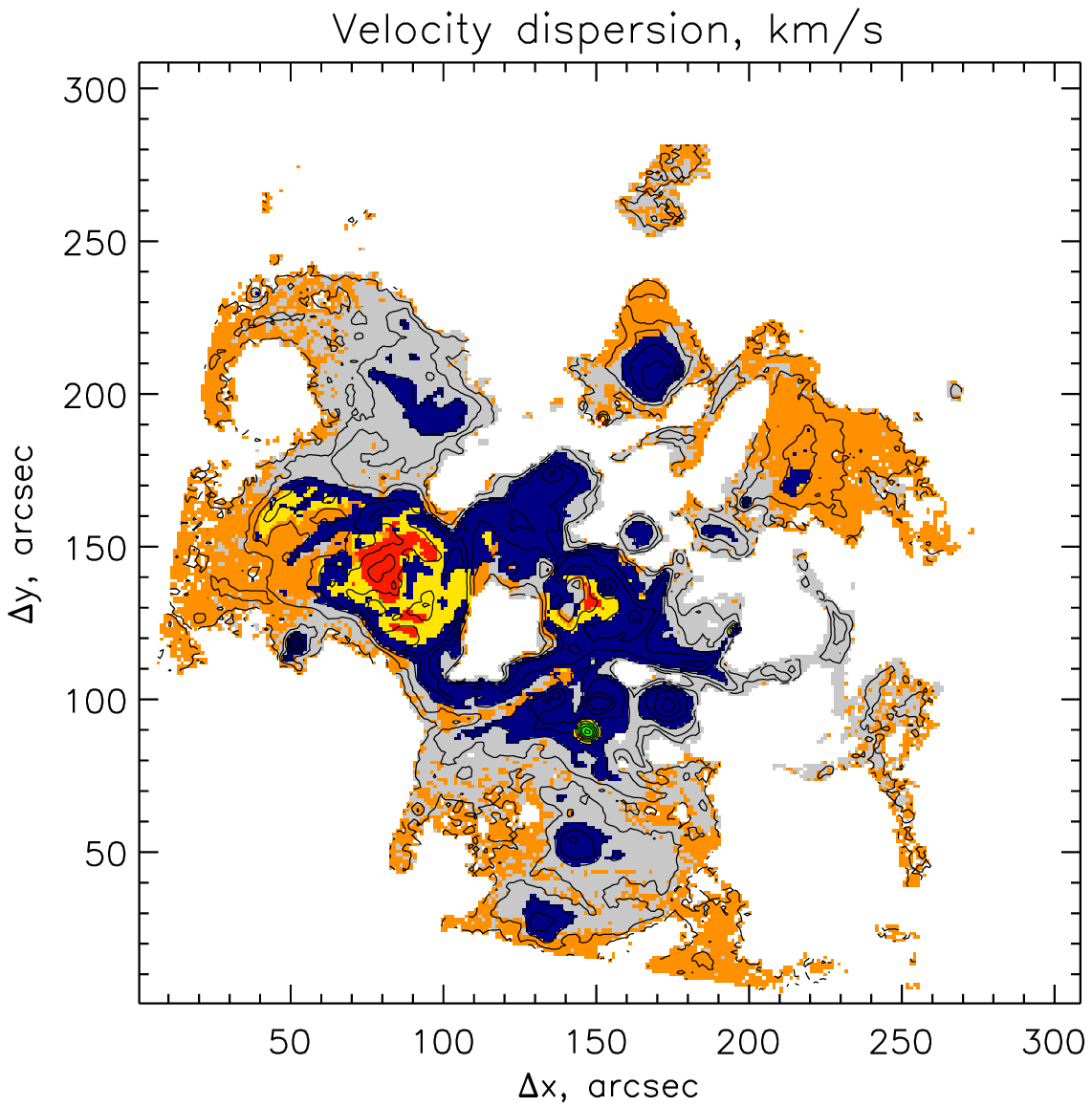}
}
\centerline{
\includegraphics[width=0.5\textwidth]{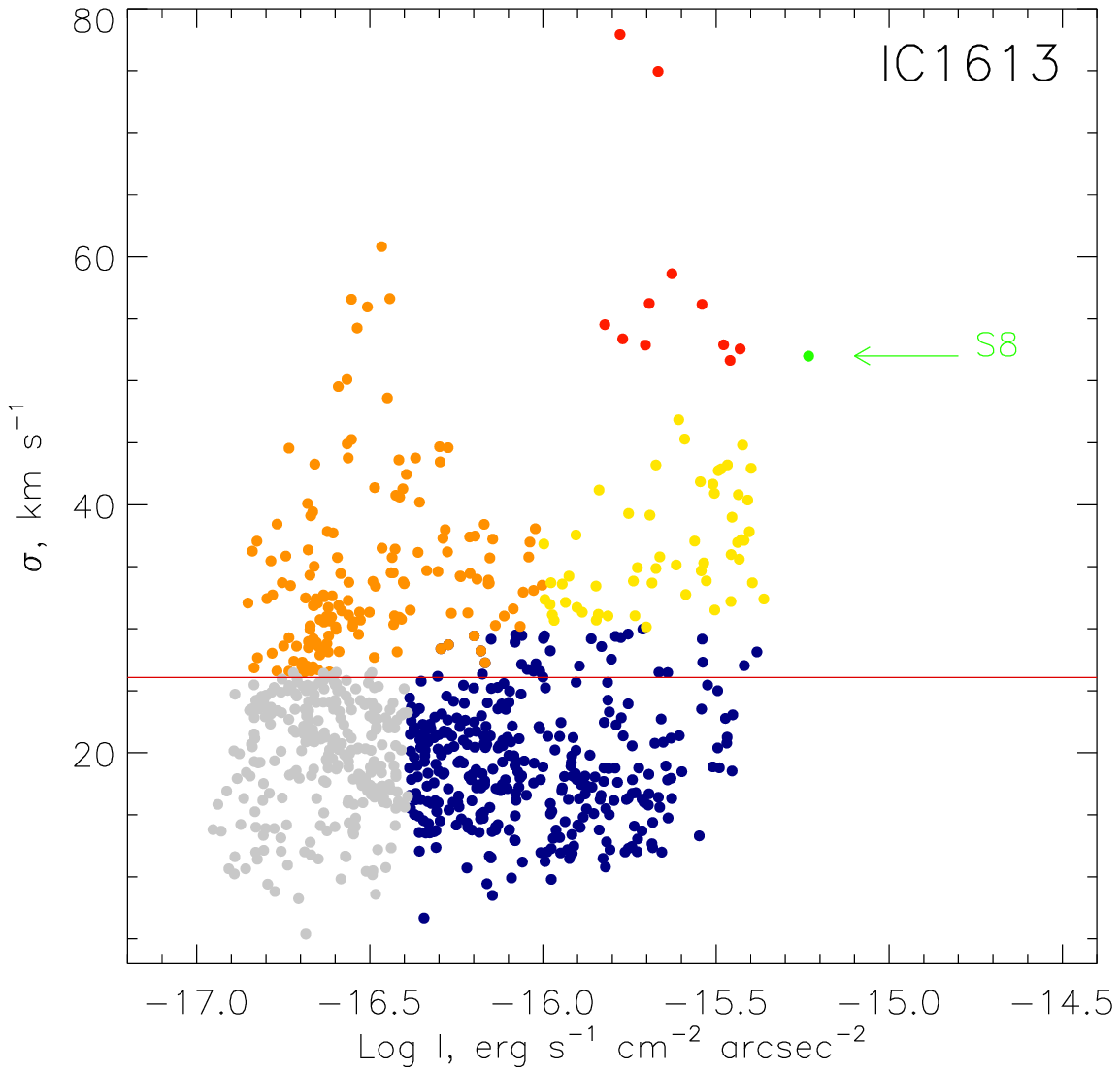}
\includegraphics[width=0.5\textwidth]{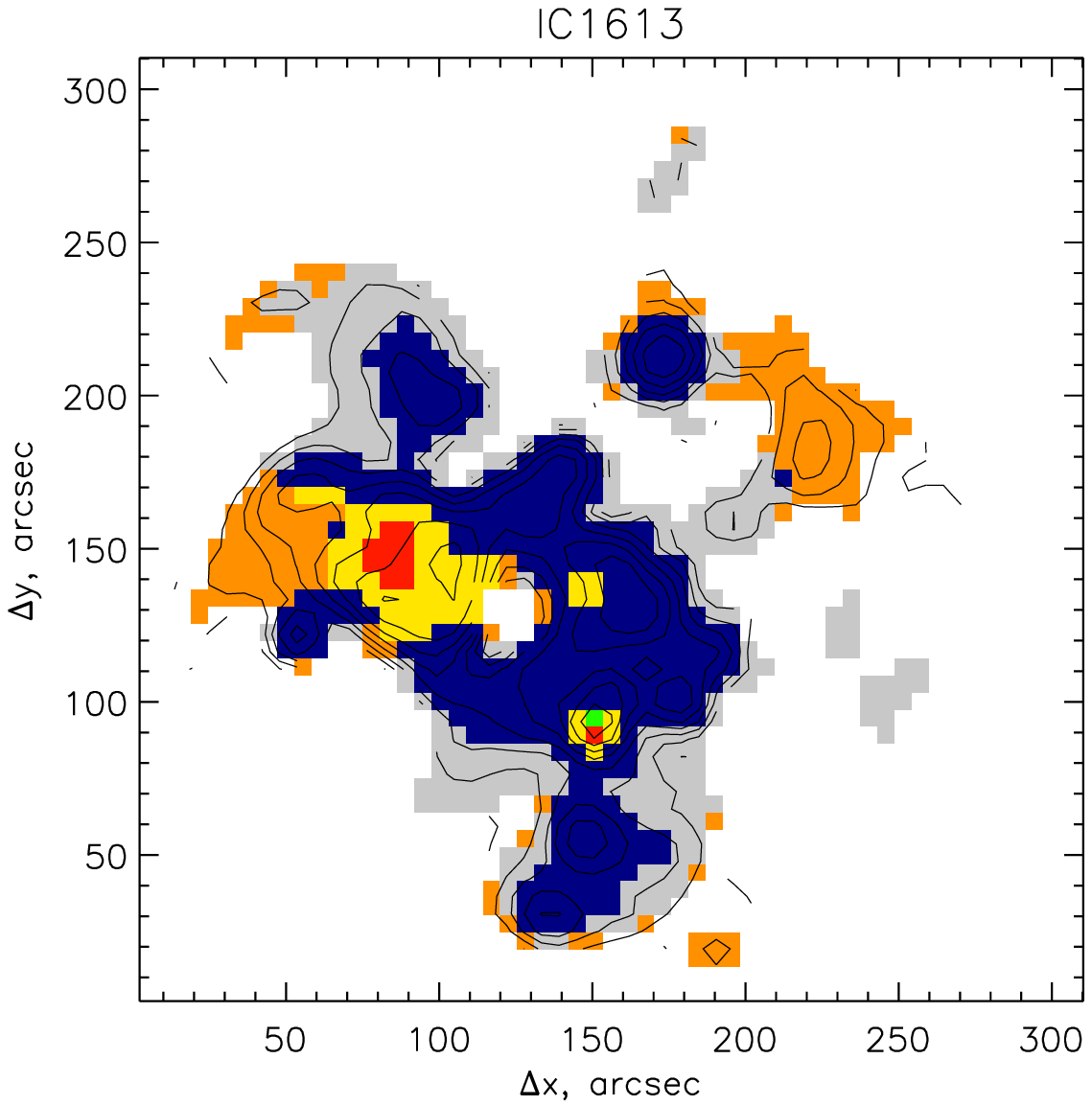}
} \caption{IC1613:  the original (top) and smoothed (bottom) data. The left plot shows
the $I-\sigma$ diagrams. The red horizontal line marks the value of the
intensity-weighted mean  velocity dispersion  $\sigma_m$. The arrow with a marking
directs to the point, corresponding to the known supernova remnant. The right plot
shows the localization of regions, marked by different colours on the $I-\sigma$
diagram. The contours correspond to the isophotes in the \Ha{} line.}
\label{fig_IC1613}
\end{figure*}

To imitate the effect of low resolution, the original data cubes
for IC 10 and IC 1613 in the \Ha\, line were first smoothed by the
two-dimensional Gaussians, and then binned, so that the resulting
pixel size amounted to  $\sim5.6$ arcsec. Table~\ref{tab_ic} lists
the distance to the galaxy, and the initial ($\omega $) and
smoothed ($\omega_{smo}$) spatial resolution. The smoothed data
show  how these two galaxies would look like when
observed with a resolution of $\omega_{smo}\approx40$ pc, i.e.
under the same conditions as the objects of the main sample. The
velocity dispersion maps and the diagrams for the original and
smoothed cubes were constructed, using the method described in
Section~\ref{obs}. Figure~\ref{fig_IC} shows the original and
smoothed images of both galaxies in the \Ha{} line, and the
corresponding $I-\sigma$ diagrams are shown in
Figs.~\ref{fig_IC10} and \ref{fig_IC1613}.

\begin{table}
\caption{Smoothing of the observational data on galaxies of the
Local Group}\label{tab_ic}
\begin{tabular}{lcccccc}
\hline
\\
Name          & $D$     &\multicolumn{2}{c}{$\omega$}&   \multicolumn{2}{c}{$\omega_{smo}$}  \\
                   &  (Mpc)  &        (arcsec)        &          (pc)&    (arcsec)                     & (pc)                 \\
\hline
IC~10         & 0.73     &         1.5               &      5.3     &   11                         & 40     \\
IC~1613     & 0.80     &         2.2               &      8.5     &   11                         & 43     \\
  \hline
\end{tabular}
\end{table}

The $I-\sigma$ diagrams, constructed from the non-smoothed data, reveal a fine
structure that reflects the data on gas motions on the scales of a few tens of parsecs.
It is obviously lost in the smoothed data, since the construction of the corresponding
diagrams involves a significantly smaller  (by 100--200 times) number of points. A
detailed review of the diagrams and the $\sigma$ distribution maps allows to make the
following conclusions:

\begin{itemize}
\item smoothing significantly (by 5--10 times) decreases the
observed range of surface brightness  $I$ by `smearing' the
compact bright HII regions. This is particularly notable in the
case of IC10 (Fig.\ref{fig_IC10});

\item similarly, the observed spread of $\sigma$ values decreases
almost twice due to the  averaging of small-scale
perturbations. On the other hand, smoothing has almost no
effect on the weighted average value of the  velocity dispersion
$\sigma_m$, since  the variations of radial velocity inside a sampled element (`beam-smearing effect') are
small, compared with the magnitude of turbulent velocities;

\item  the shape of the distribution of points on the $I-\sigma$ diagram preserves
its old `triangular' form -- a horizontal lane with $\sigma \le \sigma_m$ (coloured
dark blue) and the regions of low  surface brightness with high velocity dispersion. It is  also
well seen how the deteriorating resolution results in the merger of separate regions
with high $\sigma$, located on the periphery of HII regions, into a common structure,
embordering the star formation regions. That is exactly the picture we often see in the
main sample galaxies.
\end{itemize}

Large-scale kinematically isolated structures remain  visible in the smoothed images as
well. In IC10 this is primarily the so-called synchrotron superbubble, located on the
southern boundary of the brightest (eastern) star formation region, and prominent by
its increased velocity dispersion. We have previously shown \citep{LozMois2007} that
this shell   centering in the  X-ray source IC10 X-1  is probably a remnant of an
hypernova explosion.   IC 1613 data reveal   a    complex of multiple shells, located
in the eastern part of the star-forming region (coloured red and orange in
Fig.\ref{fig_IC1613}). Here, the  \Ha\, line profiles have a clear two-peak structure
with the velocity difference between the components of  $\Delta V\approx100\km$. In
\citet{Lozinskaya2003} we thoroughly examined the P--V diagrams for this region and
showed that the observed kinematic features are associated with the effect of expanding
and possibly colliding shells of ionized gas. Among the galaxies from the main
sample, similar kinematic components at congruent spatial scales was detected only in
UGCA92. The second interesting feature that attracts attention in the $I-\sigma$
diagrams in IC1613 is the brightest in the \Ha\, line compact region at the centre
of the field (with coordinates $(\Delta x=150'',\Delta y=90'')$), also characterized by a high velocity dispersion
($\sigma\approx50-60\km$). In Fig.\ref{fig_IC1613} it is marked in green. This is a
remnant of the supernova S8, which stands out by its unusually high luminosity
simultaneously in the X-ray (a marker of a young remnant) and optical (a marker of an
old remnant) ranges  \citep{Lozinskaya1998}. The emission line profile has  a multi-component
 structure here.  
A object with a similar position in $I-\sigma$ diagram and also with multicomponent profile   from the main sample of galaxies is a LBV candidate in UGC~8508.

\section{Discussion}

\label{sec_discuss}

In all of the above galaxies there exists a clear link between the flux in the \Ha\,
line and   velocity dispersion of ionized gas. The brightest HII regions reveal small
line widths ($\sigma\le\sigma_m$), the scatter of observed values increases with
decreasing surface brightness, so that in most regions of low brightness the velocity
dispersion significantly exceeds the mean ($\sigma\gg\sigma_m$),  although   individual
points may still show a low velocity dispersion ($\sigma\approx3-7\km$). In general,
the shape of the $I-\sigma$ diagram for dwarf irregular galaxies bears resemblance to
the diagram constructed by \citet{MunozTunon1996} for the star formation regions NGC
588 and NGC 604. Such a resemblance  is not surprising in objects of different scales,
since in both cases the kinematics of gas must be primarily determined by the input of
mechanical energy of wind of massive young stars and multiple supernovae explosions,
the interaction of which with the circumambient interstellar medium depends on its
density.

However, there is a number of differences between the kinematic properties of giant HII
regions and Irr galaxies;  the sketch presented in Fig.~\ref{fig_model} illustrates
the diference.

The full scale of the regions in NGC~588 and NGC~604, emitting in the \Ha\, line  is
not larger than 130--150 pc, hence, at the corresponding $I-\sigma$ diagrams the entire
region of increased velocity dispersion splits into several inclined lanes, which
\citet{MunozTunon1996} identify with expanding ionized shells that  differ by  age. In
the scheme they propose, which is illustrated in our Fig.~\ref{fig_model} (inset b),
the centre of the expanding thin shell, when projected onto the sky plane, has a low surface 
brightness  and a large velocity dispersion, determined by the expansion velocity.
With increasing distance from the  centre of the shell the Balmer emission lines intensity increases (the
line-of-sight intersects an even thicker layer), and $\sigma$ declines, since the
projection of expansion velocity onto the line-of-sight decreases.

In the galaxies    considered above the pattern is more complex. Here the
total spatial size of the regions, where we examine the motions of ionized gas is
significantly (8--10 times) larger. In 5 out of 7 galaxies of the main sample (DDO53,
DDO125, UGC 8508, UGCA 92, and VII Zw 403) we were able to identify the expanding
shells of about 80--350 pc in size. The points belonging to these shells are
forming inclined lanes on the diagrams as well. However, we were unable to associate
the bulk of points having a high velocity dispersion with such structures. This is not
surprising, since the formation of giant shells in itself requires a number of specific
conditions (a sufficient initial gas density, a simultaneous onset of starburst), and
in most cases, it can be the result of the influence of several generations of star
groups on the interstellar medium \citep{McQuinn2010, Warren2011}.

Perhaps we simply do not notice small shells sized 5--50 pc due to the low spatial
resolution? The   analysis of the original and smoothed data  provided for the nearby
galaxies IC10 and IC 1613 shows that an insufficient spatial resolution can not explain
the fact fact that most of the regions with high $\sigma$ are not related to the
expanding shells. More important fact is that the points with high $\sigma$, occupying
the top left part of the $I-\sigma$ diagram, belong to the diffuse emission of low surface 
brightness and are spatially clustered around the star formation regions and on the
periphery of the disc of ionized gas in the galaxies. Note that despite the low surface
brightness, the signal-to-noise ratio here is sufficient for reliable measurements of
$\sigma$.

\begin{figure*}
\centerline{
\includegraphics[width=0.5\textwidth]{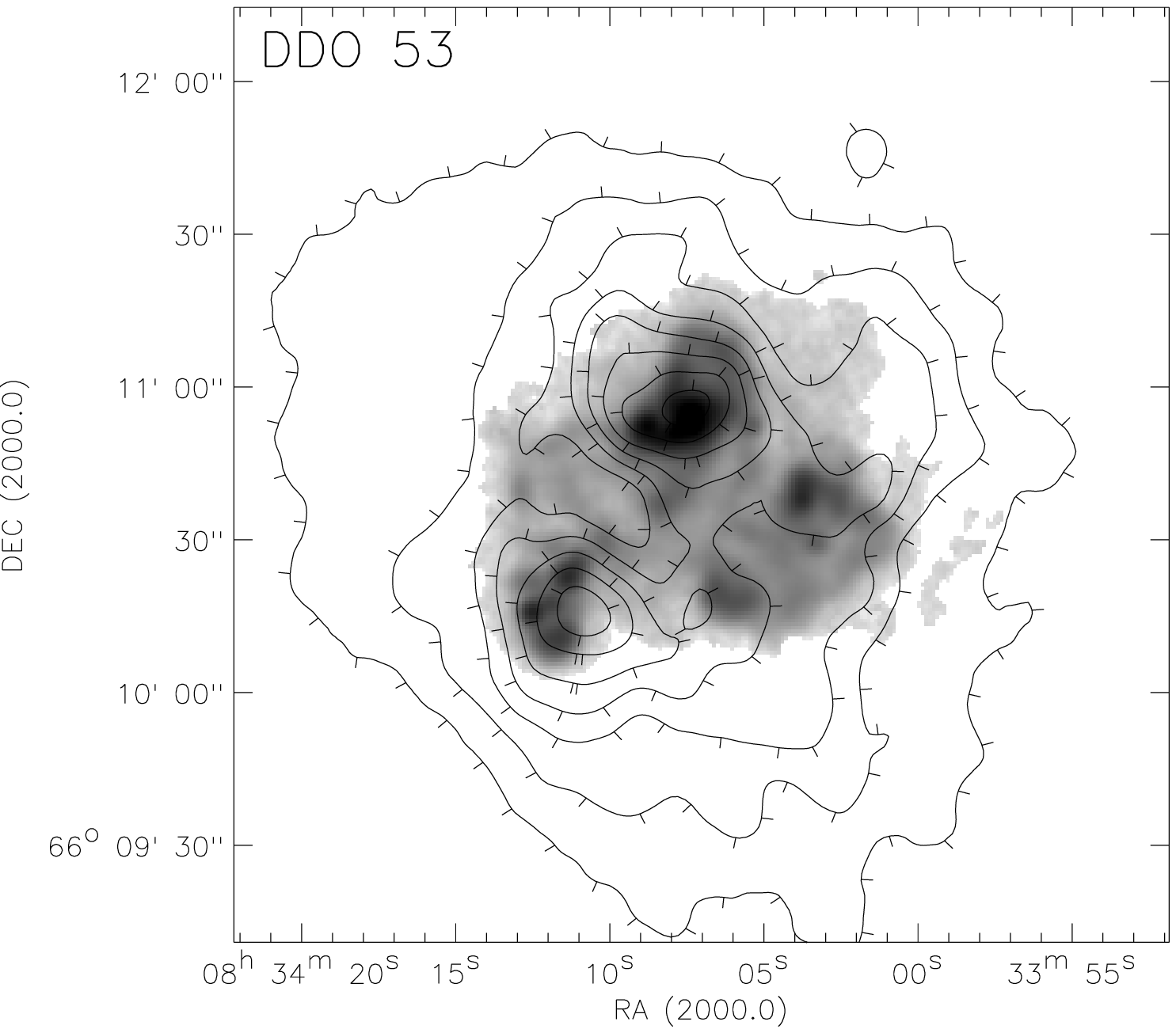}
\includegraphics[width=0.5\textwidth]{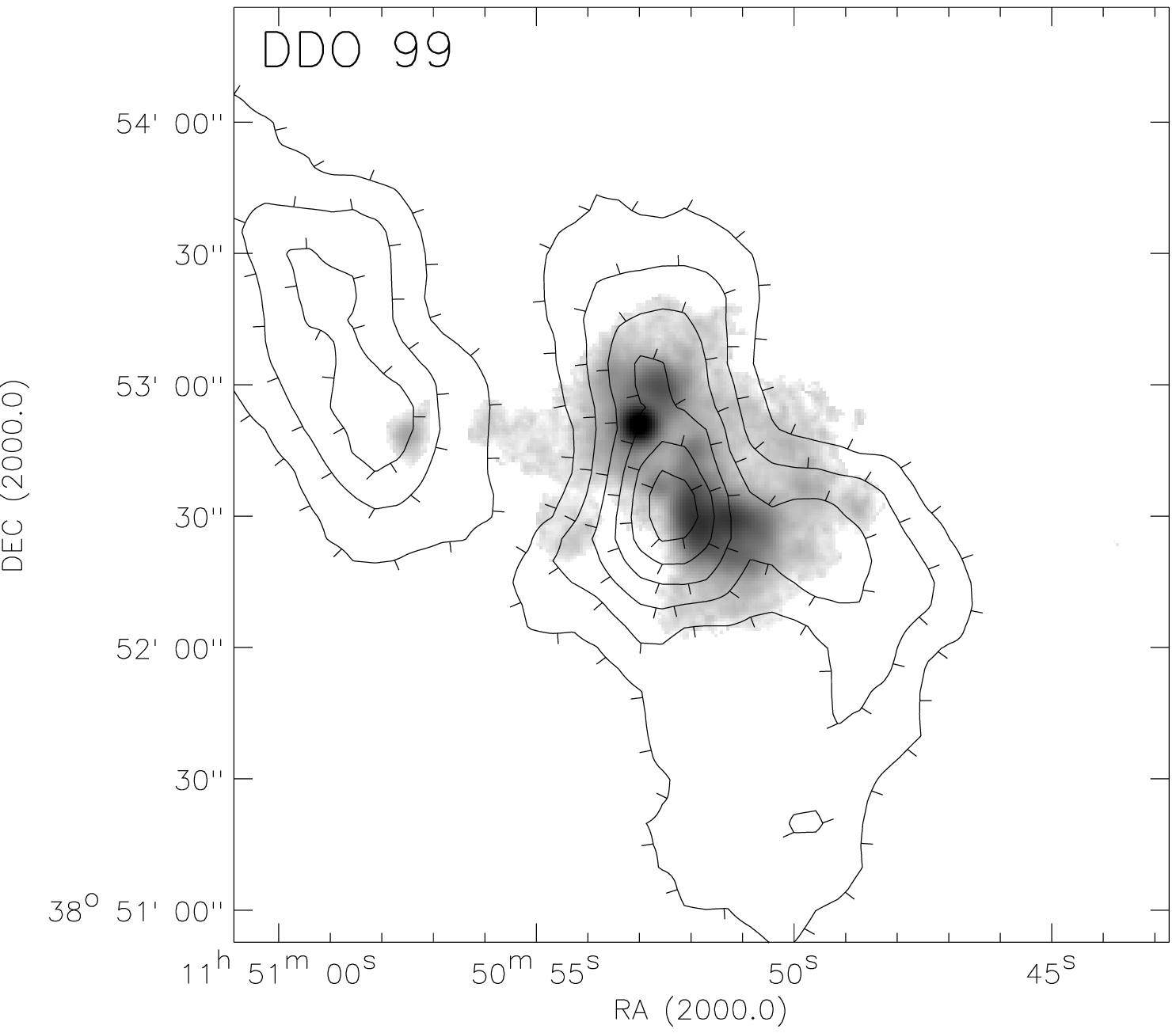}
}
\centerline{
\includegraphics[width=0.5\textwidth]{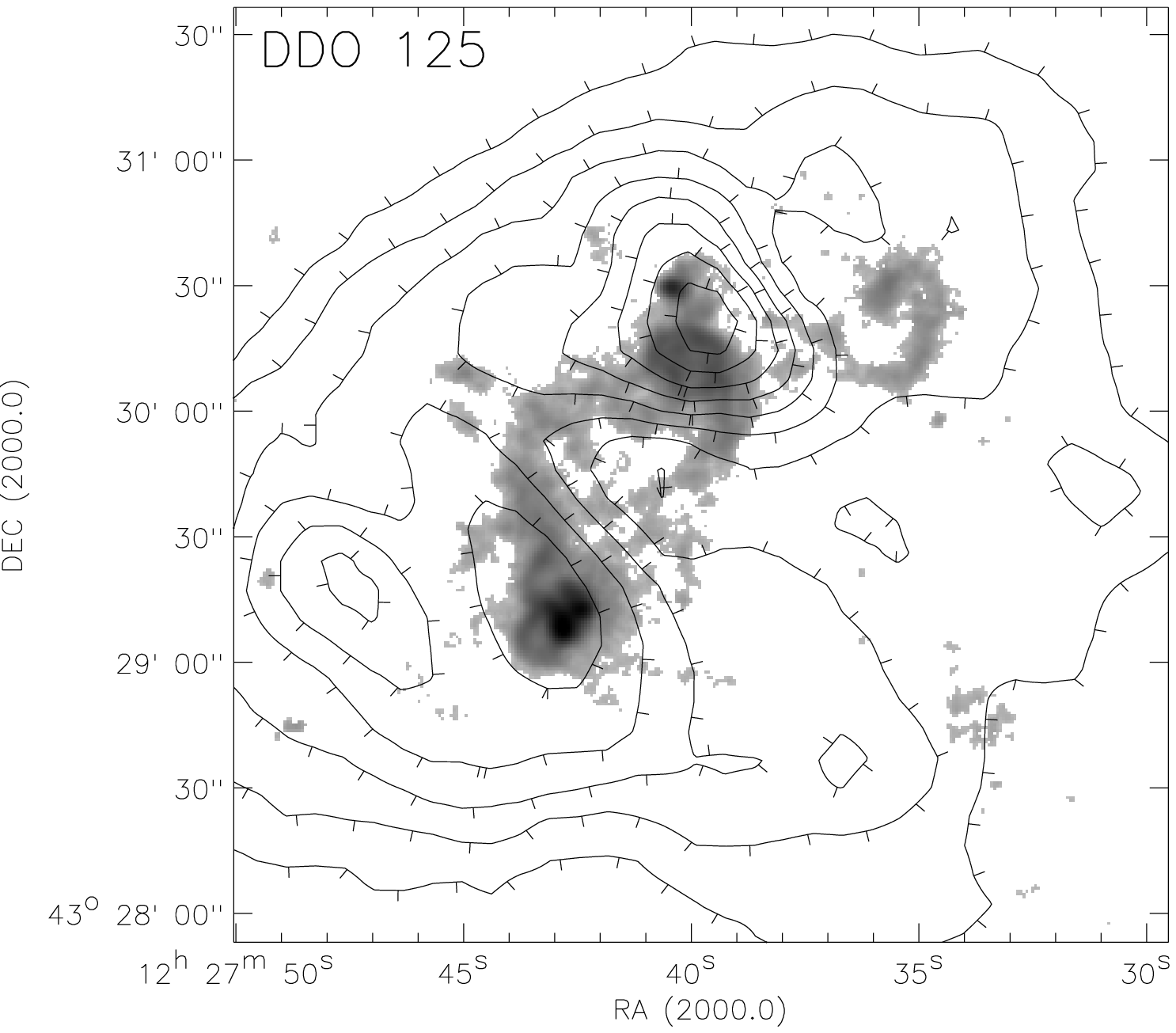}
\includegraphics[width=0.5\textwidth]{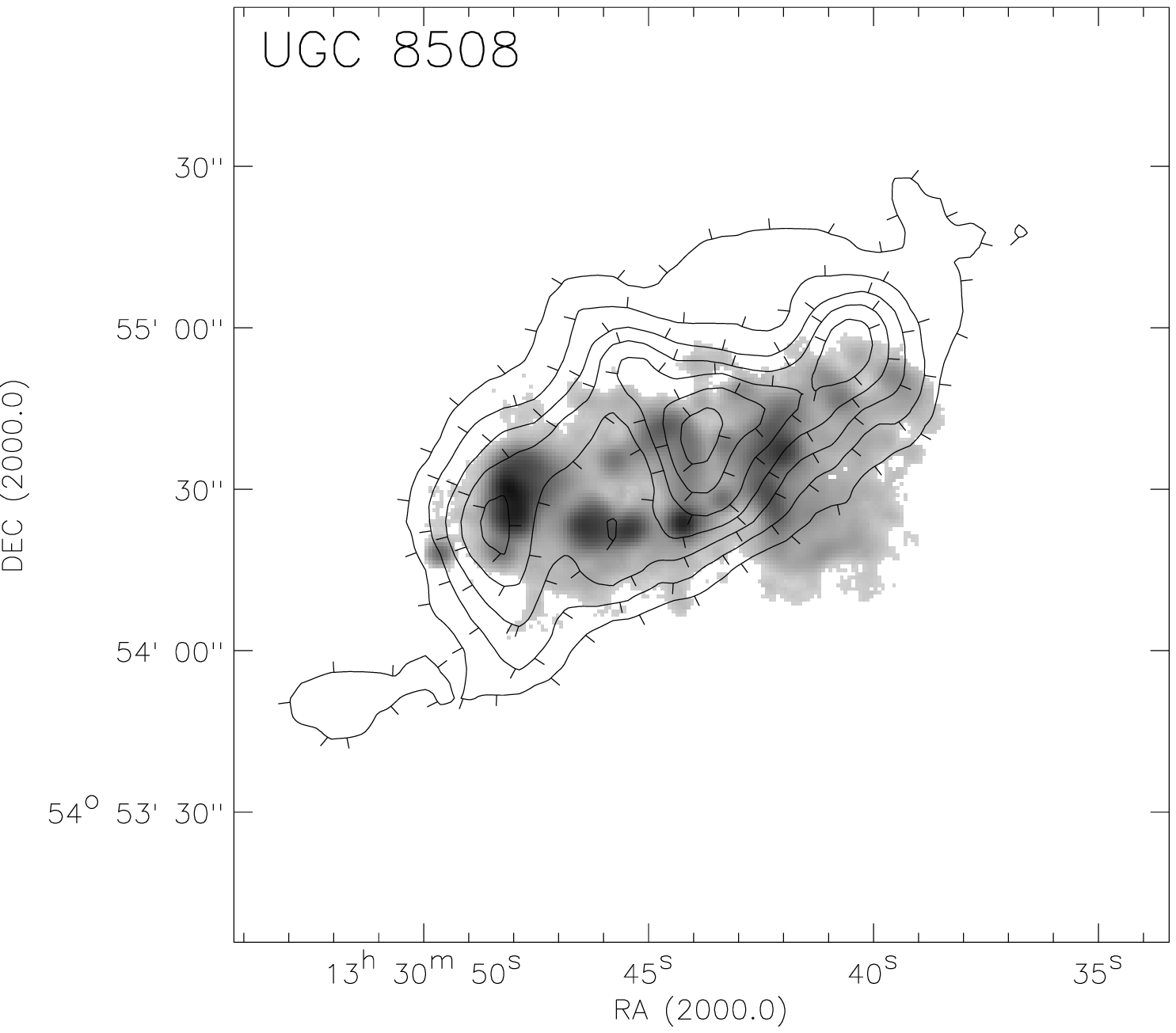}
} \caption{A comparison of surface  brightness distributions in \Ha\,
(shades of gray) and HI density (contours). The data on the
neutral hydrogen for DDO 53 are adopted from the THING survey
\citep{Oh2011}, for the other of galaxies --- from the FIGGS survey
\citep{Begum2008}.}  \label{fig_HI}
\end{figure*}

It seems that the observed distribution of points with a high velocity dispersion of
ionized gas is not bound to specific shells, but rather to the integrated  effect of
young stellar groups feedback. Meanwhile, both the photoionization radiation of OB
stars, and kinetic energy of the supernova explosions and winds of young stars lead to
an increase of chaotic, turbulent velocities of  ionized gas. At the same time, a less
dense gas has a larger amplitude of velocities than the gas in denser clouds.
The same   input of energy in the low-density medium provides higher gas motion
velocities due to the dependence of the shock wave velocity on the density of the
surrounding gas. And we have to take into account that the gas, shock-heated by stellar winds and SNe, is
only partially confined and must be leaking out of the pores in the HII shells \citep{Lopez2011}.

Given the above, to be able to interpret the observed $I-\sigma$
diagrams  of dwarf galaxies, in addition to the thin shells,
considered by \citet{MunozTunon1996} we have to bring in some more
extended structures: the HII regions, surrounded by the coronas
 of perturbed gas of low density, with low surface
brightness in the  Balmer emission  lines. This is  schematically shown in
Fig.\ref{fig_model} (a). When we look at the centers of bright HII regions,  there
prevails a contribution of the inner regions  both into \Ha\, line intensity, and into the
velocity dispersion. On the periphery, however, we mainly observe high-velocity
turbulent motions of the interstellar medium with low density, and, consequently, with
high velocity dispersion. The region of the diagram, occupied by these points has a
characteristic triangular shape because its right border is determined by the effect of
declining observed velocity spread with increasing surface brightness.

It is clear that talking about density, we should bear in mind the total density of
gas, including not only ionized, but also the molecular and neutral species. First of all, because the main part if the gas in dwarf galaxies  disc is in HI state.  Figure~\ref{fig_HI} gives a comparison of the distributions of HI and HII in several galaxies of our sample, for which \citet{Begum2008} and \citet{Oh2011} have
published maps of the distribution of neutral hydrogen. It can be clearly seen that the
bright HII regions locate in places  with high density of neutral hydrogen.
Really, there are some small-scale deviations, in particular the peaks in the
distribution of HI are often offset by a few hundred parsecs from the centres of
current star formation. This effect is well known and associated  both with the
depletion of gas the stars form from, and with the mechanical impact of young stellar
groups on the interstellar medium. See, for example, the discussions in
\citet*{Thuan2004,Simpson2011}, where the HI maps for another galaxy from our sample,
VII~Zw~403, are presented. In some cases the bright shell structures, visible in the
\Ha, lie at the inner boundary of the `holes' in the HI distribution, what is caused by  the sweep-out and depletion of neutral gas \citep[see examples
in][]{Lozinskaya2003, Begum2006}. Note that now the idea of formation of giant neutral
supershells by several generations of stars during hundreds of millions of years
becomes generally accepted. In this case,  local sites of star formation occur in the dense   walls of giant HI supershells formed by multiple generations of stars  \citep[see ][and references therein]{McQuinn2010,Warren2011}.

Therefore, a comparison of the HI distribution with our HII maps
confirms that the gas (including both neutral and ionized states) in the regions with high $\sigma$ has a low surface (and hence volume) density. \citet{Thuan2004}  explain the observed features of the distribution of surface density
and HI dispersion velocity in dwarf galaxies within the supposition that the neutral
component of the interstellar medium has two phases in an approximate pressure balance.
A `cooler' phase is characterized by a relatively small spatial scale, higher density
and low velocity dispersion, while the diffuse `warm' phase is distinguished by
increased velocity dispersion. Such ideas on {\it neutral} gas are consistent with our
explanation of the state of the {\it ionized} medium of dwarf galaxies. In this case,
the ionized gas of low density, characterized by high-velocity  turbulent motions,
is a kind of an `energy reservoir', maintaining a high velocity dispersion of the warm
phase of HI. However, speaking about the pressure balance of different  phases of gas,
it should be borne in mind that the early ideas of the dominant role of thermal
pressure are in fact simplistic views at the state of the interstellar medium, in which
an important role is played by the turbulent pressure \citep{Burkert2006}.

The question of what determines the velocity dispersion for the
points of the `horizontal branch' in the $I-\sigma$ diagram
requires a further detailed consideration.  This `branch' corresponds to
 the bright HII regions, for which $\sigma\le\sigma_m$. \citet{MunozTunon1996} refer to these areas
as the `kinematic core', in agreement with the model proposed by
\citet*{Tenorio-Tagle1993} to explain the nature of the supersonic turbulence of
ionized gas in star-forming regions. These authors believe that the mean velocity
dispersion of ionized gas in the `kinematic core' is directly related to the virial
motions, i.e. is approximately equal to the velocity dispersion of stellar population.
The latter, in turn, is determined mainly by the mass and size of the star forming
regions. Later, this model was used by \citet*{Melnick2000} to explain the physical
nature of the correlation of mean gas velocity dispersion $\sigma_m$ with the total
luminosity in the  H$\beta$ line. It should however be noted that the model by
\citet{Tenorio-Tagle1993} describes the evolution of isolated relaxed spheroidal
stellar systems. It is hence not applicable to the behaviour of gas in an entire dwarf
galaxy, the star formation regions of which are in the gravitational potential of the
disc and the dark halo. An alternative view on the nature of the H$\beta$ (or \Ha)
luminosity--$\sigma_m$ correlation is given by the authors, arguing that the main
contribution in the observed velocity dispersion of ionized gas in different types of
galaxies is provided by the current star formation \citep{Dib2006,Green2010}. A
detailed analysis of the nature of the `luminosity--ionized gas velocity dispersion'
correlation is considered in our next paper \citep{Moiseev2012}.

Finally, note another interesting feature we detected on the $I-\sigma$ diagrams for
the IC~1613 and UGC~8508 galaxies. These are points that are grouped in a horizontal
lane with a relatively large velocity dispersion  $\sigma>30-50\km$. These points are
very well detached from the other diagram areas, and belong to the isolated remnants of
supernova explosions or other expanding nebulae around the young massive objects, such
as the Wolf-Rayet stars,  ultra-bright X-ray sources,  LBV stars, etc. (see Fig.\ref{fig_model}).
Therefore, the $I-\sigma$ diagrams  can also be used to search for unique and
interesting objects in the emission galaxies.

\begin{figure}
\includegraphics[width=0.5\textwidth]{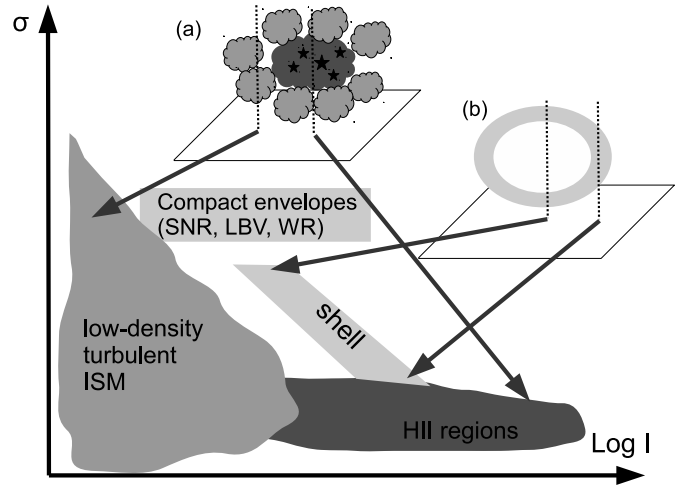}
\caption{The scheme, illustrating the location of points on the
$I-\sigma$ diagram. The insets show how we projected onto the
sky plane the surface brightness distribution and velocity dispersion
(a) from dense HII regions, surrounded by low-density gas with
considerable turbulent motions, and (b) from the expanding shell
within the model by \citet{MunozTunon1996}. The dotted line shows
the lines of sight, passing through different spatial regions.}
\label{fig_model}
\end{figure}

\section{Conclusion}
\label{sec:conclusion}

Using a scanning Fabry-Perot interferometer mounted at the 6-m
telescope of the SAO RAS we have carried out the \Ha\ line
observations of seven nearby ($D=1.8-4.3$ Mpc)  irregular dwarf
galaxies of the Local Volume. To study the features of the
distribution of radial velocity dispersion over the discs of
galaxies, characterizing the magnitude of chaotic turbulent
motions of ionized gas, we used the $I-\sigma$ diagrams. A
combination of these diagrams and  two-dimensional maps of radial
velocity dispersion has allowed to identify some common patterns,
pointing to the relation between the magnitude of chaotic motions
of gas and the ongoing processes of star formation.

Since the spatial resolution of these observations amounted to several tens of parsecs,
we tested our results using the previously obtained high-resolution data for two
galaxies of the Local Group, IC~10 and IC~1613, smoothed to the resolution of 40 pc.
The main conclusions of our work are as follows:

\begin{itemize}
\item There is a clear link between the surface brightness in the
\Ha\, line and the dispersion of line-of-sight velocities: with
decreasing surface  brightness the range of $\sigma$ values is growing, the
maximum velocity dispersion is always observed in the regions of
low surface  brightness.

\item  In  five galaxies (DDO53, DDO125, UGC 8508, UGCA 92 and VII Zw 403) we have
identified expanding  shells of ionized gas, sized $80-350$ pc, formed as a result
of collective  action of stellar groups on the gaseous medium of galaxies.
 Characteristic kinematic age of the shells is $1-4$ Myr, indicating a relation with
current star formation.

\item We demonstrate that the $I-\sigma$ diagrams may be useful
for the search of supernova remnants or other compact expanding
shells (nebulae around WR stars, etc.) in nearby galaxies. Based on $I-\sigma$ diagnostic we have found LBV candidate in
 in the UGC 8508 galaxy.

\item  The model, previously proposed by \citet{MunozTunon1996} to
explain the shapes of the $I-\sigma$ diagram  of individual star
formation regions, requires substantial additions in the case of
dwarf galaxies, where the spatial scales are substantially larger.
The most important addition here is that most of regions with high
velocity dispersion are not related to specific expanding shells,
but rather belong to the diffuse   low brightness emission,
surrounding the star formation regions. We explain this behaviour of the observed $\sigma$ distributions by the presence in HII regions of the coronas of perturbed gas
with low density and high turbulent velocities. This supposition
is consistent with current knowledge of turbulence in the
interstellar medium .

\end{itemize}

We hope that our results will be useful in the interpretation of
the ionized gas velocity dispersion maps in dwarf galaxies. All
the more so, since an increase in the number of such observations
is expected in the coming years. We consider a direct comparison
of velocity dispersion distributions for neutral and ionized
hydrogen as a challenging here, although the difference in spatial
resolution would present a certain problem.

\section*{ACKNOWLEDGMENTS}
We have made use of the NASA/IPAC Extragalactic Database (NED)
which is operated by the Jet Propulsion Laboratory, California
Institute of Technology, under the contract with the National
Aeronautics and Space Administration.  This work was supported by
the  Federal Target Programm Scientific and Scientific-Pedagogical
Cadre of Innovative Russia (contract no. 14.740.11.0800) and by
the Russian Foundation for Basic Research (project
no.~10-02-00091). AVM is also grateful for the financial support
of the `Dynasty' Foundation. We thank Prof. Jayaram Chengalur who provided us the digital maps of the
FIGGS galaxies  and anonymous Reviewer for constructive advice, which   has helped us to improve the paper.  Also we thank Sergei Fabrika, Olga Sholukhova and Vera Arkhipova for their discussion on  LBV--candidate spectrum.

Funding for the SDSS has been provided by the Alfred P. Sloan
Foundation, the Participating Institutions, the National Science
Foundation, the U.S. Department of Energy, the National
Aeronautics and Space Administration, the Japanese Monbukagakusho,
the Max Planck Society, and the Higher Education Funding Council
for England. The SDSS Web Site is http://www.sdss.org/.

The SDSS is managed by the Astrophysical Research Consortium (ARC)
for the Participating Institutions. The Participating Institutions
are The University of Chicago, Fermilab, the Institute for
Advanced Study, the Japan Participation Group, The Johns Hopkins
University, Los Alamos National Laboratory, the
Max-Planck-Institute for Astronomy (MPIA), the
Max-Planck-Institute for Astrophysics (MPA), New Mexico State
University, University of Pittsburgh, Princeton University, the
United States Naval Observatory, and the University of Washington.

{}

\appendix

\section[]{The LBV candidate in UGC 8508}
\label{sec_app}

\begin{figure*}
\centerline{
\includegraphics[height=8.5 cm]{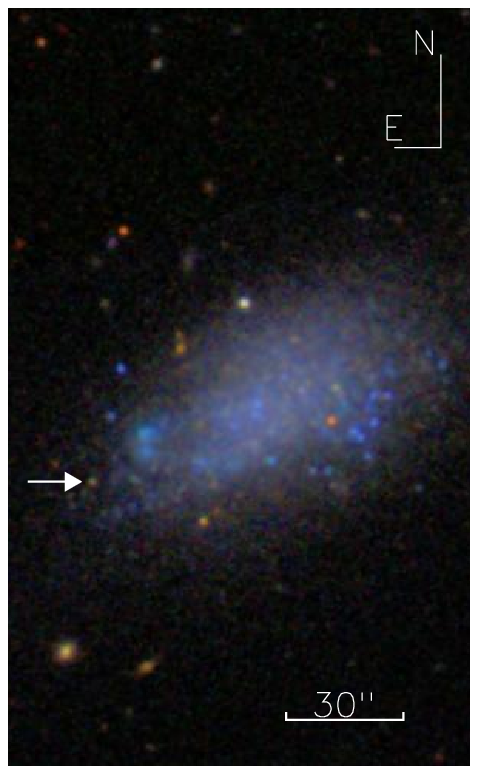}
\includegraphics[height=8.5 cm]{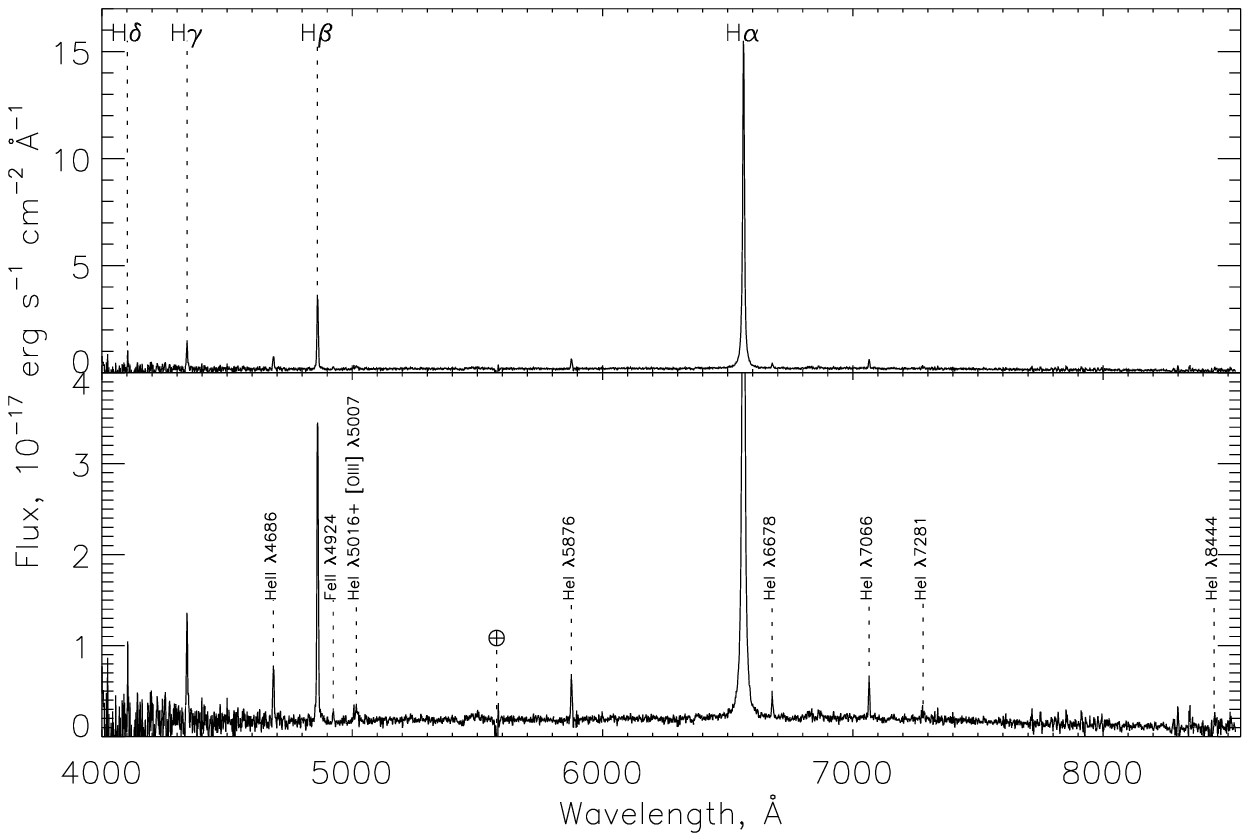}
}
\caption{UGC 8508. Colour  $gri$ image  provided by   the SDSS DR8, arrow marks the LBV candidate   J133049.80+545419.2 (the left panel). Right panels show the spectra of this star in two intensity scales. The main emission lines are labeled.}
\label{fig_LBV}
\end{figure*}

After the first version of the paper was submitted we have performed  new spectral
observations of the emission knot with a double-horned \Ha\, emission line profile in
UGC 8508. On the Sloan Digital Sky Survey  Data Release 8 (SDSS DR8) image which partly
resolved the galaxy on individual stars this knot corresponds to  one of the brightest
stars on the UGC 8508 periphery (Fig.~\ref{fig_LBV}, left). In SDSS DR8 it has name
J133049.80+545419.2.  The spectroscopic observations were carried out 2012 Feb 20/21  at
the 6-m telescope of the SAO RAS with SCORPIO-2 \citep{AfanasievMoiseev2011} focal
reducer working in the long-slit mode. We have collected 3000 sec total exposure time
with slit width of 1 arcsec  under seeing  conditions of 0.9 arcsec. The scale along
the slit amounted 0.35 arcsec/pix, the spectral resolution was about $FWHM=6.5$\AA\, in
the spectral range $3800-8500$\AA\AA. The integrated spectrum of J133049.80+545419.2 is
shown on fig.~\ref{fig_LBV} (right).

Unfortunately, the signal at the wavelength shorter than $\sim4500$\AA\, seems very
noisy,  since we used  CCD detector E2V 42-90 optimized for  ``red'' spectral
range. Nevertheless, the spectrum reveals  a lot of emission lines: strong hydrogen
Balmer series  with prominent He~I and He~II lines,  weak Fe II lines are also
detected.  All these features are characteristic for emission-line star, not for H~II
region or supernova remnant as we preliminary supposed from  FPI data. This star has a
huge  equivalent widths of the Balmer lines ($EW($\Ha$)=770$\AA), moreover the \Ha\,
line profile reveals multicomponent structure with broad wings. We fitted this profile
with three Gaussians having similar central velocities and with $FWHM\approx80,\,640$ and
$2030\km$ (the spectral resolution was taken into account). Remind that the narrower
component has also double--peaks structure according FPI observations with high
spectral resolution (sec. \ref{U8508}).    The emission nebula around the star is very
compact, with diameter smaller than 12--15 pc, because it was not spatially resolved in
the long-slit data.

Based on these spectral properties we suppose that  J133049.80+545419.2 is a massive
star with strong stellar wind like  luminous blue variables (LBVs). The observed
spectrum is similar to some known LBV--candidates, for example   J013332.64+304127.2 in
the M33 and J002020.35+591837.6 in the IC 10 galaxies \citep{Massey2007}.  The SDSS DR8
photometric catalogue provides for  J133049.80+545419.2 visible magnitudes
$g=22.37\pm0.13$, $r=21.14\pm0.06$, $i=20.63\pm0.05$ mag corresponded to absolute
magnitudes $M_g=-4.78$, $M_r=-6.01$, $M_i=-6.52$. The real luminosity of the star can
be significantly larger, if the possible   extinction in circumstellar envelope will
taken into account (the Galaxy interstellar extinction according NED is
inessential in this direction -- $A_V=0.05$).  The strong   Balmer decrement
($I_{H\alpha}/I_{H_\beta}=7.2$) can be connected with dust extinction as well as
with shock origin of the emission line spectrum.

Discussion about origin of this star lies outside the frameworks of our paper, new
detailed observations are needed. However,  the  presented results provide a very
good illustration of the power of $I-\sigma$ diagrams method that allows us to found a
new LBV--candidate outside the Local Group.

\label{lastpage}

\end{document}